\documentclass[longauth]{aa}
\usepackage{graphicx}
\usepackage{natbib}
\usepackage{lineno}
\usepackage{placeins}
\usepackage{subcaption}
\usepackage[font=small,labelfont=bf]{caption}
\usepackage{txfonts,textcomp}
\usepackage{hyperref}
\hypersetup{
    colorlinks=true,
    linkcolor=blue,
    filecolor=magenta,      
    urlcolor=blue,
    citecolor=blue
}
\makeatletter
\renewcommand*\aa@pageof{, page \thepage{} of \pageref*{LastPage}}
\makeatother
\usepackage{euclid}
\usepackage{orcidlink}
\let\orcid\orcidlink

\begin{document}

\title{\Euclid: Early Release Observations of diffuse stellar structures and globular clusters as probes of the mass assembly of galaxies in the Dorado group\thanks{This paper is published on behalf of the Euclid Consortium}}

\author{M.~Urbano\orcid{0000-0001-5640-0650}\thanks{\email{m.urbano@unistra.fr}}\inst{\ref{aff1}}
\and P.-A.~Duc\orcid{0000-0003-3343-6284}\inst{\ref{aff1}}
\and T.~Saifollahi\orcid{0000-0002-9554-7660}\inst{\ref{aff1}}
\and E.~Sola\orcid{0000-0002-2814-3578}\inst{\ref{aff2}}
\and A.~Lan\c{c}on\orcid{0000-0002-7214-8296}\inst{\ref{aff1}}
\and K.~Voggel\orcid{0000-0001-6215-0950}\inst{\ref{aff1}}
\and F.~Annibali\inst{\ref{aff3}}
\and M.~Baes\orcid{0000-0002-3930-2757}\inst{\ref{aff4}}
\and H.~Bouy\orcid{0000-0002-7084-487X}\inst{\ref{aff5},\ref{aff6}}
\and M.~Cantiello\orcid{0000-0003-2072-384X}\inst{\ref{aff7}}
\and D.~Carollo\orcid{0000-0002-0005-5787}\inst{\ref{aff8}}
\and J.-C.~Cuillandre\orcid{0000-0002-3263-8645}\inst{\ref{aff9}}
\and P.~Dimauro\orcid{0000-0001-7399-2854}\inst{\ref{aff10},\ref{aff11}}
\and P.~Erwin\orcid{0000-0003-4588-9555}\thanks{Deceased}\inst{\ref{aff12}}
\and A.~M.~N.~Ferguson\inst{\ref{aff13}}
\and R.~Habas\orcid{0000-0002-4033-3841}\inst{\ref{aff7}}
\and M.~Hilker\orcid{0000-0002-2363-5522}\inst{\ref{aff14}}
\and L.~K.~Hunt\orcid{0000-0001-9162-2371}\inst{\ref{aff15}}
\and M.~Kluge\orcid{0000-0002-9618-2552}\inst{\ref{aff12}}
\and S.~S.~Larsen\orcid{0000-0003-0069-1203}\inst{\ref{aff16}}
\and Q.~Liu\orcid{0000-0002-7490-5991}\inst{\ref{aff17}}
\and O.~Marchal\inst{\ref{aff1}}
\and F.~R.~Marleau\orcid{0000-0002-1442-2947}\inst{\ref{aff18}}
\and D.~Massari\orcid{0000-0001-8892-4301}\inst{\ref{aff3}}
\and O.~M\"uller\orcid{0000-0003-4552-9808}\inst{\ref{aff19},\ref{aff2},\ref{aff20}}
\and R.~F.~Peletier\orcid{0000-0001-7621-947X}\inst{\ref{aff21}}
\and M.~Poulain\orcid{0000-0002-7664-4510}\inst{\ref{aff22}}
\and M.~Rejkuba\orcid{0000-0002-6577-2787}\inst{\ref{aff14}}
\and M.~Schirmer\orcid{0000-0003-2568-9994}\inst{\ref{aff23}}
\and C.~Stone\orcid{0000-0002-9086-6398}\inst{\ref{aff24},\ref{aff25},\ref{aff26}}
\and R.~Z\"oller\orcid{0000-0002-0938-5686}\inst{\ref{aff27},\ref{aff12}}
\and B.~Altieri\orcid{0000-0003-3936-0284}\inst{\ref{aff28}}
\and S.~Andreon\orcid{0000-0002-2041-8784}\inst{\ref{aff29}}
\and N.~Auricchio\orcid{0000-0003-4444-8651}\inst{\ref{aff3}}
\and C.~Baccigalupi\orcid{0000-0002-8211-1630}\inst{\ref{aff30},\ref{aff8},\ref{aff31},\ref{aff32}}
\and M.~Baldi\orcid{0000-0003-4145-1943}\inst{\ref{aff33},\ref{aff3},\ref{aff34}}
\and A.~Balestra\orcid{0000-0002-6967-261X}\inst{\ref{aff35}}
\and S.~Bardelli\orcid{0000-0002-8900-0298}\inst{\ref{aff3}}
\and A.~Basset\inst{\ref{aff36}}
\and P.~Battaglia\orcid{0000-0002-7337-5909}\inst{\ref{aff3}}
\and E.~Branchini\orcid{0000-0002-0808-6908}\inst{\ref{aff37},\ref{aff38},\ref{aff29}}
\and M.~Brescia\orcid{0000-0001-9506-5680}\inst{\ref{aff39},\ref{aff40},\ref{aff41}}
\and S.~Camera\orcid{0000-0003-3399-3574}\inst{\ref{aff42},\ref{aff43},\ref{aff44}}
\and V.~Capobianco\orcid{0000-0002-3309-7692}\inst{\ref{aff44}}
\and C.~Carbone\orcid{0000-0003-0125-3563}\inst{\ref{aff45}}
\and J.~Carretero\orcid{0000-0002-3130-0204}\inst{\ref{aff46},\ref{aff47}}
\and S.~Casas\orcid{0000-0002-4751-5138}\inst{\ref{aff48},\ref{aff49}}
\and M.~Castellano\orcid{0000-0001-9875-8263}\inst{\ref{aff10}}
\and G.~Castignani\orcid{0000-0001-6831-0687}\inst{\ref{aff3}}
\and S.~Cavuoti\orcid{0000-0002-3787-4196}\inst{\ref{aff40},\ref{aff41}}
\and A.~Cimatti\inst{\ref{aff50}}
\and C.~Colodro-Conde\inst{\ref{aff51}}
\and G.~Congedo\orcid{0000-0003-2508-0046}\inst{\ref{aff13}}
\and C.~J.~Conselice\orcid{0000-0003-1949-7638}\inst{\ref{aff52}}
\and L.~Conversi\orcid{0000-0002-6710-8476}\inst{\ref{aff53},\ref{aff28}}
\and Y.~Copin\orcid{0000-0002-5317-7518}\inst{\ref{aff54}}
\and F.~Courbin\orcid{0000-0003-0758-6510}\inst{\ref{aff55},\ref{aff56}}
\and H.~M.~Courtois\orcid{0000-0003-0509-1776}\inst{\ref{aff57}}
\and H.~Degaudenzi\orcid{0000-0002-5887-6799}\inst{\ref{aff58}}
\and G.~De~Lucia\orcid{0000-0002-6220-9104}\inst{\ref{aff8}}
\and F.~Dubath\orcid{0000-0002-6533-2810}\inst{\ref{aff58}}
\and X.~Dupac\inst{\ref{aff28}}
\and S.~Dusini\orcid{0000-0002-1128-0664}\inst{\ref{aff59}}
\and M.~Farina\orcid{0000-0002-3089-7846}\inst{\ref{aff60}}
\and S.~Farrens\orcid{0000-0002-9594-9387}\inst{\ref{aff9}}
\and F.~Faustini\orcid{0000-0001-6274-5145}\inst{\ref{aff61},\ref{aff10}}
\and S.~Ferriol\inst{\ref{aff54}}
\and M.~Frailis\orcid{0000-0002-7400-2135}\inst{\ref{aff8}}
\and E.~Franceschi\orcid{0000-0002-0585-6591}\inst{\ref{aff3}}
\and M.~Fumana\orcid{0000-0001-6787-5950}\inst{\ref{aff45}}
\and S.~Galeotta\orcid{0000-0002-3748-5115}\inst{\ref{aff8}}
\and K.~George\orcid{0000-0002-1734-8455}\inst{\ref{aff27}}
\and B.~Gillis\orcid{0000-0002-4478-1270}\inst{\ref{aff13}}
\and C.~Giocoli\orcid{0000-0002-9590-7961}\inst{\ref{aff3},\ref{aff34}}
\and P.~G\'omez-Alvarez\orcid{0000-0002-8594-5358}\inst{\ref{aff62},\ref{aff28}}
\and A.~Grazian\orcid{0000-0002-5688-0663}\inst{\ref{aff35}}
\and F.~Grupp\inst{\ref{aff12},\ref{aff27}}
\and L.~Guzzo\orcid{0000-0001-8264-5192}\inst{\ref{aff63},\ref{aff29}}
\and S.~V.~H.~Haugan\orcid{0000-0001-9648-7260}\inst{\ref{aff64}}
\and J.~Hoar\inst{\ref{aff28}}
\and H.~Hoekstra\orcid{0000-0002-0641-3231}\inst{\ref{aff17}}
\and W.~Holmes\inst{\ref{aff65}}
\and F.~Hormuth\inst{\ref{aff66}}
\and A.~Hornstrup\orcid{0000-0002-3363-0936}\inst{\ref{aff67},\ref{aff68}}
\and P.~Hudelot\inst{\ref{aff69}}
\and K.~Jahnke\orcid{0000-0003-3804-2137}\inst{\ref{aff23}}
\and M.~Jhabvala\inst{\ref{aff70}}
\and E.~Keih\"anen\orcid{0000-0003-1804-7715}\inst{\ref{aff71}}
\and S.~Kermiche\orcid{0000-0002-0302-5735}\inst{\ref{aff72}}
\and B.~Kubik\orcid{0009-0006-5823-4880}\inst{\ref{aff54}}
\and M.~K\"ummel\orcid{0000-0003-2791-2117}\inst{\ref{aff27}}
\and M.~Kunz\orcid{0000-0002-3052-7394}\inst{\ref{aff73}}
\and H.~Kurki-Suonio\orcid{0000-0002-4618-3063}\inst{\ref{aff74},\ref{aff75}}
\and D.~Le~Mignant\orcid{0000-0002-5339-5515}\inst{\ref{aff76}}
\and S.~Ligori\orcid{0000-0003-4172-4606}\inst{\ref{aff44}}
\and P.~B.~Lilje\orcid{0000-0003-4324-7794}\inst{\ref{aff64}}
\and V.~Lindholm\orcid{0000-0003-2317-5471}\inst{\ref{aff74},\ref{aff75}}
\and I.~Lloro\orcid{0000-0001-5966-1434}\inst{\ref{aff77}}
\and E.~Maiorano\orcid{0000-0003-2593-4355}\inst{\ref{aff3}}
\and O.~Mansutti\orcid{0000-0001-5758-4658}\inst{\ref{aff8}}
\and S.~Marcin\inst{\ref{aff78}}
\and O.~Marggraf\orcid{0000-0001-7242-3852}\inst{\ref{aff79}}
\and K.~Markovic\orcid{0000-0001-6764-073X}\inst{\ref{aff65}}
\and M.~Martinelli\orcid{0000-0002-6943-7732}\inst{\ref{aff10},\ref{aff80}}
\and N.~Martinet\orcid{0000-0003-2786-7790}\inst{\ref{aff76}}
\and F.~Marulli\orcid{0000-0002-8850-0303}\inst{\ref{aff81},\ref{aff3},\ref{aff34}}
\and R.~Massey\orcid{0000-0002-6085-3780}\inst{\ref{aff82}}
\and E.~Medinaceli\orcid{0000-0002-4040-7783}\inst{\ref{aff3}}
\and S.~Mei\orcid{0000-0002-2849-559X}\inst{\ref{aff83},\ref{aff84}}
\and M.~Melchior\inst{\ref{aff78}}
\and M.~Meneghetti\orcid{0000-0003-1225-7084}\inst{\ref{aff3},\ref{aff34}}
\and E.~Merlin\orcid{0000-0001-6870-8900}\inst{\ref{aff10}}
\and G.~Meylan\inst{\ref{aff19}}
\and L.~Moscardini\orcid{0000-0002-3473-6716}\inst{\ref{aff81},\ref{aff3},\ref{aff34}}
\and R.~Nakajima\orcid{0009-0009-1213-7040}\inst{\ref{aff79}}
\and C.~Neissner\orcid{0000-0001-8524-4968}\inst{\ref{aff85},\ref{aff47}}
\and R.~C.~Nichol\orcid{0000-0003-0939-6518}\inst{\ref{aff86}}
\and S.-M.~Niemi\inst{\ref{aff87}}
\and C.~Padilla\orcid{0000-0001-7951-0166}\inst{\ref{aff85}}
\and S.~Paltani\orcid{0000-0002-8108-9179}\inst{\ref{aff58}}
\and F.~Pasian\orcid{0000-0002-4869-3227}\inst{\ref{aff8}}
\and K.~Pedersen\inst{\ref{aff88}}
\and W.~J.~Percival\orcid{0000-0002-0644-5727}\inst{\ref{aff89},\ref{aff90},\ref{aff91}}
\and V.~Pettorino\inst{\ref{aff87}}
\and S.~Pires\orcid{0000-0002-0249-2104}\inst{\ref{aff9}}
\and G.~Polenta\orcid{0000-0003-4067-9196}\inst{\ref{aff61}}
\and M.~Poncet\inst{\ref{aff36}}
\and L.~A.~Popa\inst{\ref{aff92}}
\and L.~Pozzetti\orcid{0000-0001-7085-0412}\inst{\ref{aff3}}
\and F.~Raison\orcid{0000-0002-7819-6918}\inst{\ref{aff12}}
\and A.~Renzi\orcid{0000-0001-9856-1970}\inst{\ref{aff93},\ref{aff59}}
\and J.~Rhodes\orcid{0000-0002-4485-8549}\inst{\ref{aff65}}
\and G.~Riccio\inst{\ref{aff40}}
\and E.~Romelli\orcid{0000-0003-3069-9222}\inst{\ref{aff8}}
\and M.~Roncarelli\orcid{0000-0001-9587-7822}\inst{\ref{aff3}}
\and E.~Rossetti\orcid{0000-0003-0238-4047}\inst{\ref{aff33}}
\and R.~Saglia\orcid{0000-0003-0378-7032}\inst{\ref{aff27},\ref{aff12}}
\and D.~Sapone\orcid{0000-0001-7089-4503}\inst{\ref{aff94}}
\and B.~Sartoris\orcid{0000-0003-1337-5269}\inst{\ref{aff27},\ref{aff8}}
\and R.~Scaramella\orcid{0000-0003-2229-193X}\inst{\ref{aff10},\ref{aff80}}
\and P.~Schneider\orcid{0000-0001-8561-2679}\inst{\ref{aff79}}
\and A.~Secroun\orcid{0000-0003-0505-3710}\inst{\ref{aff72}}
\and G.~Seidel\orcid{0000-0003-2907-353X}\inst{\ref{aff23}}
\and S.~Serrano\orcid{0000-0002-0211-2861}\inst{\ref{aff95},\ref{aff96},\ref{aff97}}
\and C.~Sirignano\orcid{0000-0002-0995-7146}\inst{\ref{aff93},\ref{aff59}}
\and L.~Stanco\orcid{0000-0002-9706-5104}\inst{\ref{aff59}}
\and J.~Steinwagner\orcid{0000-0001-7443-1047}\inst{\ref{aff12}}
\and P.~Tallada-Cresp\'{i}\orcid{0000-0002-1336-8328}\inst{\ref{aff46},\ref{aff47}}
\and A.~N.~Taylor\inst{\ref{aff13}}
\and I.~Tereno\inst{\ref{aff98},\ref{aff99}}
\and R.~Toledo-Moreo\orcid{0000-0002-2997-4859}\inst{\ref{aff100}}
\and F.~Torradeflot\orcid{0000-0003-1160-1517}\inst{\ref{aff47},\ref{aff46}}
\and I.~Tutusaus\orcid{0000-0002-3199-0399}\inst{\ref{aff101}}
\and T.~Vassallo\orcid{0000-0001-6512-6358}\inst{\ref{aff27},\ref{aff8}}
\and G.~Verdoes~Kleijn\orcid{0000-0001-5803-2580}\inst{\ref{aff21}}
\and Y.~Wang\orcid{0000-0002-4749-2984}\inst{\ref{aff102}}
\and J.~Weller\orcid{0000-0002-8282-2010}\inst{\ref{aff27},\ref{aff12}}
\and O.~R.~Williams\orcid{0000-0003-0274-1526}\inst{\ref{aff103}}
\and E.~Zucca\orcid{0000-0002-5845-8132}\inst{\ref{aff3}}
\and M.~Bolzonella\orcid{0000-0003-3278-4607}\inst{\ref{aff3}}
\and C.~Burigana\orcid{0000-0002-3005-5796}\inst{\ref{aff104},\ref{aff105}}
\and A.~Mora\orcid{0000-0002-1922-8529}\inst{\ref{aff106}}
\and V.~Scottez\inst{\ref{aff107},\ref{aff108}}}

\institute{Universit\'e de Strasbourg, CNRS, Observatoire astronomique de Strasbourg, UMR 7550, 67000 Strasbourg, France\label{aff1}
\and
Institute of Astronomy, University of Cambridge, Madingley Road, Cambridge CB3 0HA, UK\label{aff2}
\and
INAF-Osservatorio di Astrofisica e Scienza dello Spazio di Bologna, Via Piero Gobetti 93/3, 40129 Bologna, Italy\label{aff3}
\and
Sterrenkundig Observatorium, Universiteit Gent, Krijgslaan 281 S9, 9000 Gent, Belgium\label{aff4}
\and
Institut universitaire de France (IUF), 1 rue Descartes, 75231 PARIS CEDEX 05, France\label{aff5}
\and
Laboratoire d'Astrophysique de Bordeaux, CNRS and Universit\'e de Bordeaux, All\'ee Geoffroy St. Hilaire, 33165 Pessac, France\label{aff6}
\and
INAF - Osservatorio Astronomico d'Abruzzo, Via Maggini, 64100, Teramo, Italy\label{aff7}
\and
INAF-Osservatorio Astronomico di Trieste, Via G. B. Tiepolo 11, 34143 Trieste, Italy\label{aff8}
\and
Universit\'e Paris-Saclay, Universit\'e Paris Cit\'e, CEA, CNRS, AIM, 91191, Gif-sur-Yvette, France\label{aff9}
\and
INAF-Osservatorio Astronomico di Roma, Via Frascati 33, 00078 Monteporzio Catone, Italy\label{aff10}
\and
Observatorio Nacional, Rua General Jose Cristino, 77-Bairro Imperial de Sao Cristovao, Rio de Janeiro, 20921-400, Brazil\label{aff11}
\and
Max Planck Institute for Extraterrestrial Physics, Giessenbachstr. 1, 85748 Garching, Germany\label{aff12}
\and
Institute for Astronomy, University of Edinburgh, Royal Observatory, Blackford Hill, Edinburgh EH9 3HJ, UK\label{aff13}
\and
European Southern Observatory, Karl-Schwarzschild-Str.~2, 85748 Garching, Germany\label{aff14}
\and
INAF-Osservatorio Astrofisico di Arcetri, Largo E. Fermi 5, 50125, Firenze, Italy\label{aff15}
\and
Department of Astrophysics/IMAPP, Radboud University, PO Box 9010, 6500 GL Nijmegen, The Netherlands\label{aff16}
\and
Leiden Observatory, Leiden University, Einsteinweg 55, 2333 CC Leiden, The Netherlands\label{aff17}
\and
Universit\"at Innsbruck, Institut f\"ur Astro- und Teilchenphysik, Technikerstr. 25/8, 6020 Innsbruck, Austria\label{aff18}
\and
Institute of Physics, Laboratory of Astrophysics, Ecole Polytechnique F\'ed\'erale de Lausanne (EPFL), Observatoire de Sauverny, 1290 Versoix, Switzerland\label{aff19}
\and
Visiting Fellow, Clare Hall, University of Cambridge, Cambridge, UK\label{aff20}
\and
Kapteyn Astronomical Institute, University of Groningen, PO Box 800, 9700 AV Groningen, The Netherlands\label{aff21}
\and
Space physics and astronomy research unit, University of Oulu, Pentti Kaiteran katu 1, FI-90014 Oulu, Finland\label{aff22}
\and
Max-Planck-Institut f\"ur Astronomie, K\"onigstuhl 17, 69117 Heidelberg, Germany\label{aff23}
\and
Department of Physics, Universit\'{e} de Montr\'{e}al, 2900 Edouard Montpetit Blvd, Montr\'{e}al, Qu\'{e}bec H3T 1J4, Canada\label{aff24}
\and
Ciela Institute - Montr{\'e}al Institute for Astrophysical Data Analysis and Machine Learning, Montr{\'e}al, Qu{\'e}bec, Canada\label{aff25}
\and
Mila - Qu{\'e}bec Artificial Intelligence Institute, Montr{\'e}al, Qu{\'e}bec, Canada\label{aff26}
\and
Universit\"ats-Sternwarte M\"unchen, Fakult\"at f\"ur Physik, Ludwig-Maximilians-Universit\"at M\"unchen, Scheinerstrasse 1, 81679 M\"unchen, Germany\label{aff27}
\and
ESAC/ESA, Camino Bajo del Castillo, s/n., Urb. Villafranca del Castillo, 28692 Villanueva de la Ca\~nada, Madrid, Spain\label{aff28}
\and
INAF-Osservatorio Astronomico di Brera, Via Brera 28, 20122 Milano, Italy\label{aff29}
\and
IFPU, Institute for Fundamental Physics of the Universe, via Beirut 2, 34151 Trieste, Italy\label{aff30}
\and
INFN, Sezione di Trieste, Via Valerio 2, 34127 Trieste TS, Italy\label{aff31}
\and
SISSA, International School for Advanced Studies, Via Bonomea 265, 34136 Trieste TS, Italy\label{aff32}
\and
Dipartimento di Fisica e Astronomia, Universit\`a di Bologna, Via Gobetti 93/2, 40129 Bologna, Italy\label{aff33}
\and
INFN-Sezione di Bologna, Viale Berti Pichat 6/2, 40127 Bologna, Italy\label{aff34}
\and
INAF-Osservatorio Astronomico di Padova, Via dell'Osservatorio 5, 35122 Padova, Italy\label{aff35}
\and
Centre National d'Etudes Spatiales -- Centre spatial de Toulouse, 18 avenue Edouard Belin, 31401 Toulouse Cedex 9, France\label{aff36}
\and
Dipartimento di Fisica, Universit\`a di Genova, Via Dodecaneso 33, 16146, Genova, Italy\label{aff37}
\and
INFN-Sezione di Genova, Via Dodecaneso 33, 16146, Genova, Italy\label{aff38}
\and
Department of Physics "E. Pancini", University Federico II, Via Cinthia 6, 80126, Napoli, Italy\label{aff39}
\and
INAF-Osservatorio Astronomico di Capodimonte, Via Moiariello 16, 80131 Napoli, Italy\label{aff40}
\and
INFN section of Naples, Via Cinthia 6, 80126, Napoli, Italy\label{aff41}
\and
Dipartimento di Fisica, Universit\`a degli Studi di Torino, Via P. Giuria 1, 10125 Torino, Italy\label{aff42}
\and
INFN-Sezione di Torino, Via P. Giuria 1, 10125 Torino, Italy\label{aff43}
\and
INAF-Osservatorio Astrofisico di Torino, Via Osservatorio 20, 10025 Pino Torinese (TO), Italy\label{aff44}
\and
INAF-IASF Milano, Via Alfonso Corti 12, 20133 Milano, Italy\label{aff45}
\and
Centro de Investigaciones Energ\'eticas, Medioambientales y Tecnol\'ogicas (CIEMAT), Avenida Complutense 40, 28040 Madrid, Spain\label{aff46}
\and
Port d'Informaci\'{o} Cient\'{i}fica, Campus UAB, C. Albareda s/n, 08193 Bellaterra (Barcelona), Spain\label{aff47}
\and
Institute for Theoretical Particle Physics and Cosmology (TTK), RWTH Aachen University, 52056 Aachen, Germany\label{aff48}
\and
Institute of Cosmology and Gravitation, University of Portsmouth, Portsmouth PO1 3FX, UK\label{aff49}
\and
Dipartimento di Fisica e Astronomia "Augusto Righi" - Alma Mater Studiorum Universit\`a di Bologna, Viale Berti Pichat 6/2, 40127 Bologna, Italy\label{aff50}
\and
Instituto de Astrof\'isica de Canarias, Calle V\'ia L\'actea s/n, 38204, San Crist\'obal de La Laguna, Tenerife, Spain\label{aff51}
\and
Jodrell Bank Centre for Astrophysics, Department of Physics and Astronomy, University of Manchester, Oxford Road, Manchester M13 9PL, UK\label{aff52}
\and
European Space Agency/ESRIN, Largo Galileo Galilei 1, 00044 Frascati, Roma, Italy\label{aff53}
\and
Universit\'e Claude Bernard Lyon 1, CNRS/IN2P3, IP2I Lyon, UMR 5822, Villeurbanne, F-69100, France\label{aff54}
\and
Institut de Ci\`{e}ncies del Cosmos (ICCUB), Universitat de Barcelona (IEEC-UB), Mart\'{i} i Franqu\`{e}s 1, 08028 Barcelona, Spain\label{aff55}
\and
Instituci\'o Catalana de Recerca i Estudis Avan\c{c}ats (ICREA), Passeig de Llu\'{\i}s Companys 23, 08010 Barcelona, Spain\label{aff56}
\and
UCB Lyon 1, CNRS/IN2P3, IUF, IP2I Lyon, 4 rue Enrico Fermi, 69622 Villeurbanne, France\label{aff57}
\and
Department of Astronomy, University of Geneva, ch. d'Ecogia 16, 1290 Versoix, Switzerland\label{aff58}
\and
INFN-Padova, Via Marzolo 8, 35131 Padova, Italy\label{aff59}
\and
INAF-Istituto di Astrofisica e Planetologia Spaziali, via del Fosso del Cavaliere, 100, 00100 Roma, Italy\label{aff60}
\and
Space Science Data Center, Italian Space Agency, via del Politecnico snc, 00133 Roma, Italy\label{aff61}
\and
FRACTAL S.L.N.E., calle Tulip\'an 2, Portal 13 1A, 28231, Las Rozas de Madrid, Spain\label{aff62}
\and
Dipartimento di Fisica "Aldo Pontremoli", Universit\`a degli Studi di Milano, Via Celoria 16, 20133 Milano, Italy\label{aff63}
\and
Institute of Theoretical Astrophysics, University of Oslo, P.O. Box 1029 Blindern, 0315 Oslo, Norway\label{aff64}
\and
Jet Propulsion Laboratory, California Institute of Technology, 4800 Oak Grove Drive, Pasadena, CA, 91109, USA\label{aff65}
\and
Felix Hormuth Engineering, Goethestr. 17, 69181 Leimen, Germany\label{aff66}
\and
Technical University of Denmark, Elektrovej 327, 2800 Kgs. Lyngby, Denmark\label{aff67}
\and
Cosmic Dawn Center (DAWN), Denmark\label{aff68}
\and
Institut d'Astrophysique de Paris, UMR 7095, CNRS, and Sorbonne Universit\'e, 98 bis boulevard Arago, 75014 Paris, France\label{aff69}
\and
NASA Goddard Space Flight Center, Greenbelt, MD 20771, USA\label{aff70}
\and
Department of Physics and Helsinki Institute of Physics, Gustaf H\"allstr\"omin katu 2, 00014 University of Helsinki, Finland\label{aff71}
\and
Aix-Marseille Universit\'e, CNRS/IN2P3, CPPM, Marseille, France\label{aff72}
\and
Universit\'e de Gen\`eve, D\'epartement de Physique Th\'eorique and Centre for Astroparticle Physics, 24 quai Ernest-Ansermet, CH-1211 Gen\`eve 4, Switzerland\label{aff73}
\and
Department of Physics, P.O. Box 64, 00014 University of Helsinki, Finland\label{aff74}
\and
Helsinki Institute of Physics, Gustaf H{\"a}llstr{\"o}min katu 2, University of Helsinki, Helsinki, Finland\label{aff75}
\and
Aix-Marseille Universit\'e, CNRS, CNES, LAM, Marseille, France\label{aff76}
\and
NOVA optical infrared instrumentation group at ASTRON, Oude Hoogeveensedijk 4, 7991PD, Dwingeloo, The Netherlands\label{aff77}
\and
University of Applied Sciences and Arts of Northwestern Switzerland, School of Engineering, 5210 Windisch, Switzerland\label{aff78}
\and
Universit\"at Bonn, Argelander-Institut f\"ur Astronomie, Auf dem H\"ugel 71, 53121 Bonn, Germany\label{aff79}
\and
INFN-Sezione di Roma, Piazzale Aldo Moro, 2 - c/o Dipartimento di Fisica, Edificio G. Marconi, 00185 Roma, Italy\label{aff80}
\and
Dipartimento di Fisica e Astronomia "Augusto Righi" - Alma Mater Studiorum Universit\`a di Bologna, via Piero Gobetti 93/2, 40129 Bologna, Italy\label{aff81}
\and
Department of Physics, Institute for Computational Cosmology, Durham University, South Road, Durham, DH1 3LE, UK\label{aff82}
\and
Universit\'e Paris Cit\'e, CNRS, Astroparticule et Cosmologie, 75013 Paris, France\label{aff83}
\and
CNRS-UCB International Research Laboratory, Centre Pierre Binetruy, IRL2007, CPB-IN2P3, Berkeley, USA\label{aff84}
\and
Institut de F\'{i}sica d'Altes Energies (IFAE), The Barcelona Institute of Science and Technology, Campus UAB, 08193 Bellaterra (Barcelona), Spain\label{aff85}
\and
School of Mathematics and Physics, University of Surrey, Guildford, Surrey, GU2 7XH, UK\label{aff86}
\and
European Space Agency/ESTEC, Keplerlaan 1, 2201 AZ Noordwijk, The Netherlands\label{aff87}
\and
DARK, Niels Bohr Institute, University of Copenhagen, Jagtvej 155, 2200 Copenhagen, Denmark\label{aff88}
\and
Waterloo Centre for Astrophysics, University of Waterloo, Waterloo, Ontario N2L 3G1, Canada\label{aff89}
\and
Department of Physics and Astronomy, University of Waterloo, Waterloo, Ontario N2L 3G1, Canada\label{aff90}
\and
Perimeter Institute for Theoretical Physics, Waterloo, Ontario N2L 2Y5, Canada\label{aff91}
\and
Institute of Space Science, Str. Atomistilor, nr. 409 M\u{a}gurele, Ilfov, 077125, Romania\label{aff92}
\and
Dipartimento di Fisica e Astronomia "G. Galilei", Universit\`a di Padova, Via Marzolo 8, 35131 Padova, Italy\label{aff93}
\and
Departamento de F\'isica, FCFM, Universidad de Chile, Blanco Encalada 2008, Santiago, Chile\label{aff94}
\and
Institut d'Estudis Espacials de Catalunya (IEEC),  Edifici RDIT, Campus UPC, 08860 Castelldefels, Barcelona, Spain\label{aff95}
\and
Satlantis, University Science Park, Sede Bld 48940, Leioa-Bilbao, Spain\label{aff96}
\and
Institute of Space Sciences (ICE, CSIC), Campus UAB, Carrer de Can Magrans, s/n, 08193 Barcelona, Spain\label{aff97}
\and
Departamento de F\'isica, Faculdade de Ci\^encias, Universidade de Lisboa, Edif\'icio C8, Campo Grande, PT1749-016 Lisboa, Portugal\label{aff98}
\and
Instituto de Astrof\'isica e Ci\^encias do Espa\c{c}o, Faculdade de Ci\^encias, Universidade de Lisboa, Tapada da Ajuda, 1349-018 Lisboa, Portugal\label{aff99}
\and
Universidad Polit\'ecnica de Cartagena, Departamento de Electr\'onica y Tecnolog\'ia de Computadoras,  Plaza del Hospital 1, 30202 Cartagena, Spain\label{aff100}
\and
Institut de Recherche en Astrophysique et Plan\'etologie (IRAP), Universit\'e de Toulouse, CNRS, UPS, CNES, 14 Av. Edouard Belin, 31400 Toulouse, France\label{aff101}
\and
Infrared Processing and Analysis Center, California Institute of Technology, Pasadena, CA 91125, USA\label{aff102}
\and
Centre for Information Technology, University of Groningen, P.O. Box 11044, 9700 CA Groningen, The Netherlands\label{aff103}
\and
INAF, Istituto di Radioastronomia, Via Piero Gobetti 101, 40129 Bologna, Italy\label{aff104}
\and
INFN-Bologna, Via Irnerio 46, 40126 Bologna, Italy\label{aff105}
\and
Aurora Technology for European Space Agency (ESA), Camino bajo del Castillo, s/n, Urbanizacion Villafranca del Castillo, Villanueva de la Ca\~nada, 28692 Madrid, Spain\label{aff106}
\and
Institut d'Astrophysique de Paris, 98bis Boulevard Arago, 75014, Paris, France\label{aff107}
\and
ICL, Junia, Universit\'e Catholique de Lille, LITL, 59000 Lille, France\label{aff108}}       

 \abstract{
Deep surveys have helped to unveil the history of past and present galaxy mergers, and, in  particular, uncovering their tidal debris and co-located globular clusters (GCs). \Euclid's unique combination of capabilities (spatial resolution, depth, and wide sky coverage) will make it a groundbreaking tool for galactic archaeology in the Local Universe, bringing low-surface-brightness (LSB) science into the era of large-scale astronomical surveys. \Euclid's Early Release Observations (ERO) demonstrate this potential with a field of view that includes several galaxies in the Dorado group. In this paper, we aim to derive from this image a mass assembly scenario for its main galaxies: NGC\,1549, NGC\,1553, and NGC\,1546. We detected their internal and external diffuse structures, and identified candidate GCs. By analysing the colours and distributions of the diffuse structures and candidate GCs, we can place constraints on the galaxies' mass assembly and merger histories. The results demonstrate that feature morphology, surface brightness, colours, and GC density profiles are consistent with galaxies that have undergone different merger scenarios. We classify NGC\,1549 as a pure elliptical galaxy that has undergone a major merger. NGC\,1553 appears to have recently transitioned from a late-type galaxy to early type, after a series of radial minor to intermediate mergers. NGC\,1546 is a rare specimen of galaxy with an undisturbed disk and a prominent diffuse stellar halo, which we infer has been fed by minor mergers and then disturbed by the tidal effect from NGC\,1553. Finally, we identify limitations specific to the observing conditions of this ERO, in particular, stray light in the visible and persistence in the near-infrared bands. Once these issues are addressed and the extended emission from LSB objects is preserved by the data-processing pipeline, the Euclid Wide Survey will allow for studies of the Local Universe to be extended to statistical ensembles over a large part of the extragalactic sky.
}

    \keywords{Galaxies: interactions; Galaxies: structure; Galaxies: groups: individual: Dorado; Galaxies: star clusters: general.}

   \titlerunning{\Euclid view of diffuse stellar structures and globular clusters in the Dorado group of galaxies }
   \authorrunning{Mathias Urbano et al}
   
   \maketitle

\section{\label{sc:Intro}Introduction}

The dark energy, cold dark matter standard model, coupled with  hierarchical mass assembly theory, suggests that galaxies originate from the agglomeration of baryonic matter within small and low-mass dark matter halos in the early Universe \citep[e.g.][]{1986ApJ...303...39D,1999ApJ...522...82K,1999ApJ...524L..19M}. According to this paradigm, one of the galaxy mass accretion channels is via the mergers they have undergone during their history \citep[e.g.][]{2014ARA&A..52..291C, 2015ARA&A..53...51S}. These events are known to have several repercussions on the resultant galaxies. Key outcomes of such mergers could include an increase in the activity of an active galactic nucleus \citep[e.g.][]{1996ARA&A..34..749S,2019MNRAS.487.2491E}, interactions between the central black holes of the progenitors leading to a binary and, possibly, a fusion \citep[e.g.][]{2023ApJ...942L..24K,2018MNRAS.479.3952B}, and the initiation of starbursts due to the collision of gas clouds from the merging galaxies \citep[e.g.][]{2004MNRAS.350..798B,2009ApJ...694L.123K,2009PASJ...61..481S,2022MNRAS.516.4922R}. These processes often result in morphological changes, primarily due to a violent relaxation of the stars and tidal effects. Those tidal effects produce a wealth of debris and extended stellar structures around galaxies.

Among those debris, tidal tails, which are large elongations on either side of a galaxy, are the most well known  through particularly illustrative systems such as the Antennae galaxies \citep[e.g.][]{2018MNRAS.475.3934L}. Such structures appear during a major merger. Tails can appear as bridges between two interacting galaxies in the early phase of the merger, or less elongated `plumes' when the progenitors are of early type \citep[e.g.][]{Arp_1966,Toomre_and_Toomre_1972,1995AAS...186.3909M}. Stellar streams are the tidal tails of satellites which have been disrupted and then accreted by their host galaxy. They are thus witnesses of minor mergers \citep[e.g.][]{Bullock_and_Johnston_2005, 2006ApJ...642L.137B, 2021ApJ...914..123I,2022MNRAS.516.5331M}. Finally, shells are shaped like arcs, usually sharing the same centre as the galaxy. They form during radial collisions, when galactic material is expelled radially \citep[e.g.][]{Quinn_1984,1990dig..book...72P}. Cosmological simulations show that tidal features disappear by phase-mixing after two to eight billion years \citep[e.g.][]{2018MNRAS.480.1715P,2019A&A...632A.122M}.

Diffuse stellar features, including extended halos, can be traced by individual compact sources that are related to them: red giant branch and asymptotic giant branch  stars, H\textsc{ii} regions, planetary nebulae, and globular clusters (GCs), enabling their chemo-dynamical mapping \citep[e.g.][]{2000AJ....119..162C,2003ApJ...582..170D,2009MNRAS.394.1249C,2012MNRAS.426.1475U,2013MNRAS.428..389P,2014A&A...562A..73G,2016A&A...586A.102V,2018A&A...616A..74K,2022A&A...657A..41R,2022A&A...663A..12H}. Recent studies observed GC alignment and dynamics along tidal features for various galactic systems in the Local Universe, including M31 \citep[]{2010ApJ...717L..11M,2013ApJ...768L..33V,2014MNRAS.442.2165H,2019Natur.574...69M},
NGC\,4651 \citep[]{2014MNRAS.442.3544F}, NGC\,0474 \citep {2017ApJ...835..123L}, and
NGC\,5128 \citep[]{2021ApJ...914...16H}. These works have led to the detection of several GCs associated with tails, streams, and shells. Nevertheless, the use of GC clustering for detecting tidal features remains to be proven.

For more distant Local Universe galaxies not resolved into stars, tidal features appear in the form of diffuse structures that may be faint in terms of stellar surface brightness. Noteworthy contributions for unveiling the low-surface-brightness (LSB) Universe include both amateur \citep[e.g.][]{2010AJ....140..962M,2015AstBu..70..379K,2016A&A...588A..89J,2020MNRAS.494.1751M} and professional facilities \citep[e.g.][]{2014ApJ...787L..37M,2018ApJ...857..104G,2019A&A...621A.133B,2019A&A...624L...6M,2022ApJ...933...47C,2023ApJS..265...57P}. Examples worth mentioning are the Dragonfly telephoto array \citep[e.g.][]{2014PASP..126...55A}, the Hyper Suprime-Cam Subaru Strategic Program (HSC-SSP) \citep[e.g.][]{10.1093/pasj/psx066}, the Large Binocular Telescope (Smallest Scale of Hierarchy survey: \citealp{2020MNRAS.491.5101A}; LBT Imaging of Galactic Halos and Tidal Structures: \citealp{2021A&A...654A..40T}), the Sloan Digital Sky Survey \citep[SDSS; e.g.][]{2000AJ....120.1579Y,2006ApJ...642L.137B,2018A&A...614A.143M}, the Dark Energy Spectroscopic Instrument (DESI) Legacy imaging surveys (e.g. the Dark Energy Camera Legacy Survey [DECaLS]; \citealt{2019AJ....157..168D}), the 2m Fraunhofer Wendelstein Telescope \citep{2014SPIE.9145E..2DH,2020ApJS..247...43K,2024ApJS..271...52Z}, and the VLT Survey Telescope (VST) Early-type GAlaxy Survey (VEGAS; \citealt{2015A&A...581A..10C}) as well as research conducted at the Canada-France-Hawaii Telescope (Mass Assembly of early-Type GaLAxies with their fine  Structures, MATLAS: \citealp{2015MNRAS.446..120D,Bilek_et_al_2020}, Canada-France Imaging Survey CFIS: \citealp{2017ApJ...848..128I}, Next Generation Virgo Cluster Survey NGVS: \citealp{2012ApJS..200....4F}). Several future projects will be compatible with LSB studies, taking this science into the era of large-scale surveys. These include the \textit{Vera Rubin} Observatory \citep{2019ApJ...873..111I,2020arXiv200111067B},  \textit{Nancy Roman Grace} Space Telescope \citep{2023AAS...24210102K,2023arXiv230609414M}, and ARRAKIHS \citep{2024eas..conf.1990G}.

\Euclid \citep{2025A&A...697A...1E}, launched in July 2023, is leading a six-year cosmological mission to observe the extragalactic sky in a unique combination of spatial resolution, coverage, photometric bands, and depth. Its capabilities, particularly in terms of its detection limits in surface brightness and its high image resolution associated with a sharp point spread function (PSF), offer promising prospects of both tracing tidal features \citep[]{Borlaff-EP16} and studying the distribution of GCs \citep[e.g.][]{2021sf2a.conf..447L,Euclid_Voggel}. The Euclid Wide Survey (EWS; \citealt{Scaramella-EP1}) thus raises the possibility to extend the study of both tidal features and GCs in numerous galactic systems and across an extensive area of the sky (nearly $14\,000 \text{ deg}^{2}$).

\Euclid's Early Release Observations (ERO) programme \citep{2024arXiv240513496C} managed by the European Space Agency (ESA) provides a new perspective on numerous objects within the Local Universe, showcasing this telescope's potential in LSB science and galactic archaeology across various scales. It spans from detecting tidal tails around Milky Way GCs \citep{2025A&A...697A...8M} and dwarf galaxies of the Perseus galaxy cluster \citep{2025A&A...697A..12M} to investigating the intricacies of intra-cluster light \citep{2025A&A...697A..13K}. Among the images acquired as part of the ERO programme, one pointing is well suited for investigation of surroundings of massive galaxies in Dorado, a rich galaxy group located at a distance of 17.7 Mpc \citep[e.g.][]{1982ApJ...257..423H,2006MNRAS.372.1856F,2017ApJ...843...16K}. In the observed field of view (FoV), several early-type galaxies (ETGs; NGC\,1549, NGC\,1553, and NGC\,1546) are particularly striking showcases of tidal distortions. In particular, NGC\,1549 and NGC\,1553 are known to form an interacting pair, and NGC\,1546 displays also a system of tidal features \citep[e.g.][]{Malin_and_Carter_1983,1987IAUS..127..469D,1993ASPC...48..629W}. 

Already on photographic plates, the NGC\,1553\,/\,NGC\,1549 (S0\,/\,E) pair was identified as hosting streamer and ripple features located at around and beyond a radius of $5\,\arcminute$ from the centre of these galaxies \citep{1975IAUS...69..367F,1987MNRAS.226..747J,1990AJ.....99.1100B}. The study of the central regions of those galaxies, presented in \cite{2023MNRAS.522.2207R}, reveals that NGC\,1553 has a Low-Ionization Nuclear Emission-line Region (LINER) featuring a broad component, an X-ray core \citep{2017ApJ...835..223S,2020ApJ...900..124B}, and coronal [Ne\,\textsc{v}] emission in the mid-infrared \citep{2013MNRAS.432..374R}.
The Sa galaxy NGC\,1546 \citep{2014A&A...562A.121C} is located at a projected distance about 140\,kpc from the pair and its tidal feature system has not been described in detail.

\cite{2020A&A...643A.176R,2021JApA...42...31R,2022A&A...664A.192R} investigated the star formation of the Dorado galaxies, comparing  their emission in H$\alpha$ (Las Campanas Observatory H$\alpha$[\ion{N}{ii}] narrow bands observations) and far-ultraviolet (Astrosat-UVIT FUV.CaF2 observations) with earlier research on their emission in H\textsc{i} \citep{1991rc3..book.....D,2005MNRAS.356...77K}, respectively. Those studies clarify why, among the three ETGs, only NGC\,1546 has H\textsc{i}: dissipative events would have initiated star formation and exhausted the H\textsc{i} in NGC\,1549 as in NGC\,1553, whose far-ultraviolet emission are each associated with a disk structure. Additionally, far-ultraviolet imaging reveal NGC\,1549, NGC\,1553, and NGC\,1546 inner and resonance rings and H\textsc{ii} regions for NGC\,1546 and NGC\,1553. Unlike NGC 1549, which appears as a purely quenched elliptical galaxy, NGC\,1553 has a star-formation rate above the average for ETGs, and presents star-forming clumps and rotation. Finally, \textit{Hubble} Space Telescope and recent JWST images of NGC\,1546 revealed its flocculent spiral galaxy nature.
 
\begin{figure*}[htbp]
  \centering
  \includegraphics[width=\textwidth,trim={0cm 1cm 0cm 0cm},clip]{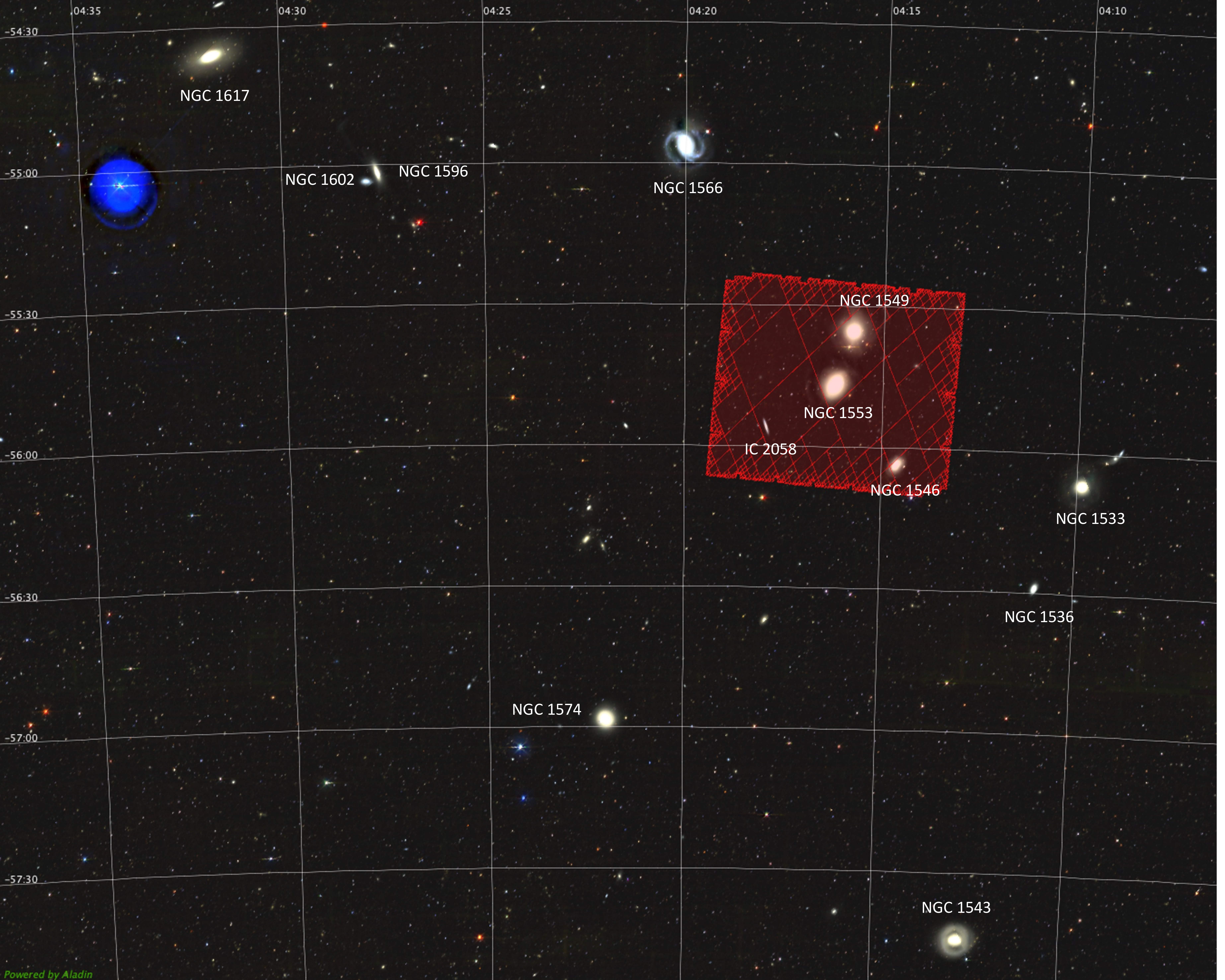}
  \caption{DECaLS colour image of the Dorado group. The brightest members are labeled in the figure. The HEALPix Multi-Order Coverage map \citep{2014ivoa.spec.0602F} of the \Euclid ERO-D FoV is added to the image as a  red-transparent overlay. The coordinates are given in RA, Dec.}
  \label{fig:dorado_coverage}
\end{figure*}

\begin{figure*}[htbp]
  \centering
  \includegraphics[width=0.89\textwidth,trim={0 0.25cm 0 0},clip]{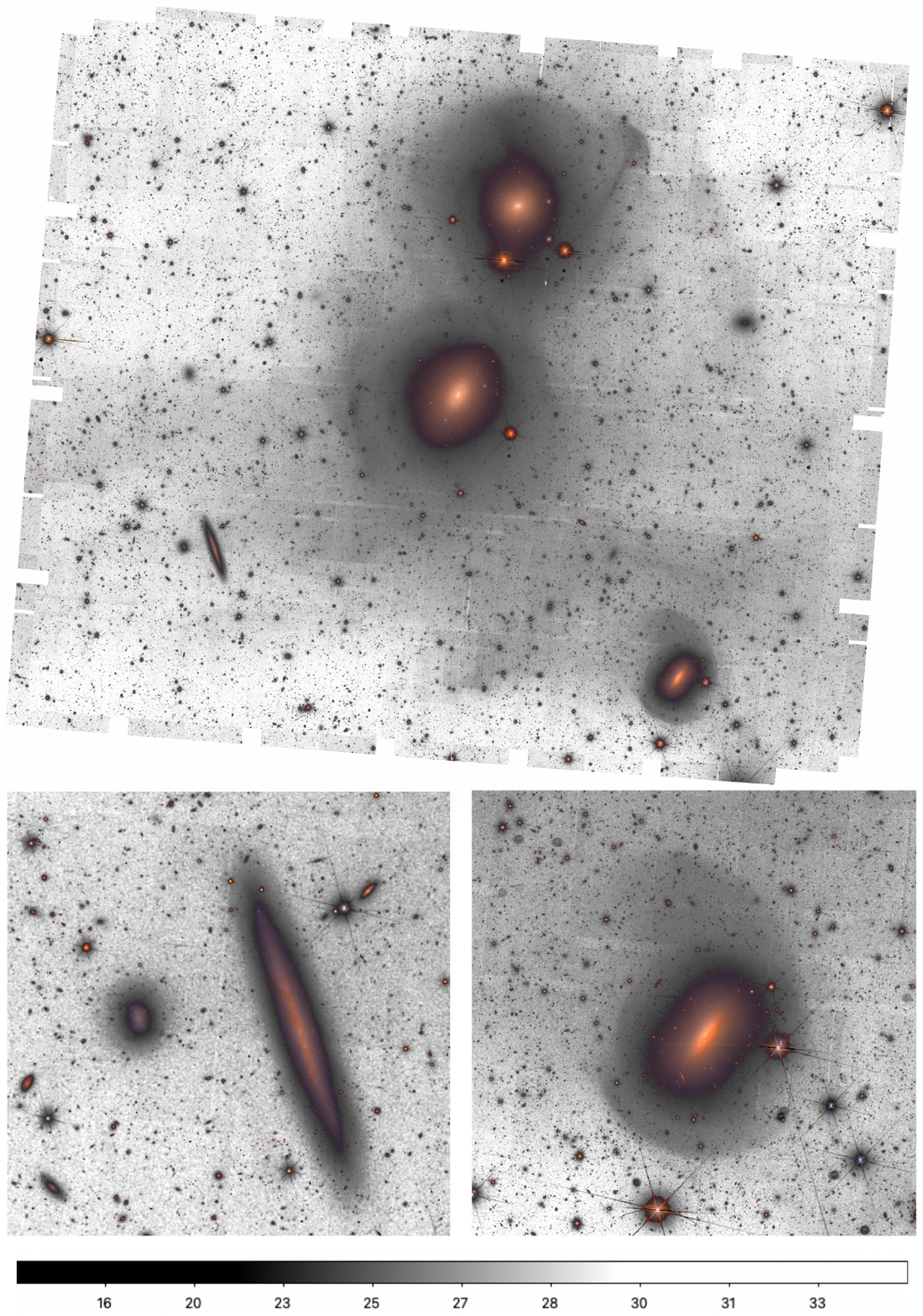}
  \caption{\Euclid surface brightness maps in the $\IE$ band with a scale in $\text{mag\,arcsec}^{-2}$ indicated to the bottom. Colour images made by co-adding the $\IE$, the $\YE$, and the $\HE$ bands using \texttt{Astropy} \citep{astropy:2013,astropy:2018,astropy:2022} are super-imposed in the inner regions of the main galaxies.  \emph{Top}: Whole FoV ($\approx56\,\text{arcmin}^2$) of the ERO-D observations. 
  \emph{Lower-left}: $300\arcsec \times 300\arcsec$ cutout around IC\,2058 and the dwarf galaxy PGC\,75125. \emph{Lower-right}: $600\arcsec \times 600\arcsec$ cutout around NGC\,1546. In this figure, north is up, east is to the left.}
  \label{fig:dorado_general}
\end{figure*}

Several hypotheses have been proposed to explain the position of these galaxies in the structure of the Dorado group (and its connection with the nearby clusters, Fornax and Eridanus; \citealp[e.g.][]{2024A&A...690A..92R}). \cite{1993A&AS..100...47G} considered that NGC\,1549 was associated with an independent group centred around NGC\,1553, while later \cite{2005MNRAS.356...77K} estimated both galaxies to be part of a group with NGC\,1566 at its core. Also, \cite{2011MNRAS.412.2498M} envisioned that NGC\,1553 is one of the centres of two independent groups forming Dorado. Finally, \cite{2002AJ....124.2471I} presents NGC\,1549, NGC\,1553, and NGC\,1546 as forming, with the edge-on late-type galaxy (LTG) IC\,2058, the compact group SCG\,0414$-$5559 at the barycentre of Dorado.

This paper focuses on the merging histories and current interactions of these galaxies, which are studied based on the analysis of their LSB stellar structures together with their GC populations. This study allows for the evaluation of the substantial benefits and limitations of \Euclid for detection of LSB Local Universe objects that are not resolved into individual stars. The article is organised as follows. The \Euclid ERO Dorado (ERO-D) data and the methods employed to investigate the galaxies' past history are described in Sect. \ref{Data_Methods}. Section \ref{sc:Results} outlines the results associated with these analyses, with their limitations discussed in Sect. \ref{sc:Discussion}. A proposition of recent merger history for the ERO-D galaxies is also provided in this section, as well as prospects on \Euclid detection. Our findings are summarised in Sect. \ref{Summary}.

\begin{figure}[ht!]
  \centering
    \includegraphics[width=\linewidth,trim=2cm 8cm 2cm 5cm, clip]{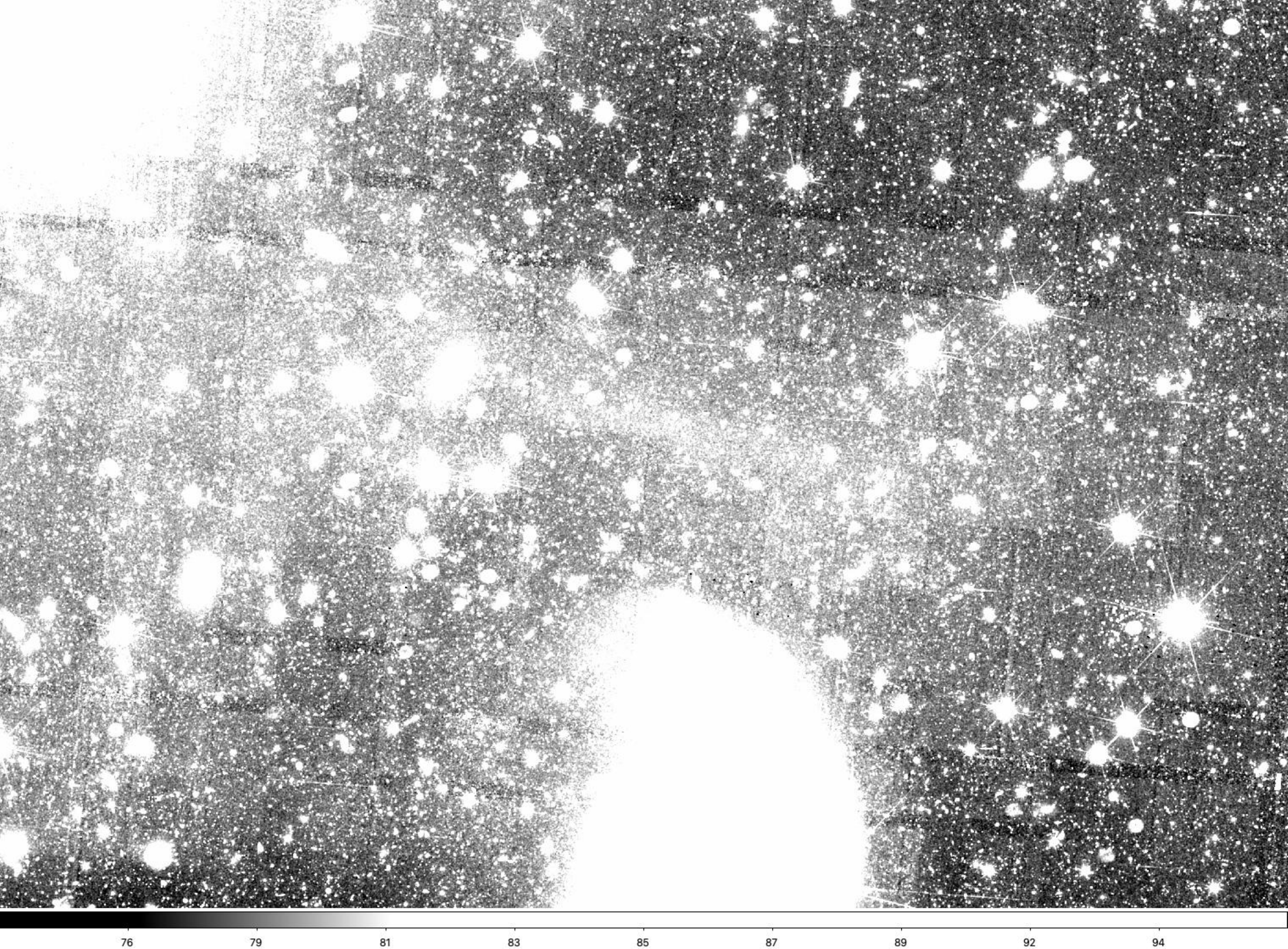}
    \includegraphics[width=\linewidth,trim=5cm 20cm 5cm 12.5cm, clip]{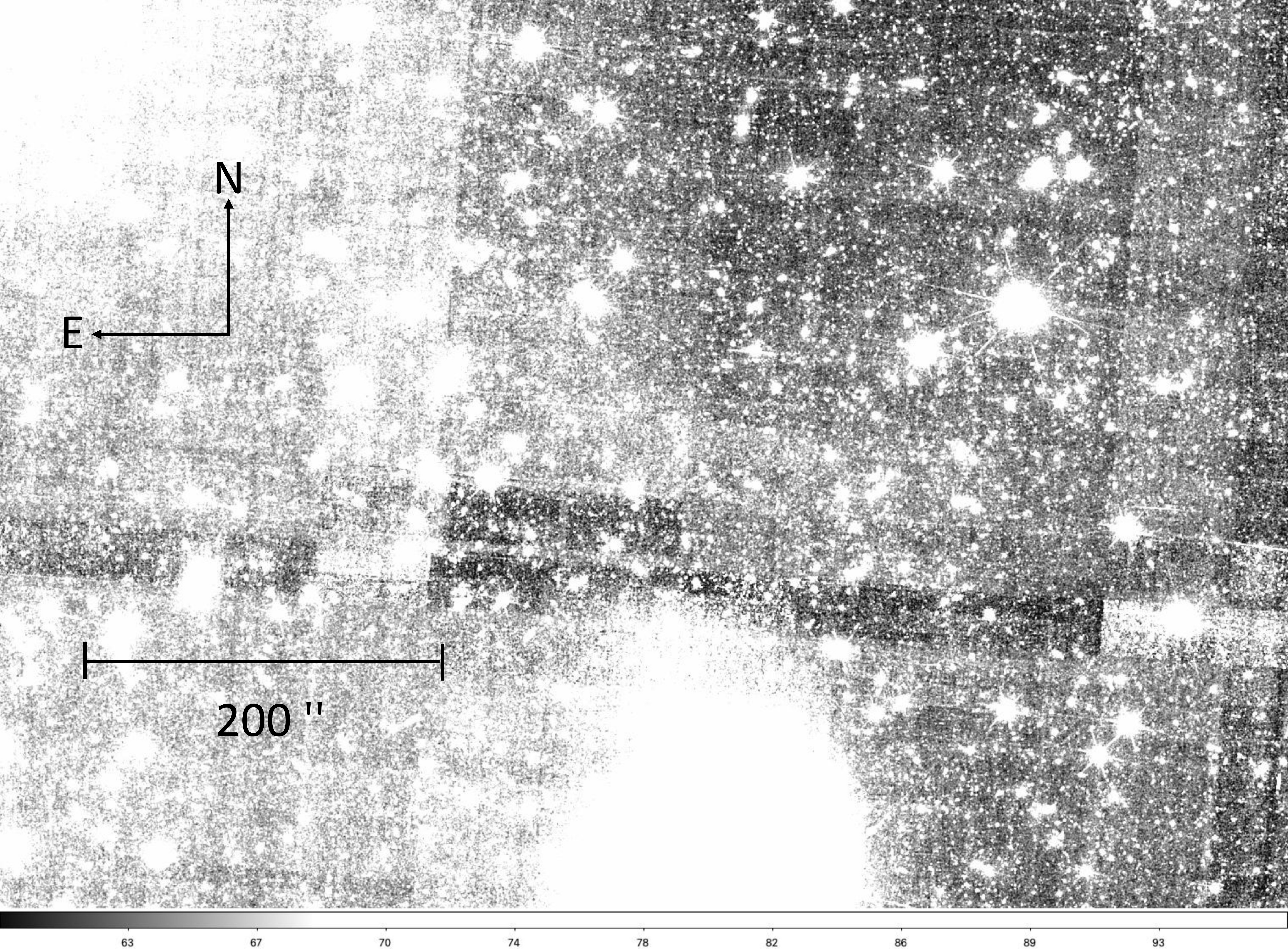}
  \caption{Impact of varying detector sensitivities and persistence on tidal feature detection in NIR bands. \emph{Top:} $\IE$ image centred on a diffuse feature at the north of NGC\,1546. \emph{Bottom:} Same field in $\HE$, where the feature is much less visible.}
  \label{fig:stream_VIS_vs_H}
\end{figure}

\section{\label{Data_Methods}Data}

\subsection{The ERO-D dataset}

We used imaging data from the \Euclid visible (VIS) instrument \citep{2018SPIE10698E..28C,2025A&A...697A...2E}, employing the $\IE$ optical filter with a pixel scale of $\ang{;;0.1}$ and from the Near Infrared Spectrometer and Photometer (NISP) instrument \citep{2022A&A...662A..92E,2025A&A...697A...3E}, using the near-infrared (NIR) filters $\YE$, $\JE$, and $\HE$, each with a pixel scale of $\ang{;;0.3}$. In this paper, we use the AB system and adopt the following convention: apparent magnitudes are denoted as $\IE$, $\YE$, $\JE$, and $\HE$, while the absolute magnitudes are denoted as $M_{\IE}$, $M_{\YE}$, $M_{\JE}$, and $M_{\HE}$.

ERO-D was observed in November 2023 using one reference observational sequence (ROS), namely four 560-second exposures in the $\IE$ filter and four 87.2-second exposures in each of the NISP instrument's filters. This process mirrors the approach that will be adopted for the EWS, with the exception that a non-optimal position of the spacecraft with respect to the Sun makes the presence of significant stray light possible in the ERO-D $\IE$ image \citep{2024arXiv240513496C}.

The raw single exposures were processed through a dedicated pipeline, which produces the stacks used in this paper. This processing is described in \cite{2024arXiv240513496C}, along with further detail about each ERO dataset. In particular, for ERO-D, the limiting surface brightness for extended emission, measured in $\text{mag\,arcsec}^{-2}$ on a $10\arcsec\times10\arcsec$ scale \citep{2020A&A...644A..42R}, is estimated as 30.05 in $\IE$, 28.41 in $\YE$, 28.58 in $\JE$, and 28.60 in $\HE$.

In Fig. \ref{fig:dorado_coverage}, the ERO-D coverage (approximately $\ang{0.8}\times\,\ang{0.7}$ centred on $ {\rm RA} = \ra{64;0;51.613}$, ${\rm Dec} = \ang{-55;46;51.33}$) is shown in red. It notably includes four major galaxies of the Dorado group: NGC\,1553, NGC\,1549, NGC\,1546, and IC\,2058. The $\IE$ surface brightness map and the colour image of the \Euclid ERO-D field is displayed in Fig. \ref{fig:dorado_general}.

\subsection{\label{contaminants}Diffuse light contaminants}

The ERO-D observations were acquired with one single ROS and therefore share the same depth as the standard EWS. As such they may be used as a test-bed on the \Euclid capability to detect and analyse extended LSB structures like tidal debris. In this subsection, we list the potential contaminants for each \Euclid filter.

The instrument PSF is usually one of the main limiting factors in LSB studies \citep[e.g.][]{2014A&A...567A..97S,2014PASP..126...55A,2017A&A...601A..86K}. Indeed, in many studies using non-optimised cameras, the wings of the PSF introduce an additional, artificial diffuse component in the surrounding of every source, including extended galaxies. In this paper, for diffuse light photometry, we do not use deconvolution by an extended PSF of \Euclid. Instead, we make careful decisions in terms of photometric methods (as described in Sect. \ref{methods}) that minimise its impact (which is already limited due to its notable sharpness and the low level of its wings, according to \citealt{2024arXiv240513496C}). For more details regarding the impact of \Euclid's PSF and on the regimes for which omitting deconvolution when producing surface brightness profiles of nearby galaxies is possible, we refer the reader to Appendix\,B of \cite{2025arXiv250602745M}.

\Euclid's NISP instrument covers not only imaging in the $\YE$, $\JE$, and $\HE$ bands, but also acquisition of spectra using a grism, which creates artefacts on IR images. They take the form of parallel linear persistence charge features centred on bright objects. In areas of the image affected by persistence, measured photometry, and therefore colour, of LSB objects is less reliable. If masked incorrectly, these contaminants can also alter the surface brightness profiles of the outer regions of galaxies. The problem is partially corrected through modelling and subtraction during the ERO pipeline run \citep{2024arXiv240513496C}, but there are still sources that present this issue in the ERO-D FoV. In addition, the NIR images are subject to detector levels differences, which are still visible in the stacks. These two issues can hinder the detection of faint features, as shown in Fig. \ref{fig:stream_VIS_vs_H}.

\begin{figure}[h]
  \centering
  \includegraphics[width=\linewidth,trim={0cm 1.5cm 0cm 0cm},clip]{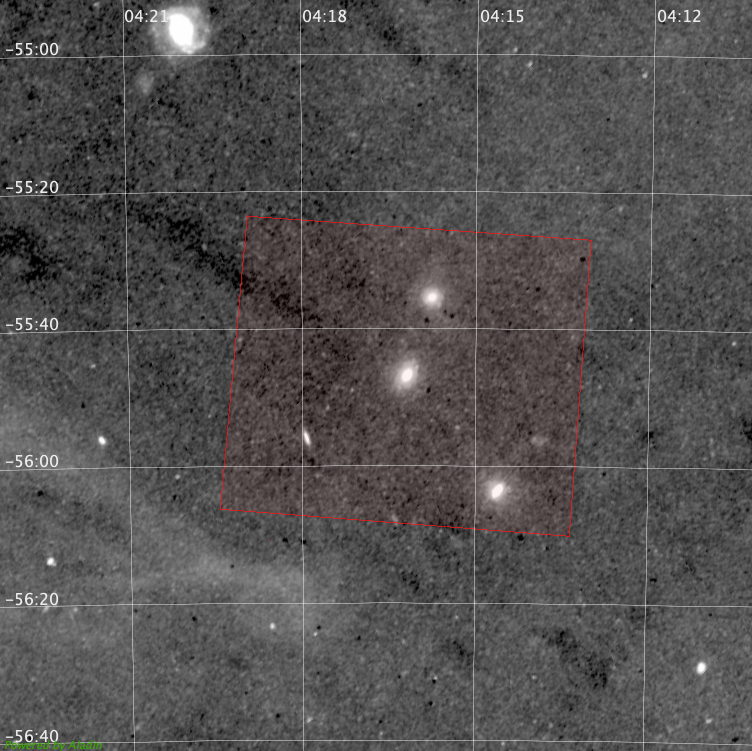}
  \caption{View of the 12\,\textmu m photometric band of the WISE IR survey (reprocessed by \citealp{2014ApJ...781....5M}), allowing  us to probe the presence of Galactic cirrus. The ERO-D FoV is displayed in semi-transparent red. The coordinates are given in RA, Dec.}
  \label{fig:dorado_cirrus}
\end{figure}

During the visual inspection of the $\IE$ image, we detect  a large diffuse light component which affects mainly the southern part of the image, and reaches IC\,2058 (see Fig. \ref{fig:dorado_general}). There is no bright pattern in NIR that matches this diffuse emission in $\IE$, which can, in principle, be explained by the lower depth of \Euclid's IR bands (and implies that no reliable colour information is available for this component). Three possible interpretation of this diffuse component were explored.
\begin{itemize}
\item The presence of an intra-group light. Nevertheless, the shape of this emission areas seemed suspicious, tracing sometimes the boundaries of the CCDs.
\item Galactic cirrus within the ERO-D FoV. This type of structure can indeed take on characteristic shapes with preferred directions. However, checking IR images of the FoV region suggests that the risk of such contamination is limited: in Fig. \ref{fig:dorado_cirrus}, the WISE \citep{2010AJ....140.1868W,2016A&A...593A...4M} data do not show such structures in the ERO-D FoV.
\item Stray light contamination. Due to observing constrains during the ERO observations, the telescope was not oriented in a position minimising stray light \citep{2024arXiv240513496C}, as will be done for the regular \Euclid survey.
\end{itemize}

The comparison and striking similarity of this diffuse emission with early, incomplete stray light maps from the first \Euclid data (Schirmer \& Soldano, priv. comm.) convinced us to favour the latter hypothesis throughout this article, although a combination of several of the previously described origins remains possible. This region should be the subject of a dedicated study on intra-group light once re-observed as part of the EWS.

\begin{figure*}[htbp]
\centering
    \includegraphics[width=\linewidth,trim={0cm 1.5cm 0cm 0cm},clip]{figures/correction_complete_v2.pdf}
    \caption{Large-scale diffuse light component correction for the $\IE$ image. \emph{Left}: Original $\IE$ image. \emph{Centre}: \texttt{sep} background map. The \texttt{MTObjects} masks are in black. \emph{Right}: \texttt{sep} local background-subtracted output $\IE$ image.}
  \label{fig:correction}
\end{figure*}

\section{\label{methods}Methods}

\subsection{Large-scale diffuse light component subtraction and comparison with ground-based data}

Regardless of its nature, the large-scale diffuse light component may interfere with the individual photometry of smaller features, prompting us to create a large-scale diffuse light subtracted version of the $\IE$ image dedicated to this analysis.

In the absence of a model to subtract this large-scale diffuse component, we are opting for a local background approach. For this, we need to mask not only the galaxies but also their faint features, so that only the sky and large-scale diffuse light component remain unmasked. This can be achieved with \texttt{MTObjects}, a segmentation algorithm optimised for faint extended sources, based on a Max-Tree method \citep{teeninga2015improved,TeeningaMoschiniTragerWilkinson+2016}. To obtain a mask, we run this software on the $\IE$ image, re-binned six times to enhance the signal-to-noise ratio (S/N) of the galaxy features (central panel of Fig. \ref{fig:correction}). During this step, \texttt{MTObject} is used with its default configuration, except for the \texttt{move\_factor} which is set to 0, ensuring extensive masks. Finally, we calculate the local background on the original grid with the resampled \texttt{MTObjects} mask using the source-extractor \citep{Bertin_and_Arnouts_1996} Python version \texttt{sep} \citep{2016JOSS....1...58B} with a cell size of $1000\,\textnormal{pixels}\times1000\,\textnormal{pixels}$ ($\approx1.7\arcmin\times1.7\arcmin$). The large-scale diffuse light component unmasked and above this scale is removed, and the background is flattened (right panel of Fig. \ref{fig:correction}). Although useful for the photometry of individual few arcminute-sized features, the corrected image might have removed real, very extended LSB stellar structures and cannot be used as a reference image for other tasks, such as calculating the photometric profile at large distances from the galaxy centre. The large-scale diffuse light subtracted image is therefore used solely for the photometry of smaller-scale individual structures and their qualitative comparison in colour maps. Additionally, in the following, we specify the use of the image with large-scale diffuse light subtracted by referring to it as the $\IE$ corrected image. We employ \texttt{Gnuastro Astwarp} \citep{2015ApJS..220....1A} to create a version of the $\IE$ corrected image, which is reprojected on NIR images, ensuring alignment with the NIR bands' pixel scale and grid.

\begin{figure*}[h!]
  \centering
  \includegraphics[width=0.495\linewidth,trim={0cm 0cm 0.5cm 0cm},clip]{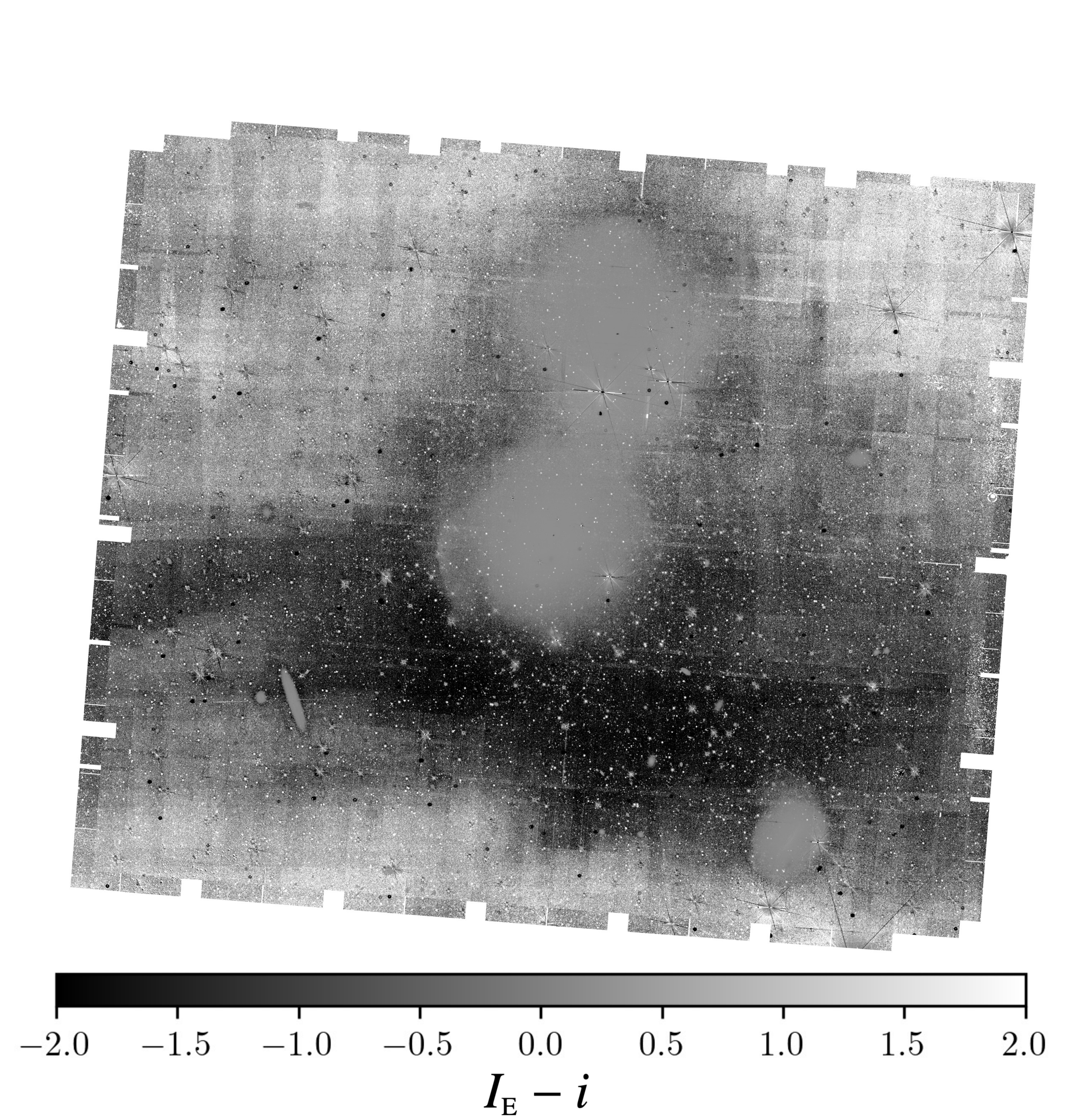}
  \includegraphics[width=0.495\linewidth,trim={0cm -2cm 0.5cm 2cm},clip]{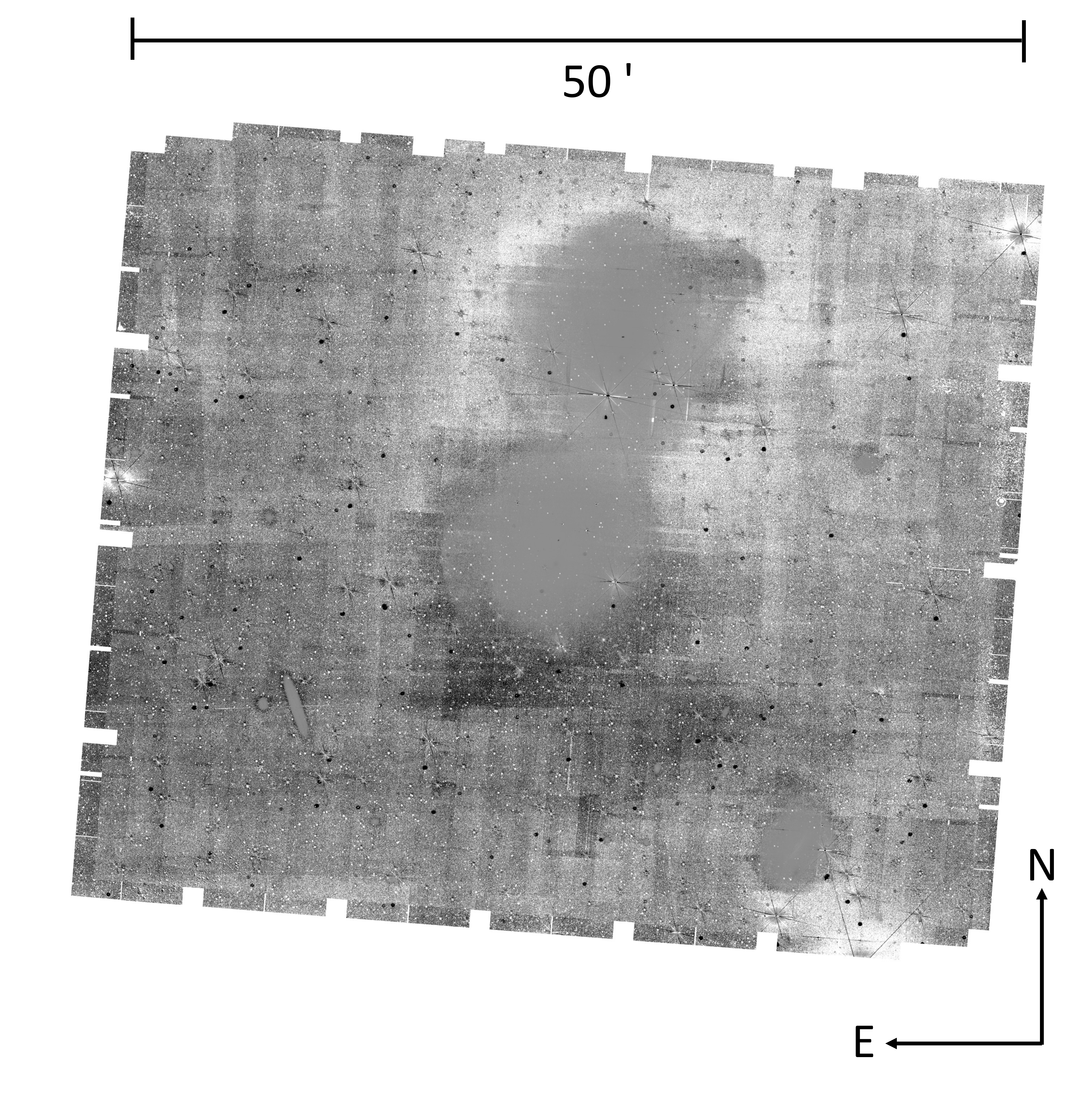}
  \caption{Residual image subtracting the DECaLS \textit{i}-band surface brightness map from the \Euclid $\IE$ surface brightness map. The scale gives the residual in $\text{mag\,arcsec}^{-2}$. \emph{Left:} Residual image using the original $\IE$ image. The extended dark area in the lower part of the image indicates the possible presence of stray light. \emph{Right:} Residual image using the $\IE$ corrected surface brightness map.}
  \label{fig:straylight}
\end{figure*}

To further investigate the instrumental contamination on the diffuse light, in particular the stray light, we compare the \Euclid original $\IE$ image to non-background-subtracted data from the ground-based DECaLS survey, which fully covers the ERO-D region (tiles 412-5540, 413-5540, 417-5540, and 418-5622). We used the \textit{i} band, which is close to the \Euclid $\IE$ band (despite the latter covering also a part of the DECaLS \textit{r} band and having a higher throughput). We subtract a constant background, and then co-add the DECaLS \textit{i} images using \texttt{GnuAstro AstWarp}. We estimate the DECaLS \textit{i} surface brightness map and finally subtract it from the \Euclid $\IE$ surface brightness map. It is worth noting that ERO-D has a surface brightness limit (30.05\,$\text{mag\,arcsec}^{-2}$ for $\IE$, \citealt{2024arXiv240513496C}) fainter than that of DECaLS (up to 29\,$\text{mag\,arcsec}^{-2}$, \citealt{2023A&A...669L..13M}). Consequently, in the left panel of Fig. \ref{fig:straylight}, the dark areas reveal both where \Euclid detects more flux and features than DECaLS, and where \Euclid is potentially still affected by stray light.

The stray light issue has been noted for several data sets in the \Euclid ERO programme \citep{2024arXiv240513496C}, leading to a redefinition of the orientation of the telescope during the EWS \citep{2025A&A...697A...1E}. As a result, the impact of this source of contamination will be drastically reduced in the future data releases. In agreement with the middle panel of Fig. \ref{fig:correction}, we show in Fig. \ref{fig:straylight}  that some diffuse halo-light was included in the local background map we used to construct our $\IE$ corrected image.

\subsection{\label{colours}Identification and characterisation of tidal features and substructures}

\subsubsection{\label{cutouts_masks}Cutouts and sources masking}

We generated cutouts centred on each bright galaxy of interest and encapsulating its visible halo, stellar features, and enough sky area to estimate the background value for modelling and photometry. Specifically, the cutouts for galaxies NGC\,1549 and NGC\,1553 are created with a dimension of $\ang{;;1200}\times\ang{;;1200}$, whereas the cutout for NGC\,1546 and IC\,2058 are smaller, measuring  $\ang{;;600}\times\ang{;;600}$ and $\ang{;;330}\times\ang{;;330}$, respectively.

We created a version of the $\IE$ original image which is reprojected on the pixel grid of the NIR bands, which is useful for the feature detection (since this process increases their S/N) and colour maps and profiles (see Sect. \ref{colours_1}). Precise galaxy modelling requires masking all foreground stars and background galaxies in the FoV. Our method consists in producing binary masks with the help of segmentation maps of both cutouts and complete FoV images using the \texttt{MTObjects} tool.

\subsubsection{\label{unsharp_masking}Unsharp masking}

We used unsharp masking to make the tidal features more prominent and to highlight internal structures. This technique involves subtracting a smoothed version of the initial image from the original. To prevent excessive subtraction in the innermost regions of the galaxies of interest, we started by creating an image with a flattened dynamic range. This was achieved by applying an $\operatorname{asinh}$ stretch to the original image. For each filter, we then subtracted a stretched, smoothed, and star-masked image from the stretched image.

\subsubsection{\label{fitting}\label{colours_1}Ellipse fitting, colours, and profiles}

To model the major galaxies of the ERO-D FoV, we opted for an ellipse fitting approach. \texttt{AutoProf} \citep{2021MNRAS.508.1870S} is an automated non-parametric ellipse fitting \textsc{Python} tool that takes as inputs the cutouts and associated masks described in Sect. \ref{cutouts_masks}. We ran the software on the non-re-binned cutouts (for the $\IE$ band, we used the non-corrected image) to subtract a constant background and extract surface brightness profiles for the galaxies NGC\,1549, NGC\,1553, NGC\,1546, and IC\,2058. During this profile extraction process, all the parameters (centre, ellipticity, position angle) were allowed to vary, in order to achieve high-precision galaxy subtraction for each band.

We note that the galaxies NGC\,1549 and NGC\,1553 partially overlap. To avoid any issues related to this overlap, we performed their ellipse fitting in two steps. We first carry out an ellipse fitting for NGC\,1553, allowing us to subtract this galaxy before performing the final ellipse fitting on NGC\,1549. The opposite approach is applied when fitting NGC\,1553. Subtracting the galaxy models, we obtain residuals images in the $\IE$ (both original and corrected), $\YE$, $\JE$, and $\HE$ bands.

Before examining the colours, we note that in this region of the sky, the extinction is negligible \citep{1998ApJ...500..525S,2011ApJ...737..103S}. To examine the colour profiles of the main galaxies, we used \texttt{AutoProf} with the prepared cutouts and masks, this time using a forced photometry mode (i.e. the centre, ellipticity, and position angle are fixed to the values obtained for the $\IE$-band profile). This choice facilitates the consistent comparison of surface brightness profiles across different bands. The final step involves subtracting the surface brightness profiles of one band from another, yielding colour profiles (in particular, $\IE-\HE$ and $\IE-\JE$) for each galaxy.

In addition to colour profiles, we also generate 2D colour maps. In order to do this, we make use of the $\IE$ corrected image. We first estimate the background level for the full FoV, both rebinned $\IE$ and original $\JE$, $\YE$, and $\HE$ images. We make use of \texttt{MTObjects} masking and a method derived from the \texttt{AutoProf} algorithm, using as background value the first prominent peak of the smoothed histogram of pixel values of the image. We then constructed detailed surface brightness maps for each band by replacing flux with surface brightness in each pixel. The colour maps are produced by subtracting the pairs of surface brightness maps. It is worth noting that for the photometric results presented in this paper (both profiles and values), we propagated the uncertainty arising from the width of the first histogram peak used to determine the background.

\subsubsection{\label{Jafar}\label{individual_flux_photometry}Detecting, classifying, and characterising features with annotations}

We used the \texttt{Jafar} annotation tool \citep{Sola_et_al_2022} to perform a visual inspection of the images and identification of tidal features. This online software enables users to navigate images, zoom in and out, and to draw the shapes of LSB features superimposed on deep images. In practice, it is necessary to precisely delineate the boundaries of each feature using polygonal shapes and assign labels, describing the type of the structure. Two other experts verify the final annotations visually. The coordinates of the contours and annotation labels are stored in a database, allowing for the subsequent retrieval of quantitative measurements and the creation of feature masks. The annotations are made taking into account the $\IE$ and colour images, but also ellipse fitting residual and unsharp masked images for probing the inner features.

The colour maps alone enable a qualitative study, but they are not representative of the individual feature photometry. Indeed, the galaxy models were not subtracted, so the colour of the features ends up mixed with that of the extended halos of these galaxies. It is then necessary to perform individual feature photometry, which heavily depends on local background variations near the studied structure. To minimise these variations, and thereby the uncertainty in the photometry of individual features, we used the $\IE$ corrected image.

We generated masks that precisely match the shape of each structure using the contours extracted from Jafar. To avoid flux contamination by the host galaxy, we used the ellipse fitting residual images to perform this photometry. We considered the remaining background to be flat.

For each feature and in each photometric band, we used a cutout that encompasses the structure and its immediate surroundings. We masked stars and distant galaxies using \texttt{MTObjects} and the tidal feature, using its contour extracted from the \texttt{Jafar} annotation tool. The background value estimation method that follows is similar to the one described in Sect. \ref{colours_1}. To measure the feature integrated flux, small sources inside the structure have to be masked. For the segmentation step, \texttt{MTObjects}, which is optimised for more extended sources masking, is then replaced by \texttt{sep}. Its default parameters were used here, except \texttt{thresh}, which is set to 1.5, and \texttt{err} to the global background RMS.

\subsection{\label{GC}Globular clusters (GCs)}

\subsubsection{\label{GC_selection}GC identification}

The spatial resolution of \Euclid $\IE$ images, combined with the optical and NIR colours of \Euclid, as well as the ground-based surveys, enables us to identify GC candidates within the ERO-D FoV. Additional information from the GCs, including their spatial distribution and colours, can provide further insights into investigating the mass assembly of the two interacting galaxies, NGC\,1549 and NGC\,1553, in this ERO field. We identified GCs in the FoV through several steps, including creating PSF models, source detection, photometry, injecting artificial GCs, and finally, selecting GC candidates. This procedure is similar to the methodology in \cite{2025A&A...697A..10S} for the \Euclid ERO Fornax galaxy cluster (ERO-F) with some modifications to the colour selection. Furthermore, in this work, we also used $g$, $r$, and $i$ band images from the Dark Energy Survey (DES; \citealp{des-dr2}). The additional colour information provided by these images helps to reduce contamination from foreground stars and background compact sources. 

\begin{figure}
    \centering
    \includegraphics[width=\linewidth]{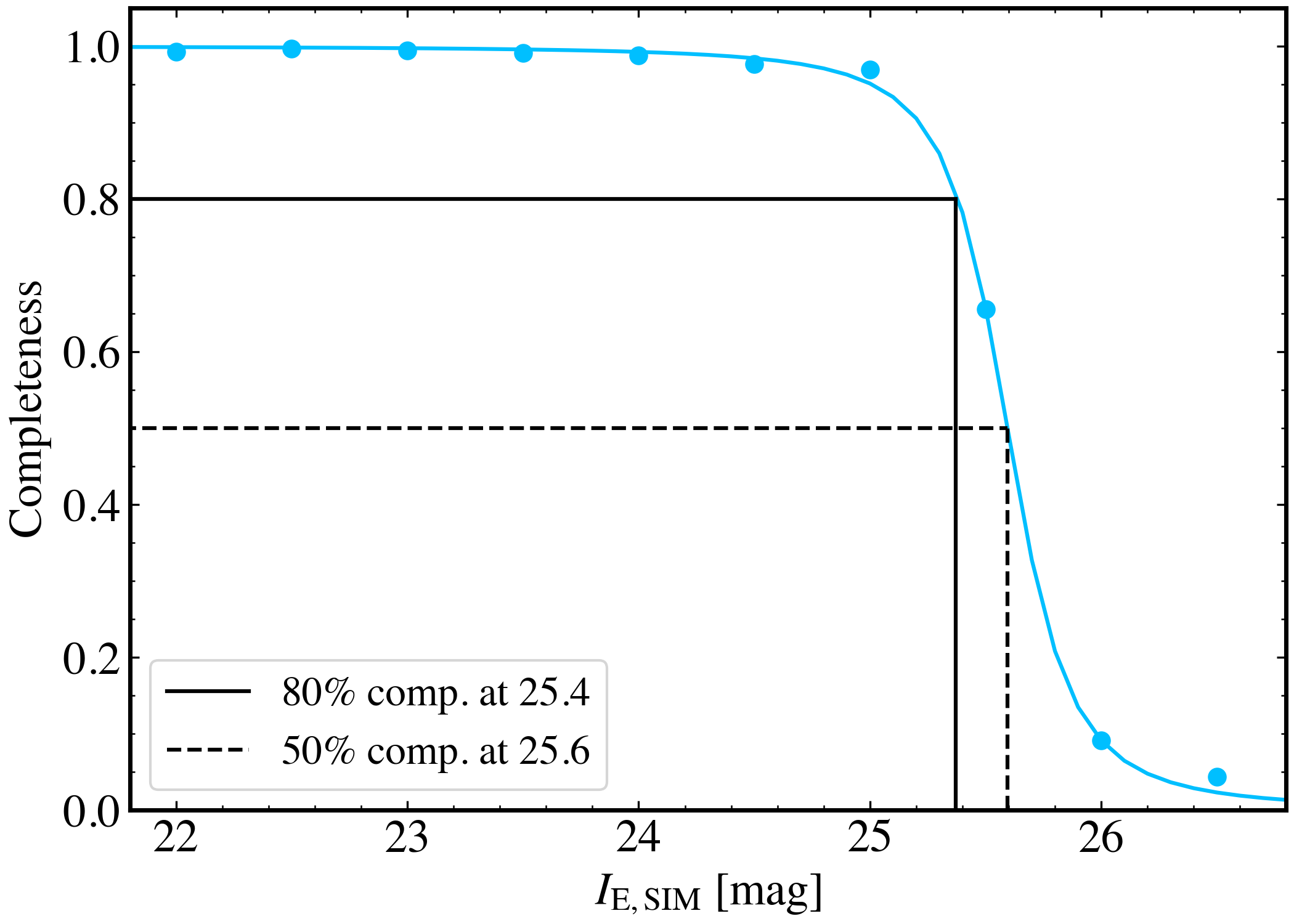}
    \caption{GC completeness of ERO-D as a function of $I_\sfont{E,\,SIM}$, the magnitude of artificial GCs injected in the data. Note: at the distance of the Dorado galaxies, real GCs are expected to be distributed between \IE\,=\,21 and \IE\,=\,25.}
    \label{completeness}
\end{figure}

We started the analysis by producing PSF models in all seven bands (four \Euclid and three DES bands) using bright, non-saturated stars within the FoV. Next, we generated a multi-wavelength source catalogue with photometry in seven bands. Sources were detected by applying unsharp-masking on the original $\IE$ image \citep{2025arXiv250316367S} and using \texttt{SExtractor} with \textsc{DETECT\_THRESH}=1.5. We used similar \texttt{SExtractor} parameters as provided in Table 3 in \citet{2025A&A...697A..10S}. Subsequently, forced aperture photometry was carried out using \texttt{photutils} \citep{2016ascl.soft09011B} in all bands at the given coordinates of the detected sources. In addition, we measured a compactness index of sources in \IE within aperture diameters of two and four pixels (\citealp{peng2011}). Here, the aperture photometry was done on background-subtracted images using a 12\,pixel $\times$ 12\,pixel background mesh size. Such a small mesh size is needed to estimate the local background close to the central regions of massive galaxies, where the slope of the light profile is changing rapidly; however, this mesh size does not influence the photometry of small sources by more than 1\%. Based on our analysis of artificial GCs, the completeness of source detection in the \IE band is above 98\% at $\IE<25$ (which corresponds to the faintest GCs for a Gaussian globular cluster luminosity function (GCLF) around massive galaxies, \citealp{2025A&A...697A..10S}), with 80\% and 50\% completeness limits at $\IE=25.4$ and $\IE=25.6$, respectively. Later, for GC selection, we apply a compactness index, ellipticity, and various colour criteria to select GC candidates that will reduce completeness by 20\%. We examined not only the completeness as a function of the magnitude, but also the completeness of source detection as a function of the location in the FoV (called spatial completeness hereafter). An initial visual inspection shows that outside the central 1\,kpc of the main galaxies our detection procedure does not miss point sources. To further evaluate this, we conducted a straightforward test aimed at assessing the completeness of source detection by selecting foreground stars. We identified these stars across the FoV based on their full width at half maximum (FWHM) and $\IE - \YE$ colour. Our analysis revealed that the distribution of stars was uniform throughout the frame. Furthermore, we evaluated the completeness of source detection within the ERO-D FoV using artificial GCs, and found a consistent level of spatial completeness for sources with $\IE<24$ in the field, including the regions around NGC\,1553 and NGC\,1549 (see Fig. \ref{completeness}).

In the next step, we performed the GC identification. We used the created PSF models and produce artificial GCs (using the King profile, \citealp{1966AJ.....71...64K}) of sizes (half-light radius) between 2\,pc and 5\,pc, absolute $\IE$ magnitude between $-11$ and $-5$, and colours $\IE-\YE=0.45$, $\YE-\JE=0.1$, $\JE-\HE=0$, $g-r=0.6$, and $g-i=0.9$. The artificial GCs are then distributed uniformly across the frames. Using the artificial GCs, we estimate the uncertainties in the measured parameters of GCs at a given magnitude. Given these uncertainties, we define flux-dependent selection criteria for compactness index (in \IE), colours, and ellipticity (in \IE) for GCs. These criteria select the artificial GCs within 5--95, 1--99, and 1--99 percentiles in compactness index, ellipticity and colours (5 colours), and overall, 80\%. Taking into account the completeness in source detection (above 97\%), our methodology selects 77\% of the artificial GCs.

The scatter in the colours of the artificial GCs only takes into account the measurement uncertainty. For colour selection of GC candidates, we broaden the range of colours of the artificial GCs by the width of the colours for old GCs (older than 7\,Gyr) with sub-solar metallicities ($Z<0.02$) from stellar population models (See Appendix\,A in \citealp{2025A&A...697A..10S}). This broadening is the same across magnitude and is about $\pm$0.3, $\pm$0.15, $\pm$0.15, $\pm$0.3, and $\pm$0.3\,mag for $\IE-\YE$, $\YE-\JE$, $\JE-\HE$, $g-r$, and $g-i$, respectively. This means that for the fainter sources that are not detected in the DES bands, only \Euclid bands and colours are used for colour selection. Once the GC candidates are selected, we continue with the analysis of GC distribution around galaxies and across the FoV of ERO-D, as well as their colours.

Lastly, considering that our analysis is based on artificial GCs with a half-light radius between 2 and 5\,pc, we expected to miss those GCs that are smaller and larger than these limits. Based on the analysis of the ERO-F data in \citet{2025A&A...697A..10S}, we expected at least 30\% of the GCs to be too small to be resolved in \Euclid \IE images. Considering this, we expect an overall completeness of about 50\% in the GC selection applied here, but this also depends on the properties of the GC population of each galaxy. For more details on the general method of GC selection and completeness estimation, we refer the reader to \cite{2025arXiv250316367S}.

\subsubsection{\label{GC_profiles}From GC distribution to GC colours in features}

To study the distribution of GCs over the ERO-D FoV, we generate density maps. We consider GC candidates verifying $\IE<24$, which is about 1\,mag fainter than the turn-over magnitude of the GCLF (based on the GCLF for massive galaxies in Fig. 14 of \citealt{2025A&A...697A..10S} which peaks at $\IE\approx23\,\text{mag}$, see also \citealt{2001stcl.conf..223H,2012Ap&SS.341..195R}). We therefore reduce the number of contaminant sources (non-GCs), which increase at the fainter magnitudes \citep{2025A&A...697A..10S}. In the following, we refer to this set as the `bright GC candidates'. Isodensity contours can then be easily obtained from the  catalogue using the \texttt{gaussian\_kde} function from the \texttt{SciPy} Python package. The kernel bandwidth to obtain the smoothed distribution respects the Scott's rule, described in Eq. \eqref{scott} as

\begin{equation}
\label{scott}
    h = n^{-1/(d+4)}
,\end{equation}
where \( h \) is the bandwidth in degree, \( n \) is the number of data points, and \( d \) is the dimensionality of the data.

\newcommand{\hms}[3]{#1\textsuperscript{h}\ #2\textsuperscript{m}\ #3\textsuperscript{s}}

By selecting GC candidates in concentric rings around each galaxy, and dividing their number by the area of these rings, we can generate radial density profiles. 
It is worth noting that the GC counts described in this paper are not adjusted for contamination or completeness and simply reflect the number of identified candidates. We used the following galaxy centre coordinates, obtained from the ellipse fitting: \ra{04;15;45.11}, \ang{-55;35;32.05} for NGC\,1549, \ra{04;16;10.48}, \ang{-55;46;48.06} for NGC\,1553, \ra{04;14;36.32}, \ang{-56;03;39.25} for NGC\,1546, and \ra{04;17;54.35}, \ang{-55;55;58.91} for IC\,2058 (ICRS coordinates, epoch J2000).
We can also isolate the associated GC candidates in each feature with the help of the contours extracted from the \texttt{Jafar} annotation tool. We then calculated the weighted average GC colour for each pair of filters. We also applied the same method on each galaxy, for determining their GC colours and compare with those of their associated features.

\begin{table}[h!]
\caption{Main properties of the brightest galaxies of the ERO Dorado FoV.}.
\centering
\renewcommand{\arraystretch}{1.2}
\setlength{\extrarowheight}{2pt}
\begin{tabular}{lcccccccc}
\hline \hline 
\noalign{\vskip 4pt}
Galaxy name & Type & $\sigma_v$ [${\textnormal{km\,s}}^{-1}$] & $v$ [${\textnormal{km\,s}}^{-1}$] \\
(1) & (2) & (3) & (4) \\
\noalign{\vskip 4pt}
\hline
\noalign{\vskip 4pt}
NGC\,1549 & E & $199\pm4$ & $1147 \pm 19$  \\
NGC\,1553 & S0 & $186 \pm 4$ & $\phantom{0}959 \pm 18$ \\
NGC\,1546 & S0-a & - & $1177 \pm 20$ \\
IC\,2058 & Scd & - & $1278 \pm 15$ \\
\noalign{\vskip 4pt}
\hline \hline
\end{tabular}
\label{tab:general}
\tablefoot{Morphological types in column 2 are from the Hyperleda database \citep{2003A&A...412...45P} and \cite{2014A&A...562A.121C}. Velocity dispersion in column 3 and radial velocity in column 4 are from \cite{2000ApJ...529..786M}.}
\end{table}

\section{\label{sc:Results}Results}

\begin{table*}[h]
\caption{\texttt{AutoProf} estimation of magnitudes of the main galaxies of the ERO Dorado FoV. }
\centering
\renewcommand{\arraystretch}{1.2}
\setlength{\extrarowheight}{2pt}
\begin{tabular}{lcccccccc}
\hline \hline 
\noalign{\vskip 4pt}
Galaxy name & $\IE$ & $\YE$ & $\JE$ & $\HE$ & $M_{\IE}$ & $M_{\YE}$ & $M_{\JE}$ & $M_{\HE}$  \\
(1) & (2) & (3) & (4) & (5) & (6) & (7) & (8) & (9) \\
\noalign{\vskip 4pt}
\hline
\noalign{\vskip 4pt}
NGC\,1549 & $\phantom{0}9.13$ & $\phantom{0}8.44$ & $\phantom{0}8.35$ & $\phantom{0}8.23$ & $-22.19$ & $-22.88$ & $-22.98$ & $-23.09$ \\
NGC\,1553 & $\phantom{0}8.89$ & $\phantom{0}8.33$ & $\phantom{0}8.36$ & $\phantom{0}8.23$ & $-22.43$ & $-23.00$ & $-22.96$ & $-23.10$ \\
NGC\,1546 & $11.57$ & $10.95$ & $10.81$ & $10.72$ & $-19.75$ & $-20.38$ & $-20.52$ & $-20.61$ \\
IC\,2058 & $13.59$ & $13.30$ & $13.23$ & $13.14$ & $-17.73$ & $-18.03$ & $-18.09$ & $-18.19$ \\
\noalign{\vskip 4pt}
\hline \hline
\end{tabular}
\tablefoot{The distance to the four galaxies is set at 18.4 Mpc.}
\label{tab:magnitudes}
\end{table*}

\begin{table*}[h]
\caption{\texttt{AutoProf} estimation of effective (half-light) radii of the main galaxies of the ERO Dorado FoV in the different filters. }
\centering
\renewcommand{\arraystretch}{1.2}
\setlength{\extrarowheight}{2pt}
\resizebox{\textwidth}{!}{
\begin{tabular}{lcccccccc}
\hline \hline
\noalign{\vskip 4pt}
Galaxy name & $R_{\rm e,\IE}$ [arcsec] & $R_{\rm e,\IE}$ [kpc] & $R_{\rm e,\YE}$ [arcsec] & $R_{\rm e,\YE}$ [kpc] & $R_{\rm e,\JE}$ [arcsec] & $R_{\rm e,\JE}$ [kpc] & $R_{\rm e,\HE}$ [arcsec] & $R_{\rm e,\HE}$ [kpc] \\
(1) & (2) & (3) & (4) & (5) & (6) & (7) & (8) & (9) \\
\noalign{\vskip 4pt}
\hline
\noalign{\vskip 4pt}
NGC\,1549 & 231 & 20.6 & 237 & 21.1 & 232 & 20.7 & 235 & 20.9 \\
NGC\,1553 & 157 & 14.0 & 138 & 12.3 & 158 & 14.1 & 144 & 12.9 \\
NGC\,1546 & \phantom{0}74 & \phantom{0}6.6 & \phantom{0}69 & \phantom{0}6.2 & \phantom{0}70 & 6.2 & \phantom{0}67 & \phantom{0}6.0 \\
IC\,2058 & \phantom{0}71 & \phantom{0}6.3 & \phantom{0}67 & \phantom{0}6.0 & \phantom{0}67 & \phantom{0}6.0 & \phantom{0}66 & \phantom{0}5.9 \\
\noalign{\vskip 4pt}
\hline \hline
\end{tabular}}
\tablefoot{The distance to the four galaxies is set at 18.4 Mpc.}
\label{tab:radii}
\vspace{-0.5cm}
\end{table*}

\begin{table}[h!]
\caption{Number (non-corrected for completeness) of bright ($\IE\leq24$) GC candidates for each galaxy.}
\centering
\renewcommand{\arraystretch}{1.2}
\setlength{\extrarowheight}{2pt}
\begin{tabular}{@{\hskip 0.075in}c@{\hskip 0.075in}c@{\hskip 0.075in}c@{\hskip 0.075in}c@{\hskip 0.075in}c@{\hskip 0.075in}}
\hline \hline
\noalign{\vskip 4pt}
Galaxy name & Blue GCs & Red GCs & Other  GCs  & Total GCs \\
(1) & (2) & (3) & (4) & (5) \\
\noalign{\vskip 4pt}
\hline
\noalign{\vskip 4pt}
NGC 1549 & $86 \pm 10$  & $69 \pm 9$ & $3 \pm 2$ & $158 \pm 13$ \\
NGC 1553 & $63 \pm \phantom{0}8$ & $45 \pm 7$ & $4 \pm 2$ & $112 \pm 11$ \\
NGC 1546 & $\phantom{0}5 \pm \phantom{0}3$ & $\phantom{0}5 \pm 3$ & $1 \pm 1$ & $\phantom{0}11 \pm \phantom{0}4$ \\
\noalign{\vskip 4pt}
\hline \hline
\end{tabular}
\label{tab:gc_counts}
\tablefoot{Blue and red GCs are defined relative to the threshold $\IE\,-\,\HE=0.68$. Other GCs correspond to GCs with no available colours due to their non-detection in the NIR bands, which have a lower resolution compared to the \IE band. Poisson uncertainty is provided as well.}
\vspace{-0.3cm}
\end{table}

\subsection{\label{general_consideration}NGC\,1549, NGC\,1553, NGC\,1546, and IC\,2058 properties}

The main characteristics of the studied galaxies  extracted from the literature are given in Table \ref{tab:general}. Here, we adopt the hypothesis of \cite{2002AJ....124.2471I} asserting that IC\,2058, NGC\,1549, NGC\,1553, and NGC\,1546 are located at the same distance, forming a compact group of interacting galaxies. This assumption is consistent with both the radial velocities of the four galaxies of interest which are similar according to  Table \ref{tab:general} and the literature on the estimated distances of these galaxies. NGC\,1549 and NGC\,1553 are located at $19.7^{+1.7}_{-1.6}$\,Mpc and $18.4^{+1.2}_{-1.2}$\,Mpc according to \cite{2001MNRAS.327.1004B}, which uses surface brightness fluctuations method. The distances of NGC\,1546 and IC\,2058 are respectively $13.4^{+6.0}_{-4.2}$\,Mpc and $18.1^{+3.6}_{-3.0}$\,Mpc, taken from \cite{1988cng..book.....T} and \cite{2011yCat..35320104N}, which use the Tully-Fisher relation. Those distances agree within the uncertainties. 
We  accept  as the common distance of IC\,2058, NGC\,1549, NGC\,1553, and NGC\,1546 that  
with the smallest uncertainty (18.4\,Mpc).

\subsubsection{Light profiles}

\begin{figure*}[htbp!]
  \centering 
  \begin{minipage}{.36\textwidth}
    \centering
    \includegraphics[width=\linewidth,trim=0cm 0cm 0cm 0cm, clip]{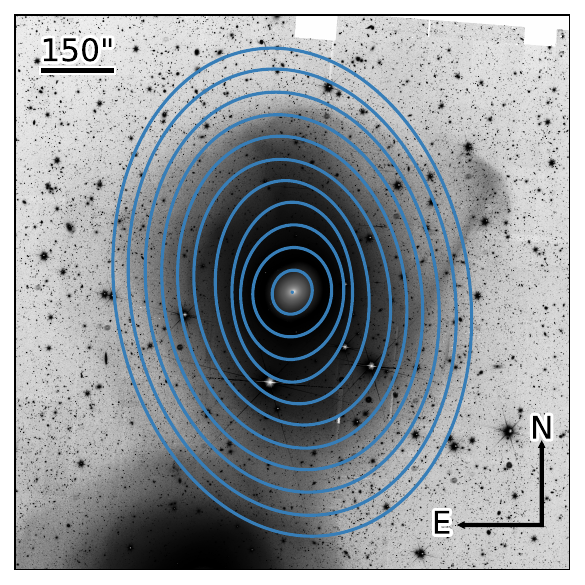}
    \subcaption*{NGC\,1549,}
  \end{minipage}\hfill
      \begin{minipage}{.57\textwidth}
    \centering
    \includegraphics[width=\linewidth]{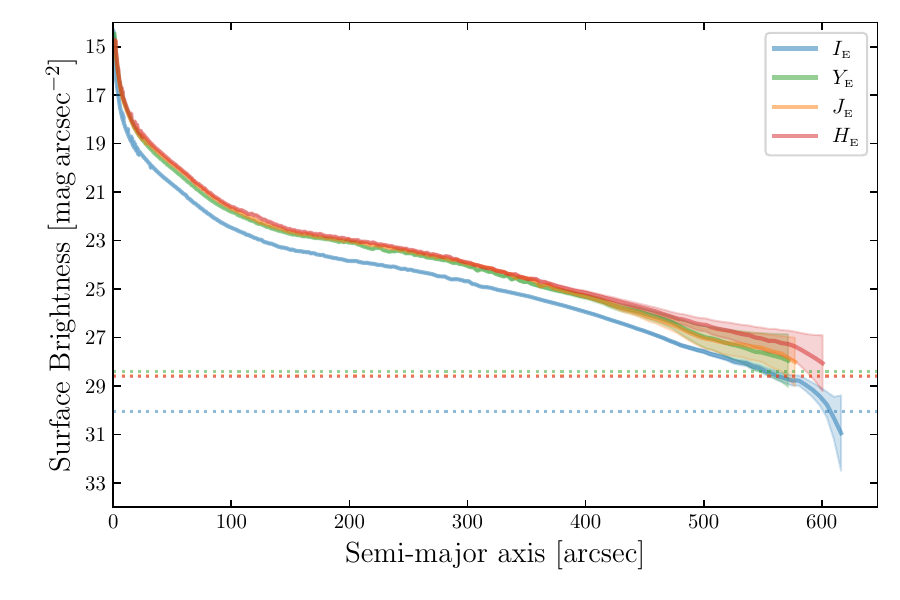}
  \end{minipage}\hfill
  \begin{minipage}{.36\textwidth}
    \centering
    \includegraphics[width=\linewidth,trim=0cm 0cm 0cm 0cm, clip]{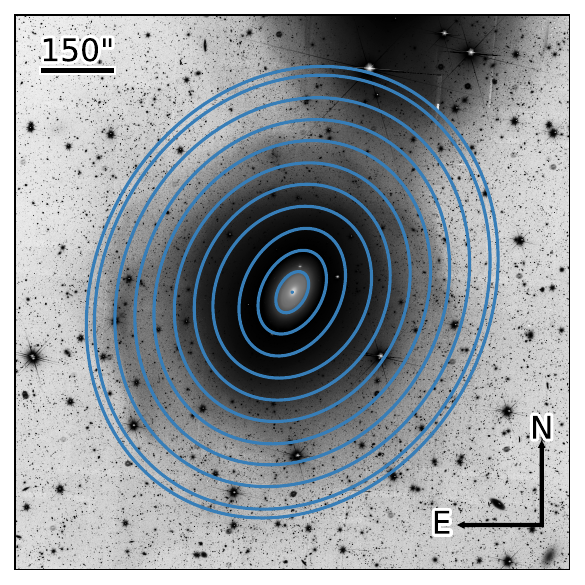}
    \subcaption*{NGC\,1553,}
  \end{minipage}\hfill
    \begin{minipage}{.57\textwidth}
    \centering
    \includegraphics[width=\linewidth]{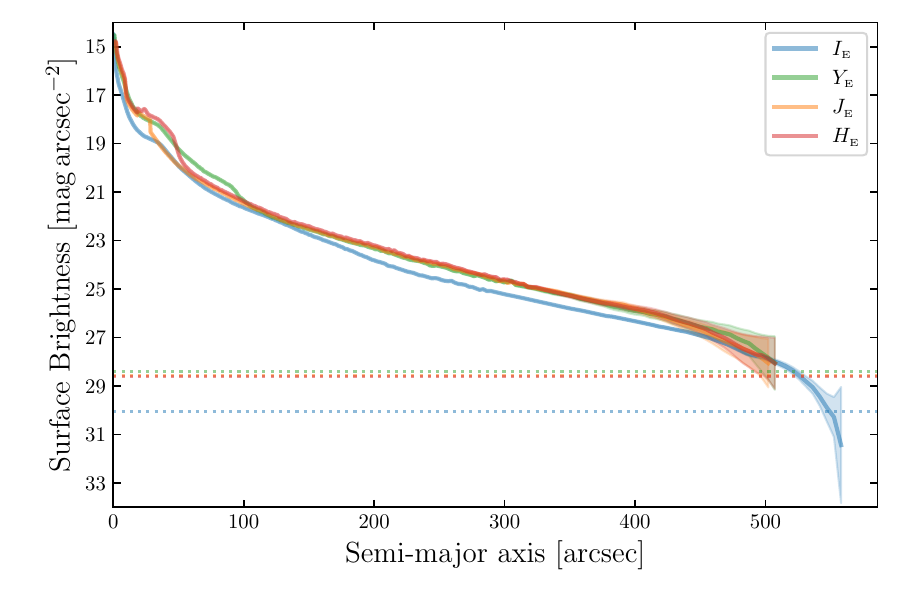}
  \end{minipage}\hfill
  \begin{minipage}{.36\textwidth}
    \centering
    \includegraphics[width=\linewidth,trim=0cm 0cm 0cm 0cm, clip]{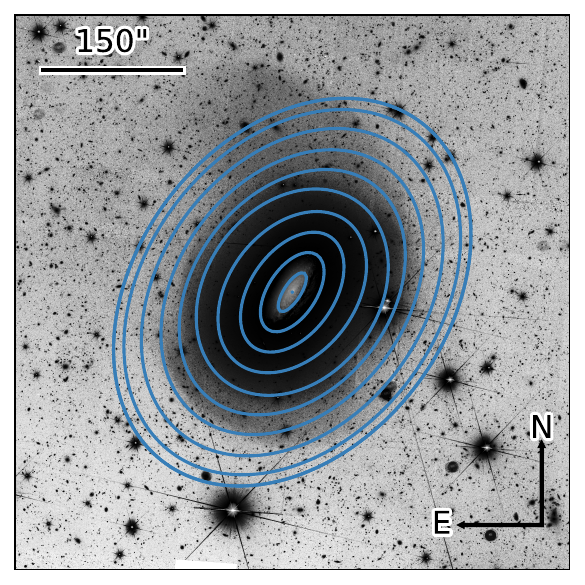} 
    \subcaption*{NGC\,1546.}
  \end{minipage}\hfill
  \begin{minipage}{.57\textwidth}
    \centering
    \includegraphics[width=\linewidth]{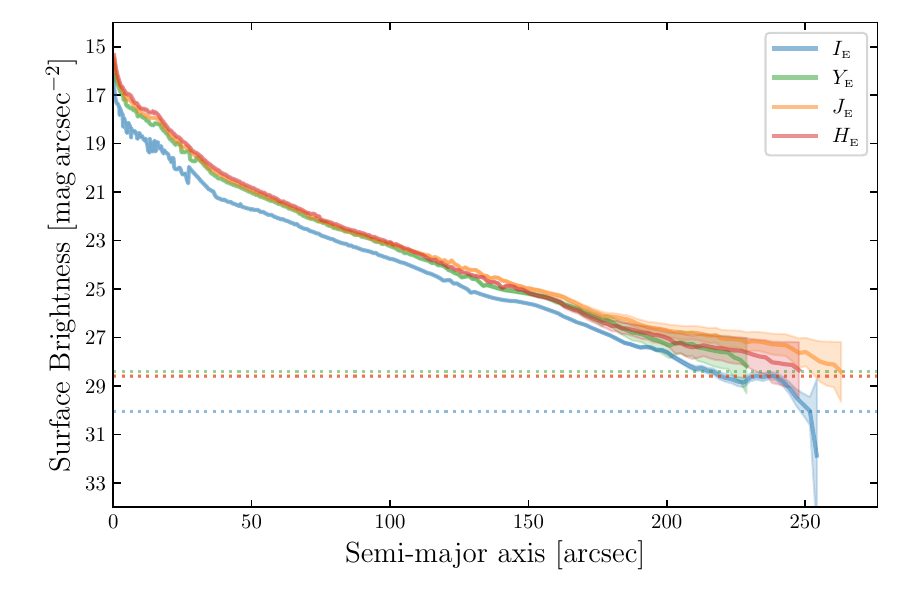} 
  \end{minipage}
  \caption{\emph{Left}: $\IE$ cutouts of the three galaxies of interest, displayed using hybrid histogram equalisation and logarithmic scale. Ellipses of the \texttt{AutoProf} profiles are displayed in blue. \emph{Right}: Corresponding surface brightness profiles (solid lines) and their uncertainties (semi-transparent areas). The limiting surface brightness for extended emission on a $10\arcsec\times10\arcsec$ scale is taken from \cite{2024arXiv240513496C} and is shown as a dotted line for each band.}
  \label{fig:autoprof_cutouts_and_profiles}
\end{figure*}

In Fig. \ref{fig:autoprof_cutouts_and_profiles}, we show the \texttt{AutoProf} ellipse fitting method applied to the three most massive galaxies in the field. IC\,2058 will not be discussed here, as its nature as an edge-on LTG makes variations in its profile less reliable. We note the presence of Type III profiles in the classification of \cite{2006A&A...454..759P} and \cite{2008AJ....135...20E}: the inner profile follows a relatively steep exponential shape, which transitions to a shallower profile beyond a break radius. This outer up-bending is also called `antitruncation'. In the $\IE$ band, the breaks for NGC\,1549, NGC\,1553 and NGC\,1546 are respectively located at around 100, 70, and 30\,arcsec, and corresponding  surface brightnesses of 23, 21.5, and 21\,$\text{mag\,arcsec}^{-2}$. Those profiles are compatible with both minor \citep{2007ApJ...670..269Y} and major \citep{2014A&A...570A.103B} mergers origin, thus confirming a tumultuous past that remains to be investigated. However, this analysis does not allow for a clear distinction between the two types of mergers based on these profiles alone.

Further out (beyond   600, 550, and 250\,arcsec, respectively; or approximately 55, 50, and 20\,kpc),  the NGC\,1549, NGC\,1553, and NGC\,1546 profiles drop sharply. The presence of this decline is confirmed by visual inspection and persists when the constant background value used by \texttt{AutoProf} is slightly varied. We extract from the surface brightness profiles the total magnitude and effective (or half-light) radius in each \Euclid filter, for the three galaxies of interest. We report our results in Table \ref{tab:magnitudes} and Table \ref{tab:radii}.

Table \ref{tab:magnitudes} shows similar NIR magnitudes for NGC\,1549 and NGC\,1553, indicating they have at first order comparable masses (since NIR bands are less sensitive to variations in the mass-to-light ratio caused by different stellar populations and ages; e.g. \citealt{2001ApJ...550..212B}). The magnitudes of  NGC\,1546 and IC\,2058 confirm their significantly lower mass, but their difference in $\IE$ magnitude imply different star-formation histories (or might also be due to differences in dust attenuation).

\subsubsection{Bright GC candidate number and distribution}

The selection process described in Sect. \ref{GC_selection} ends up with 790 bright GC candidates. Most of them are measured in both VIS and NISP images, and have associated colours. We divided this sample into two subsamples based on their $\IE-\HE$ colour. Although we did not find a clear bimodality in the distribution of this colour for our GC candidates, the spread in the observed colours is larger than the measurement uncertainties and we expect the two halves of the distributions to contain different proportions of red and blue clusters as defined in the literature (e.g. \citealt{2001AJ....121.2950K,2006ApJ...639...95P}). Hereafter, the definition of red and blue GCs used is based on whether $\IE-\HE$ is larger or smaller than the median value (0.68), which is not identical to the definitions used in those previous studies.

\begin{figure*}[h!]
    \centering
    
    \begin{subfigure}[b]{\linewidth}
    \hspace{-0.2cm}
        \includegraphics[height=6.7cm]{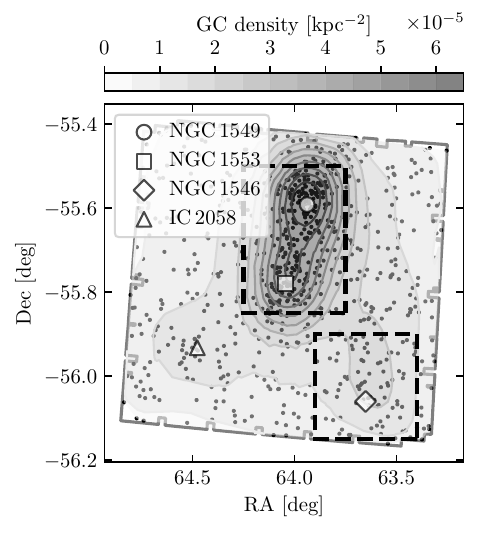}
        \hspace{-0.3cm}
        \includegraphics[height=5.5cm]{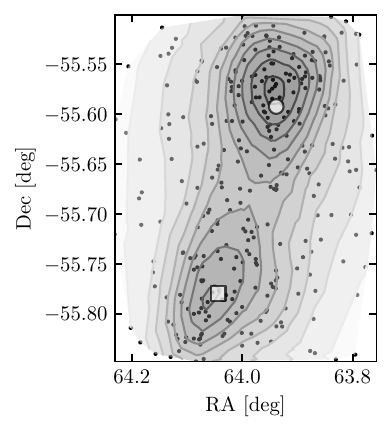}
        \hspace{-0.3cm}
        \includegraphics[height=5.5cm]{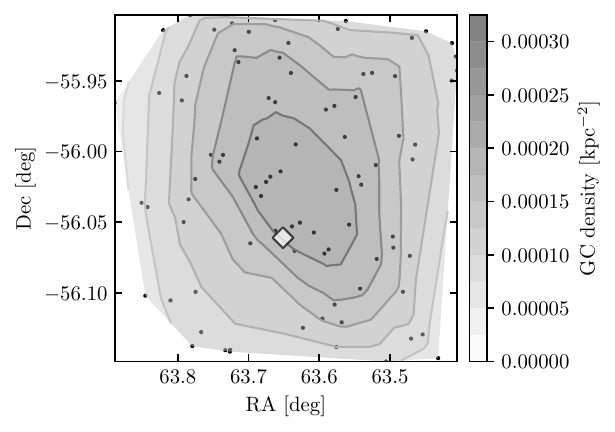}
    \end{subfigure}
    
    \begin{subfigure}[b]{\linewidth}
    \hspace{-0.2cm}
        \includegraphics[height=6.7cm]{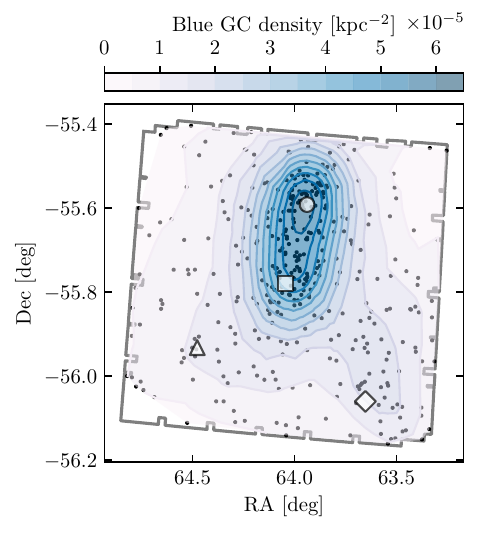}
        \hspace{-0.3cm}
        \includegraphics[height=5.5cm]{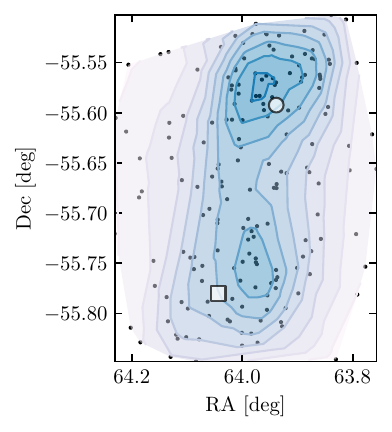}
        \hspace{-0.3cm}
        \includegraphics[height=5.5cm]{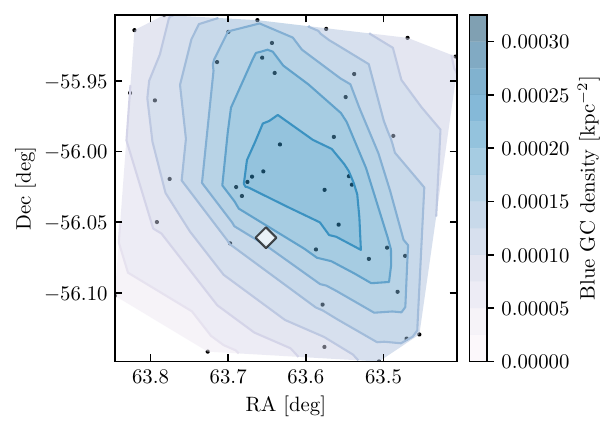}
    \end{subfigure}
    
    \begin{subfigure}[b]{\linewidth}
    \hspace{-0.2cm}
        \includegraphics[height=6.7cm]{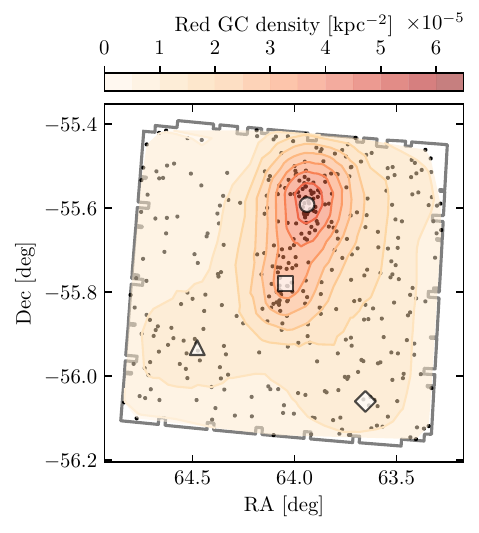}
        \hspace{-0.3cm}
        \includegraphics[height=5.5cm]{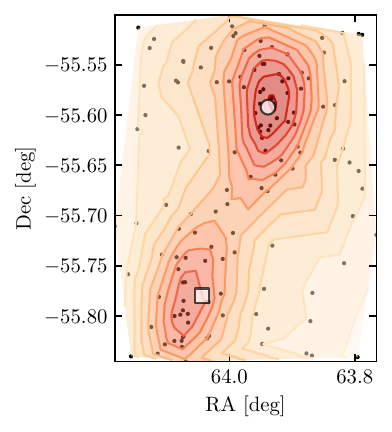}
        \hspace{-0.3cm}
        \includegraphics[height=5.5cm]{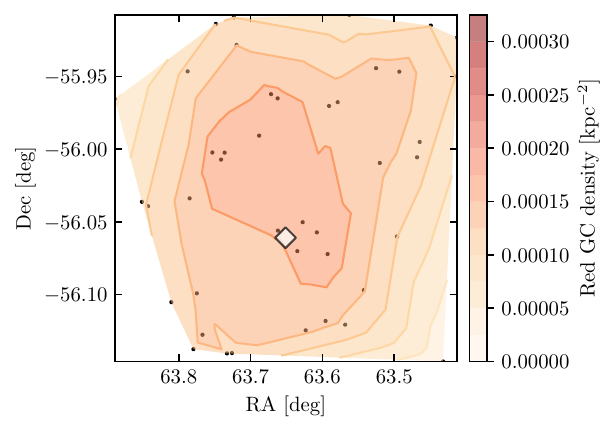}
    \end{subfigure}
    \caption{Bright GC candidates density maps of the full ERO-D FoV (left column), with zooms on the NGC\,1549-NGC\,1553 pair (middle column), and on NGC\,1546 (right column). The colour scale is the same for both zooms, where the density fields were re-evaluated locally using the samples within the cutouts. All GCs, blue GCs, and red GCs are  represented in the upper, middle, and lower rows, respectively. Blue and red GCs are defined relative to the median $\IE - \HE$ colour of the GC candidates, which is not indicative of an intrinsically bimodal population within our data.}
    \label{GC_distribution}
\end{figure*}

Table \ref{tab:gc_counts} presents the bright GC candidates counts for the three main galaxies. To cope with the overlap between NGC\,1549 and NGC\,1553 (see Figs. \ref{fig:dorado_general} and \ref{GC_distribution}), we artificially separated their GC population at an isophote of $25.5\,\text{mag\,arcsec}^{-2}$ in the $\IE$ band (roughly, 33\,kpc or 370\,arcsec for NGC\,1549, 29.5\,kpc or 330\,arcsec for NGC\,1553, and 13.5\,kpc or 150\,arcsec for NGC\,1546). We count GCs within that isophote.

To compare the numbers of expected bright GCs to those we find in Table \ref{tab:gc_counts}, we used the method described in \cite{Euclid_Voggel}. This paper describes a code to predict the total number of GCs in a given galaxy using a Monte Carlo simulation based on the empirical trend with galaxy magnitude and uses the dispersion of GC specific frequency as a function of host stellar mass to estimate the range of expected GCs (see also details in Figs. 3 and 4 of that paper). We find that the mean predicted total number of GCs in NGC\,1553, NGC\,1549, and NGC\,1546 are  475, 324, and 77, respectively, which is consistent with the total GC numbers estimated in \cite{2013ApJ...772...82H}, apart from NGC\,1546. Considering the magnitude limit $\IE=24$ we use in this study for GC detection, we find respectively 259, 178, and 42 GCs that should be brighter than that limit and be detectable. Comparing this number of detected GCs (Table \ref{tab:gc_counts}) with the expected number of GCs from the prediction method, we find that we detected 50\%, 90\%, and 20\% of all potential GCs brighter than our selection limit. The low detection fraction of GCs detected in NGC\,1546 is likely due to much of its outskirts not being covered by the \Euclid FoV, and the choice of the $I_\text{E}=25.5$ isophote as search radius. The 90\,\% of recovered GCs for NGC\,1549 is probably unrealistically high and likely means that the total GC system size prediction is underestimated for this case, and this galaxy hosts an above average number of GCs. The true efficiency of our selection method is likely somewhere between the 50\,\% of NGC\,1553 and the 90\,\% fraction of NGC\,1549.

The spatial distribution of the bright GC candidates across the FoV is presented through the density maps in Fig. \ref{GC_distribution}. Blue GCs seem to be more uniformly distributed, which aligns with expectations if this GC component is accreted from (minor) mergers \citep{2006ARA&A..44..193B}. Red GCs can form either through in situ processes \citep{2018RSPSA.47470616F} or as a result of past mergers involving metal-rich satellites. Consequently, their higher concentration compared to bluer GCs aligns with expectations.

\begin{figure}[htbp!]

\includegraphics[width=\linewidth,trim=0.25cm 0cm 0cm 0cm, clip]{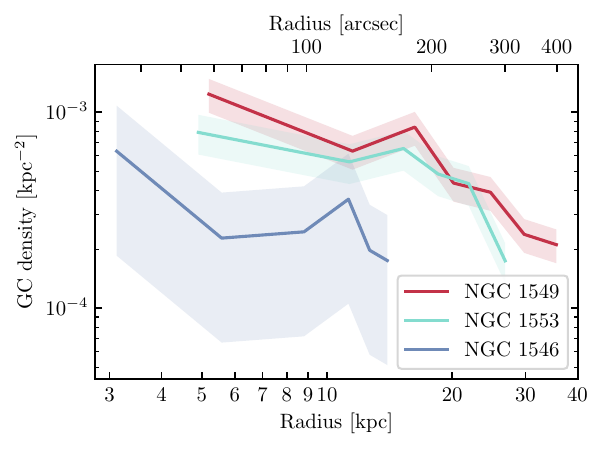}

\vspace{-0.3cm}

\caption{GC density radial profiles for the three galaxies of interest. The shaded area around each curve represents the Poisson noise in the GC counts, computed for each bin.}
\label{fig:GC_profiles}
\end{figure}

We also generated radial density profiles in Fig. \ref{fig:GC_profiles}. To determine the radius at which to stop the GC density estimation, we applied the same criterion detailed earlier. The three galaxies of interest show the expected trend: a high central concentration of GCs that decreases with increasing radius.

\subsection{\label{description} Properties of detected tidal features}

\subsubsection{\label{Detection}Detection of tidal debris}

\begin{figure*}[h]
\centering
\begin{tabular}{cc|c|c}
& NGC\,1549 & NGC\,1553 & NGC\,1546 \\
\begin{sideways}\parbox{50mm}{\centering Surface-brightness map}\end{sideways} & \includegraphics[width=0.28\linewidth]{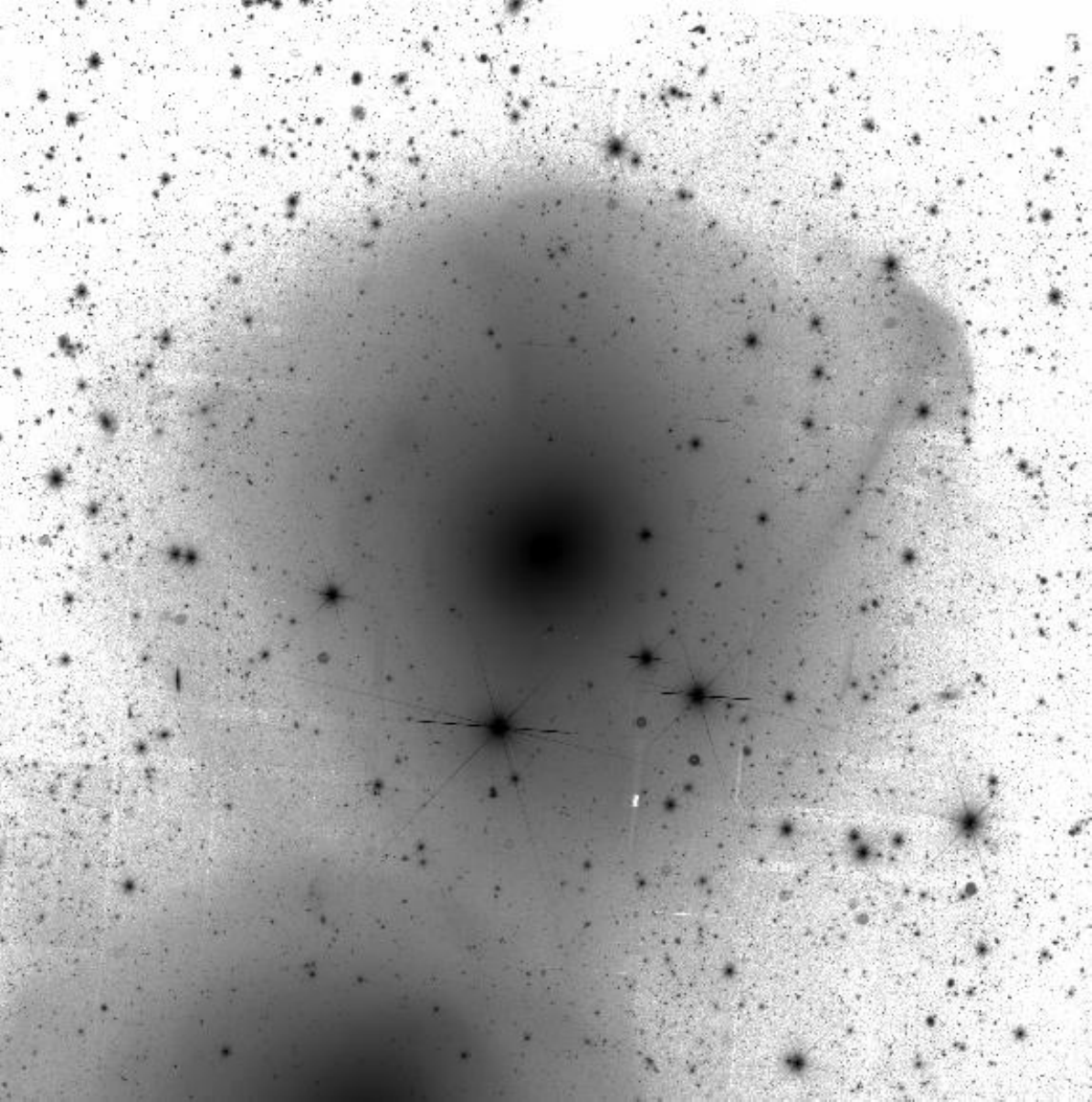} & \includegraphics[width=0.28\linewidth]{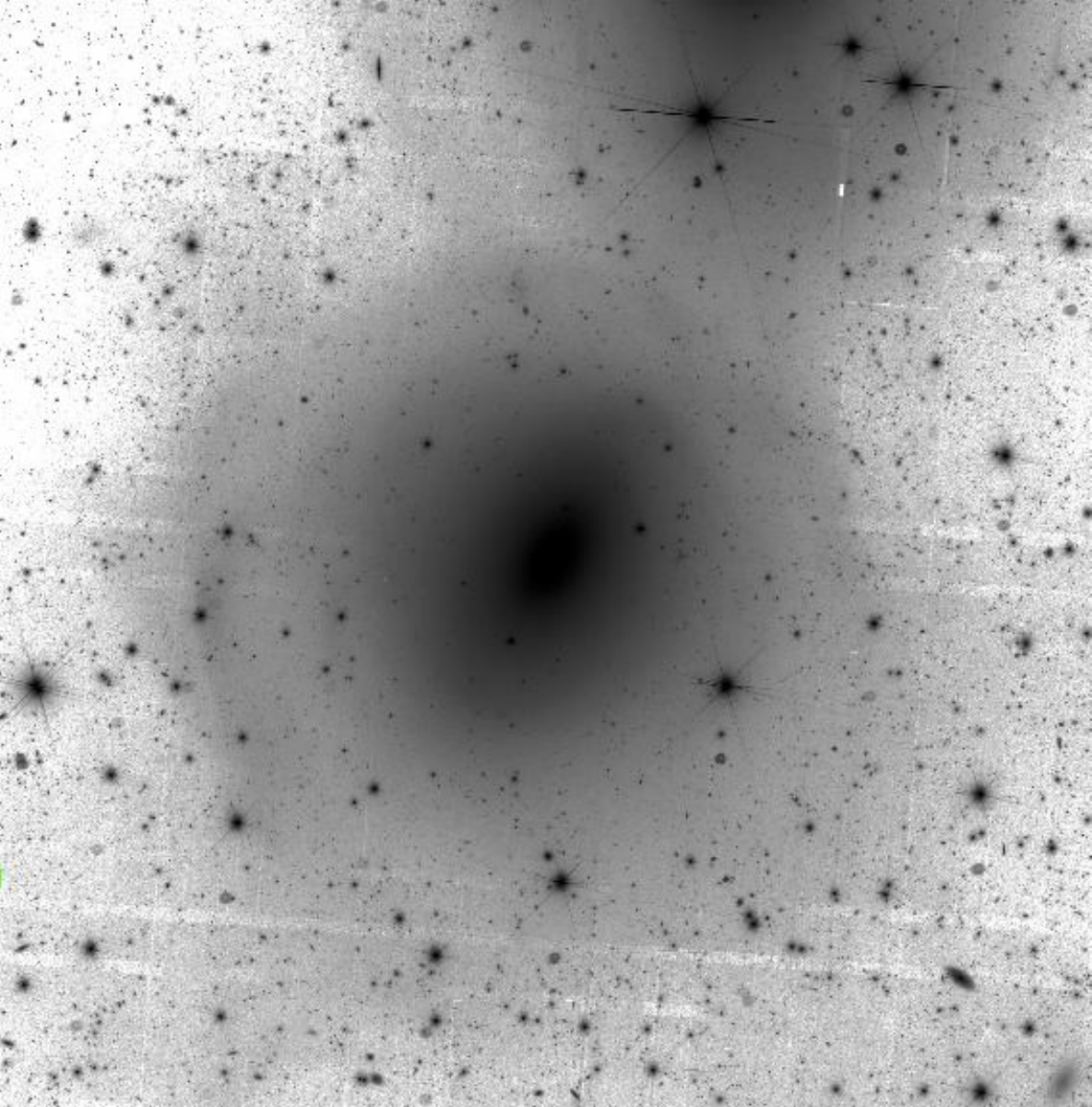} & \includegraphics[width=0.28\linewidth]{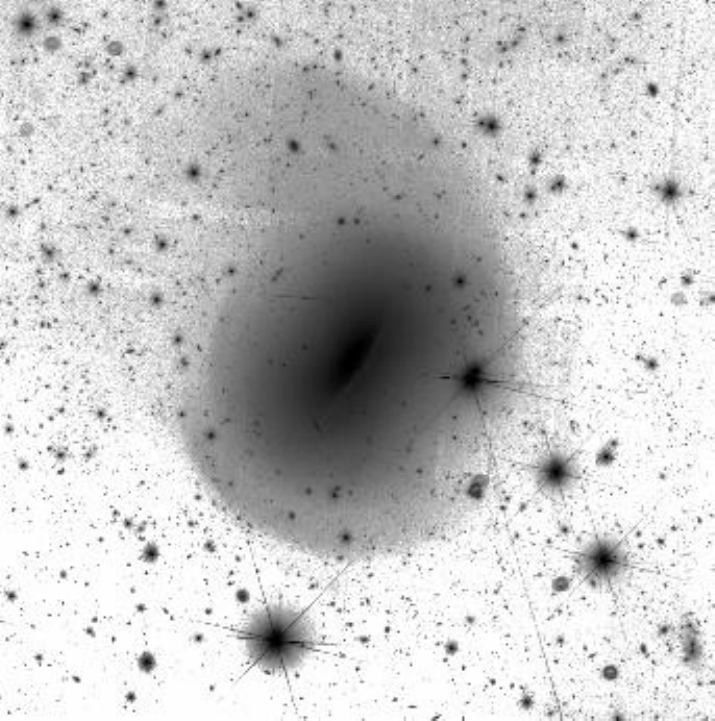} \\
& & & \\
\begin{sideways}\parbox{50mm}{\centering Ellipse fitting residuals}\end{sideways} & \includegraphics[width=0.28\linewidth]{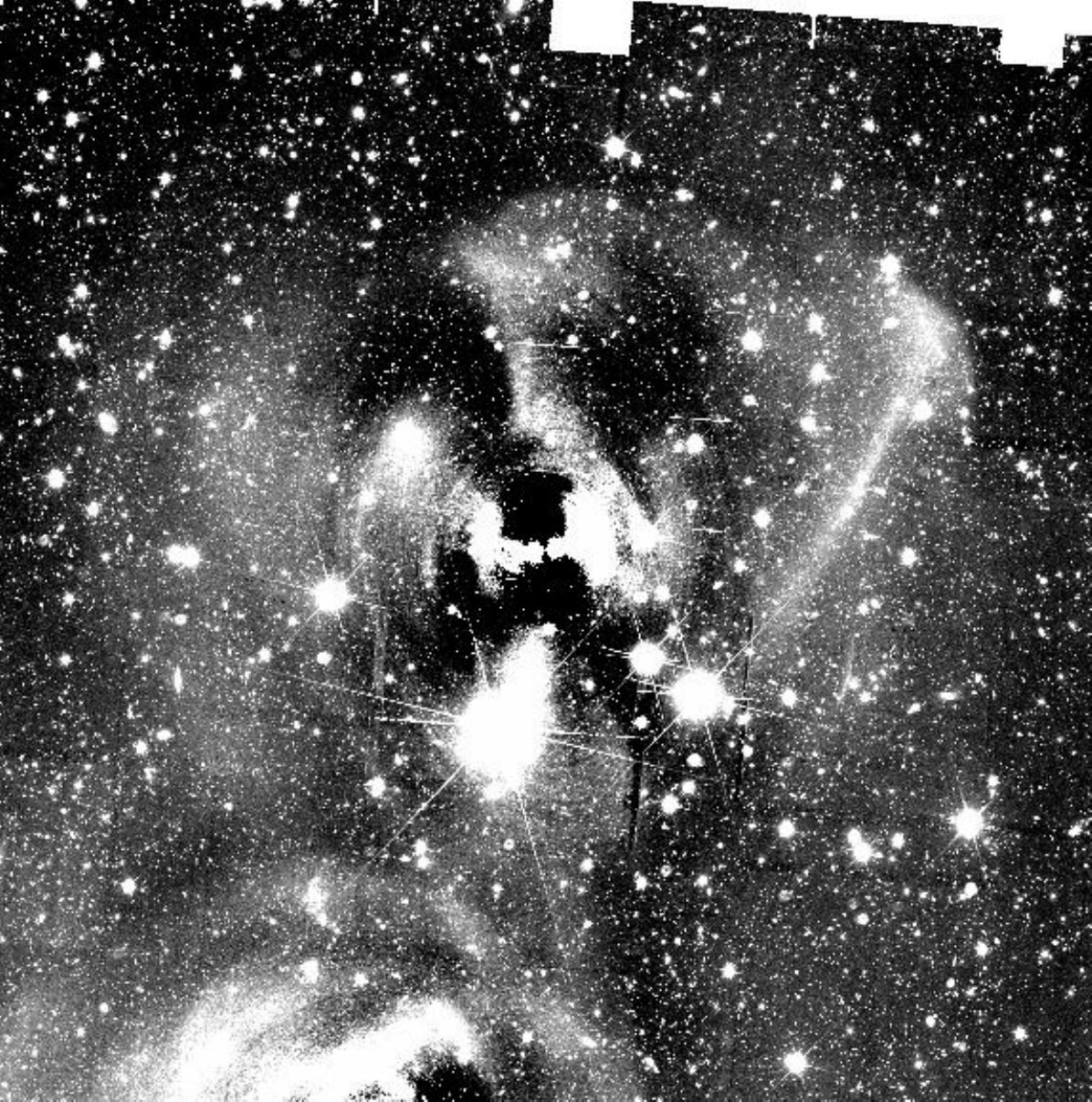} & \includegraphics[width=0.28\linewidth]{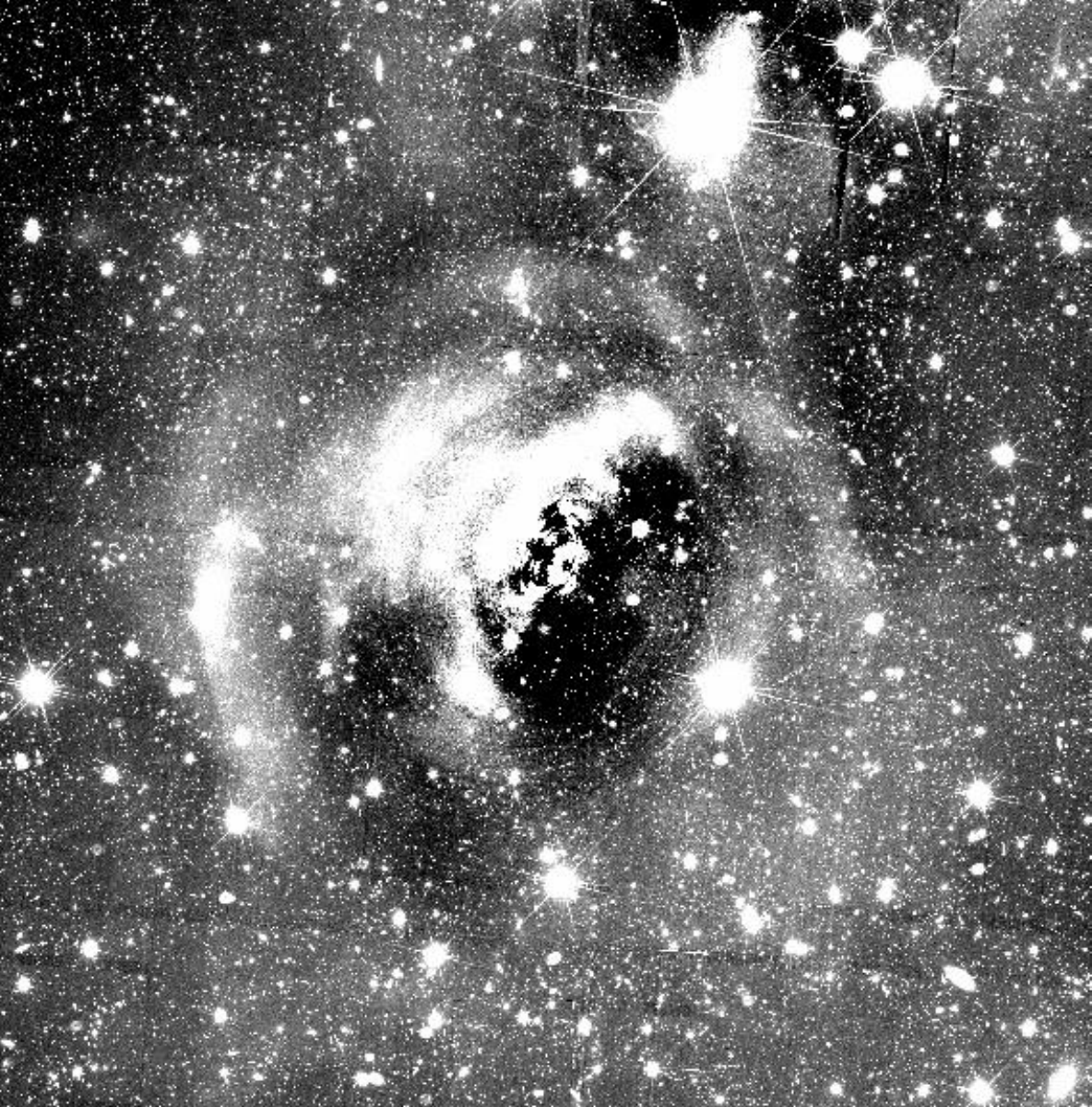} & \includegraphics[width=0.28\linewidth]{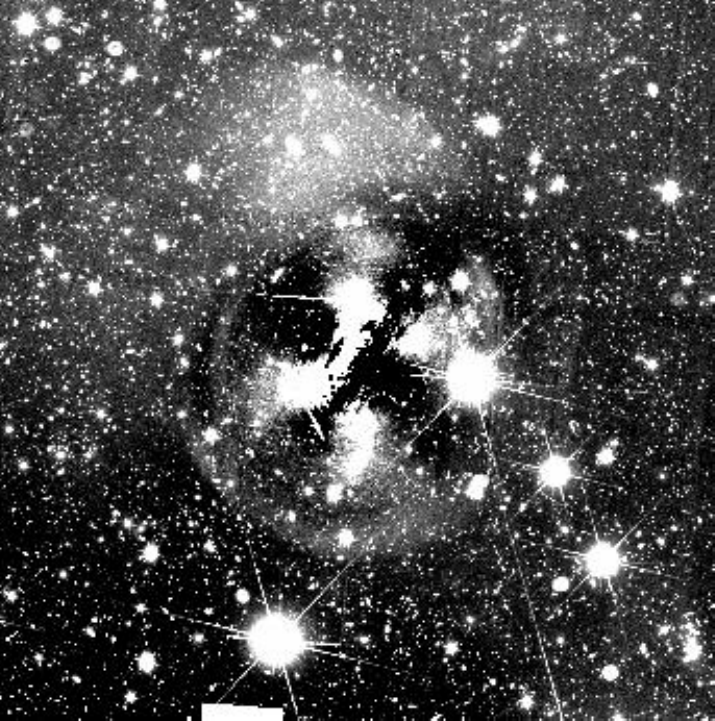} \\
& & & \\
\begin{sideways}\parbox{50mm}{\centering Unsharp masking}\end{sideways} & \includegraphics[width=0.28\linewidth]{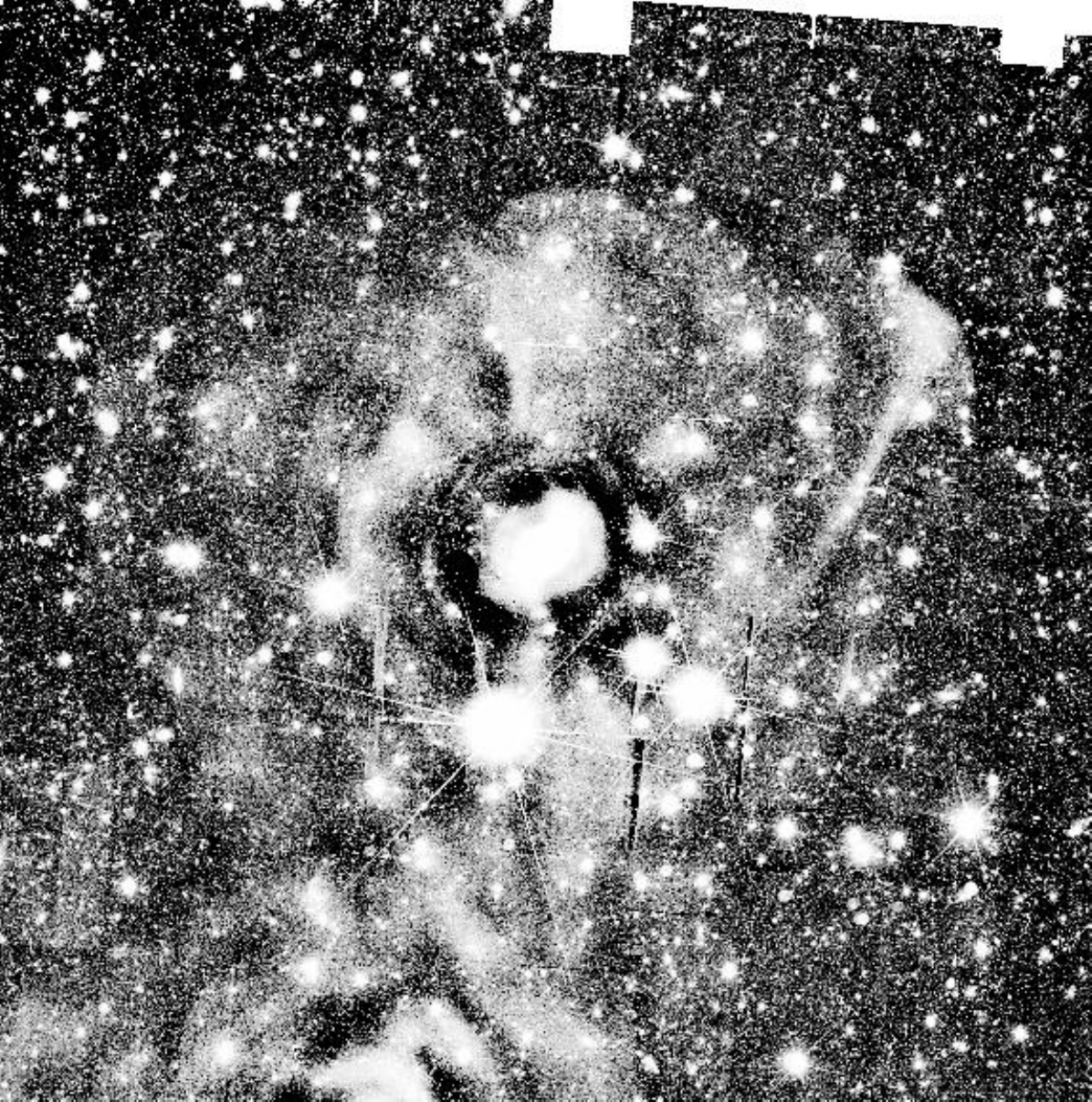} & \includegraphics[width=0.28\linewidth]{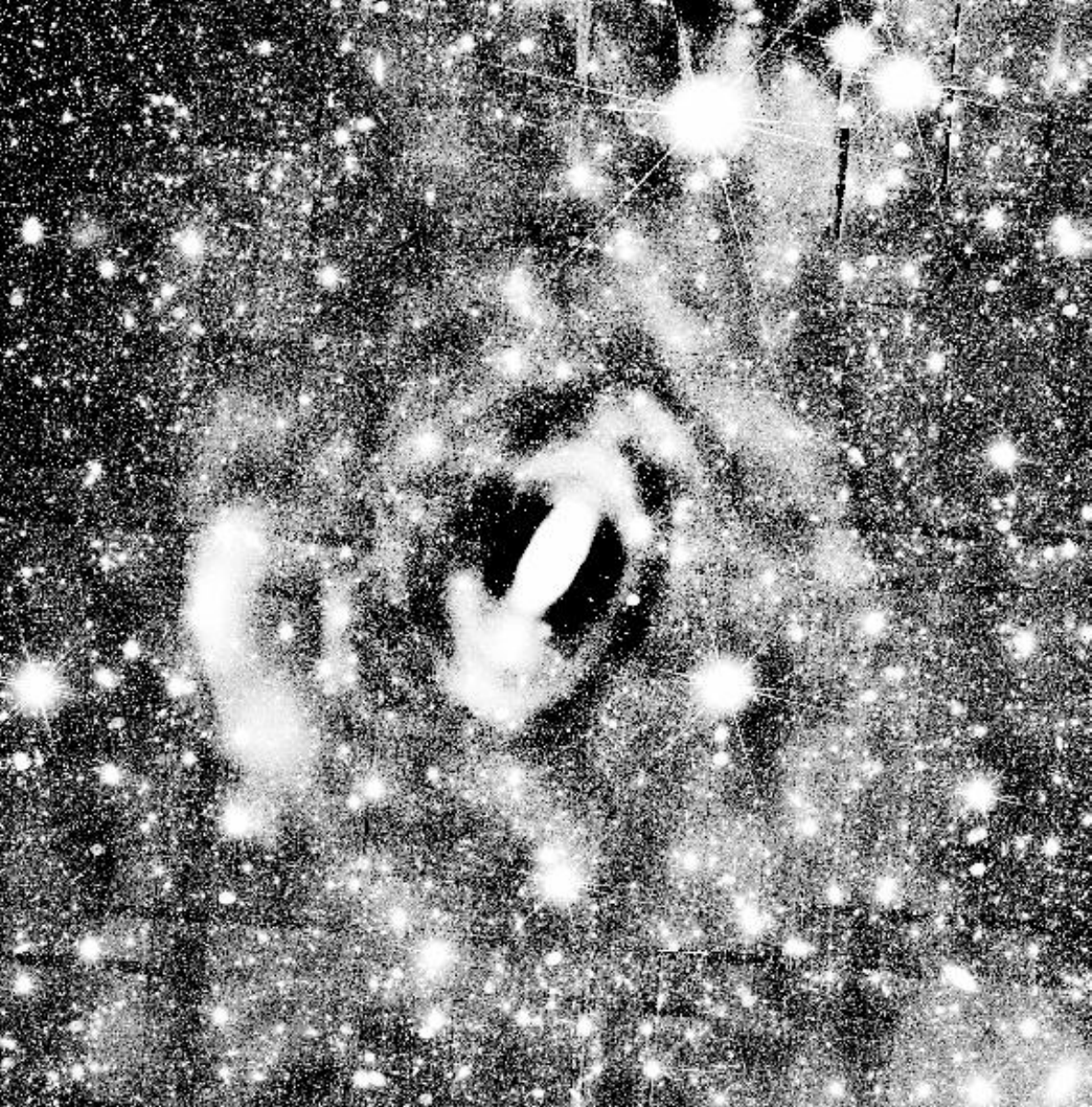} & \includegraphics[width=0.28\linewidth]{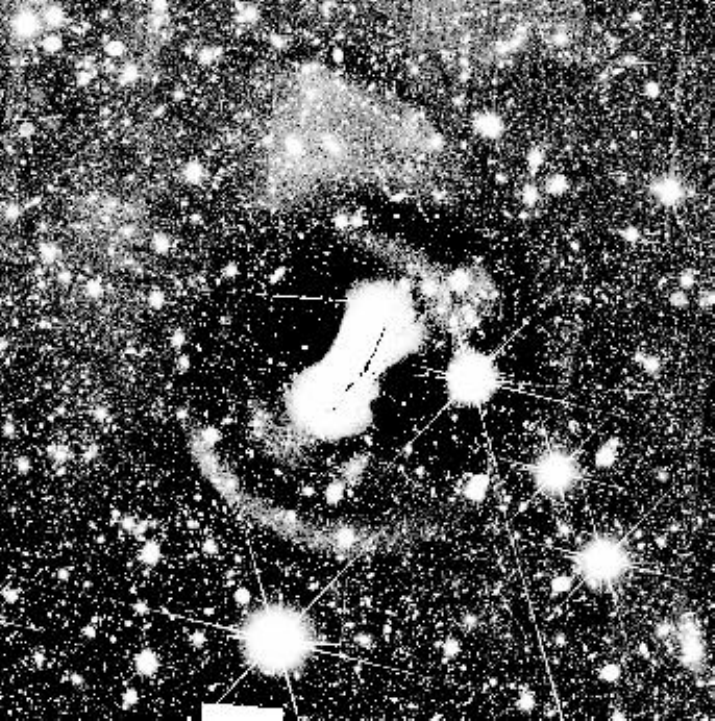} \\
\end{tabular}
\caption{$\IE$ surface brightness maps (top row), ellipse fitting residuals (middle row) and unsharp masked images (bottom row) for re-binned cutouts of the three galaxies of interest. The surface brightness image is in grayscale, black indicating where the surface brightness is below $21\,\text{mag\,arcsec}^{-2}$, and white where it exceeds $27.5\,\text{mag\,arcsec}^{-2}$. The unsharp masked image is produced using a 40 pixel standard deviation width Gaussian kernel. The scale and orientation of the cutouts are identical to those given in Fig. \ref{fig:autoprof_cutouts_and_profiles}. In each figure, North is up, east is to the left.}
\label{table:images}
\vskip -6pt
\end{figure*}

In Fig. \ref{table:images}, we show that the original image together their unsharp masking and ellipse fitting residuals versions present numerous tidal debris, demonstrating the \Euclid space telescope's ability to detect such structures. We identify and classify those structures with the \texttt{Jafar} annotation tool. We list these features, in addition to several dwarf galaxies of interest and internal substructures (like rings or spiral arms), as seen in the top left panel of Fig. \ref{features_colours}. Tidal features have been detected around NGC\,1549, NGC\,1553, and NGC 1546, but not around IC\,2058.

Among those features, on the top-right panel of Fig. \ref{features_colours}, we label with numbers those for which individual photometry was possible, at least in the $\IE$ band where they have been detected. The unlabelled structures are either too close to the galactic centre or they are located in the overlapping regions between NGC\,1549 and NGC\,1553; alternatively, they could be  shells for which obtaining the necessary photometry is challenging \citep[e.g.][]{Sola_et_al_2022}. Some of the labelled features were poorly detected in NIR bands. In particular, the $\YE$ band seems to be the most affected by contaminants specific to the NIR bands, as  discussed in Sect. \ref{sc:Discussion}. Therefore, we did not consider this filter for the individual feature photometry. We also excluded from the colour study those structures that are difficult to detect in the NIR bands. For the latter structures, the noise created by NIR-specific contaminants (including persistence) results in uncertainties ranging from over 0.2 up to several $\text{mag\,arcsec}^{-2}$ in those bands (against below $0.1\,\text{mag\,arcsec}^{-2}$ in the same bands for the structures we retained in the colour study).

Results from the photometry of the  individual features are shown in the top right and bottom left panels of Fig. \ref{features_colours}. We also calculate the median colour of the GC candidates each feature encompasses (bottom right panel of Fig. \ref{features_colours}). Figure \ref{fig:colour-colour} presents a colour-colour diagram of the detected features, with the median colours of the associated GCs overplotted. We can see on this diagram that all the feature colours fall on or near the red-end regions of the stellar population models, excluding the youngest stellar populations ($\approx$\,0.1--1\,Gyr).

When comparing the GC distribution in Fig. \ref{GC_distribution}, we can observe location differences between the peaks of GC density and the ellipse profile centre of the galaxies. While the offset seen on NGC\,1546 could be due to low-number statistics (see also the low GC count of this galaxy in Table \ref{tab:gc_counts}), the offsets found for NGC\,1549 and NGC\,1553 seem  real. Indeed, the scale of those offsets (over $\ang{1.20}$) is much larger than the central region where the galaxy is so bright that it makes detecting GCs impossible (under $\ang{0.18}$).

It is worth mentioning that different offsets are seen in the blue and red GC distribution. If we approximate the total GC distribution around each galaxy by a Gaussian of standard deviation $\sigma$, the offsets vary over $3\sigma$ in the red and blue GC sub-samples, which suggests that these variations are significant. Finally, we did not find a systematic overdensity of GCs at the location of the debris and we speculate about possible reasons below. \begin{itemize}
\item Due to the presence of contaminants, we were compelled to restrict our sample of GC candidates to only the brightest ones ($\IE\leq24$). The clustering of GCs in the vicinity of tidal features may remain undetected without including lower luminosity GCs. Higher resolution and deeper data could help overcome this limitation.
\item The detected tidal features could be formed from the material of small progenitors, such as dwarf galaxies, which do not necessarily contain GCs.
\item Phase-mixing of the GCs associated with tidal features may have erased any evidence of clustering, indicating that the mergers we observe traces of are not very recent.
\item If we consider the tidal features as remnants of the former disks of LTG progenitors that formed the galaxies observed in this study, we could hypothesise that there were few or no in situ GCs within these disks. Thus, we predominantly observe ex situ GCs here. They remain gravitationally bound to their host galaxy’s potential even during interactions with other galaxies, leaving the GC distribution unaffected and preventing any clustering from occurring.
\item For the galaxies involved in a merger, the GCs are not originally predominantly located in a disk but, rather, distributed in a halo. Consequently, they do not end up along the tidal features coming from the elongation of the former spiral arms.
\end{itemize}

\subsubsection{NGC\,1549}

In Fig. \ref{fig:colour_profiles}, we see that the colour profile of NGC\,1549 is flat, and is flatter than
NGC\,1553 in $\IE-\JE$. This is also visible with the help of the colour maps in Fig. \ref{fig:colour_maps}. An interpretation of this is discussed in  Sect. \ref{merger_history}.

The unsharp masked image of NGC\,1549 (Fig. \ref{fig:center_NGC1549}) shows an internal substructure similar to diffuse spiral arms, with no visible star-formation region. The origin of this feature is uncertain. It could be the remnant of ancient galactic disk and bar, whose dynamics has been modified by a merger. This explanation would be consistent with UV data and traces of possible ancient star formation described in the literature (\citealt{2020A&A...643A.176R,2021JApA...42...31R,2022A&A...664A.192R}; see also Sect. \ref{sc:Intro}).

A large tidal feature (6) is visible to the north, and an umbrella-shaped structure composed of another tidal feature (1) and a shell is visible to the west. (6) is one of the bluest features detected, and (1) is one of the reddest. The NGC\,1549 ellipse profile presents a slight bump from 13.38\,kpc to 44.60\,kpc (from 150 to 500\,arcsec) from the galactic centre. This coincides with the position and extent of the tidal features (1) and (6). Their GCs photometry shows that these features might host red GCs. Finally, NGC\,1549 also features a system of shells.

In the southern part of this galaxy in the direction of NGC\,1553, three tidal features are visible, notably on the images of unsharp masking and ellipse fitting residuals, but remain very uncertain. We did not label them, but we note that one of them could be the southern extension of the northern large tidal feature (6).

\subsubsection{NGC\,1553}

The colour profile trend of NGC\,1553 (see Fig. \ref{fig:colour_profiles})  varies slightly with the choice of the photometric band. A blueward slope is observed with increasing radius for colours involving the $\IE$ band. For other colours, the profile is flatter, without excluding a slope if we consider its uncertainties.

This galaxy has the second highest peak density of GCs (see Fig. \ref{GC_distribution}, Table \ref{tab:gc_counts}). The distribution of GCs appears unusual. In its central region, the eastern side seems dominated by red GCs, while the western side is primarily populated by blue GCs. This asymmetry could be attributed to several factors, detailed below.
\begin{itemize}
\item The effect could be real and due to a specific merger history, such as the accretion of a more metal-poor population along a preferred direction.
\item Depending on the galaxy orientation, there may be more galactic material along the line of sight between the observer and the GCs on one side of the galaxy compared to the other, causing varying levels of extinction across the two sides.
\item It could also be a combination of low-number statistics effects, typical central concentration of red GCs in the host galaxy, and the fact that blue GCs tend to have a shallower distribution around the host galaxy.
\end{itemize}

Figure \ref{fig:center_NGC1553} shows the NGC\,1553 core, featuring dust lanes, a bar and diffuse spiral arms without star formation. Internal regions of this galaxy have a ring system. The most internal ring is spiral shaped and is particularly enlightened in unsharp masked images, as depicted in Fig. \ref{fig:center_NGC1553}. The NGC\,1553 ellipse profile presents a bump around \ang{;;40} from the galactic centre, probably caused by this ring/spiral structure.

Further from the centre, two `lobe' features and a possible small tidal feature are observed, especially in the surface brightness and unsharp masked images (second row, and the first and third lines of Fig. \ref{table:images}). A second bump, around \ang{;;100}, is very small but more prominent in the $\YE$ band, and seems to be caused by those features. Moreover, a system of shells is detected. Finally, the halo is distorted toward the northeast and southwest, yet without forming any elongated tidal structures. We have classified those distortions as the plumes (4) and (8). The feature (8) appears to be the faintest feature of this set.

\subsubsection{NGC\,1546}

This is the first time that the tidal feature system of NGC\,1546 is described in detail. The inspection of the image reveals the singular, if not unique morphology of NGC\,1546. While Fig. \ref{fig:dorado_general} shows, in its outer regions, an extended and disturbed halo reminiscent of an elliptical galaxy, we see in Fig. \ref{fig:center_NGC155} that in its outer regions, an entirely unperturbed flocculent disk galaxy.

\begin{figure*}[h!]
\centering
\begin{tabular}{cc}
\includegraphics[width=0.487\linewidth,trim=0.25cm 0cm 0.15cm 0cm, clip]{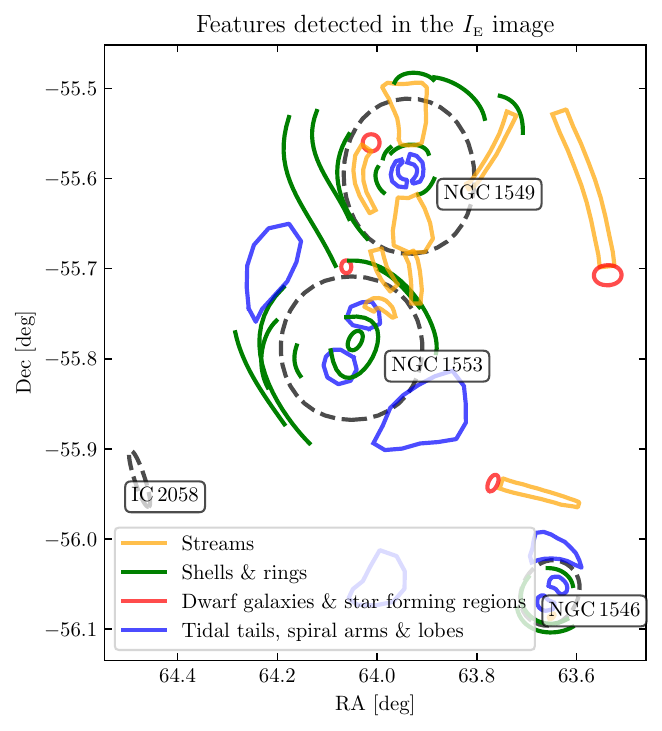} & 
\includegraphics[width=0.503\linewidth,trim=0.25cm 0cm 0.25cm 0cm, clip]{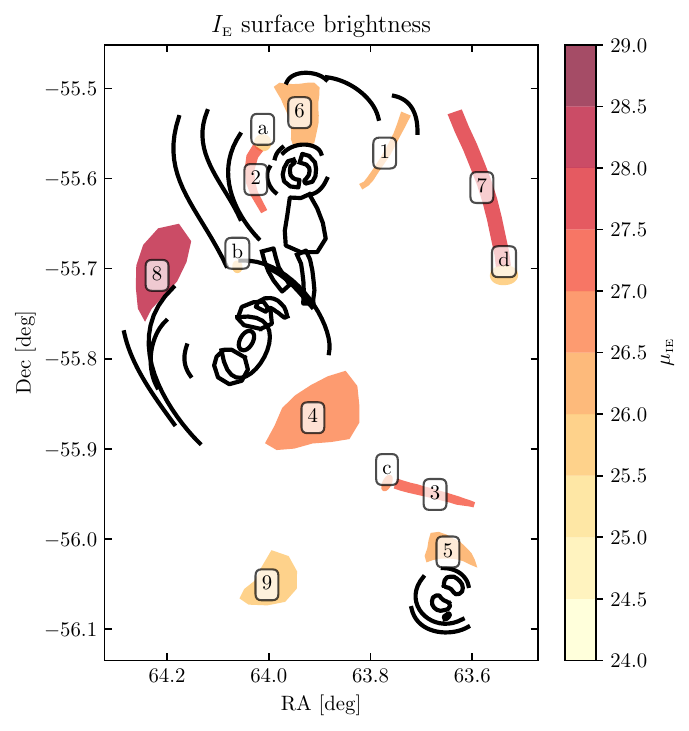} \\
\\[0.1em]
\includegraphics[width=0.495\linewidth,trim=0.25cm 0cm 0.25cm 0cm, clip]{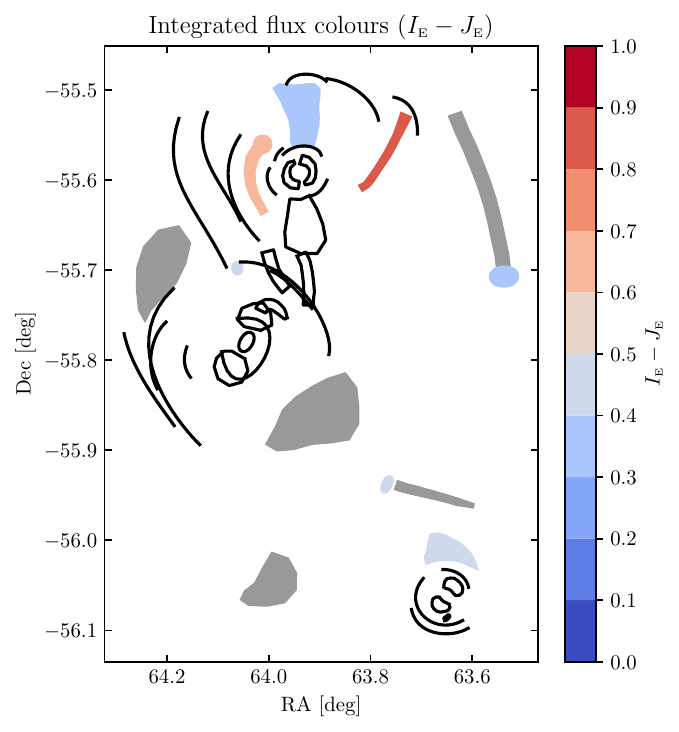} & 
\includegraphics[width=0.495\linewidth,trim=0.25cm 0cm 0.25cm 0cm, clip]{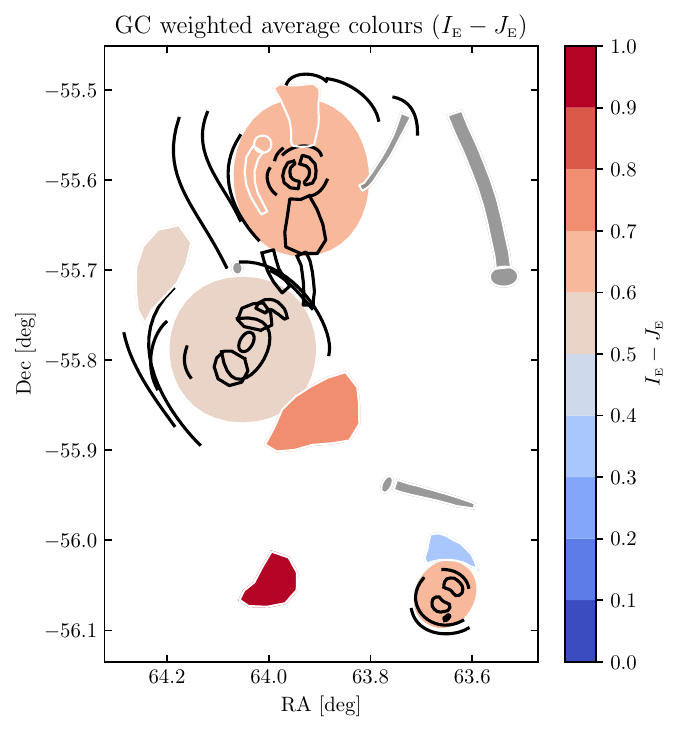} \\
\end{tabular}
\caption{\emph{Upper left}: Features detected for each galaxy with the help of the $\IE$ images and residuals. \emph{Upper right}: $\IE$ surface brightness map of the features. Those for which $\IE$ photometry can be performed are coloured according to their surface brightness and labelled. Their photometry is given in Table \ref{tab:features}. They are discussed in the context of each galaxy they belong to in Sect. \ref{description}. Questions relative to their classification and nature are addressed in Sect. \ref{merger_history}. Those for which this study is not possible are delineated in black. Especially, the photometry of the shells, of the uncertain features between NGC\,1549 and NGC\,1553, and of the features close to the galactic centres is not estimated. \emph{Lower left}: $\IE-\JE$ integrated fluxes colour map. The features with uncertain detection in the NIR bands appear in grey.  \emph{Lower right}: $\IE-\JE$ GCs colour map. The average colours are weighted based on the uncertainties of the magnitudes that contribute to them. The features that encompass less than three bright GC candidates appear in grey. Surface brightness and colours are given respectively in mag\,arcsec$^{-2}$ and in mag.}
\label{features_colours}
\end{figure*}

\begin{figure*}[h!]
  \centering
  \begin{subfigure}[t]{0.495\textwidth}
    \centering
    \includegraphics[width=\textwidth]{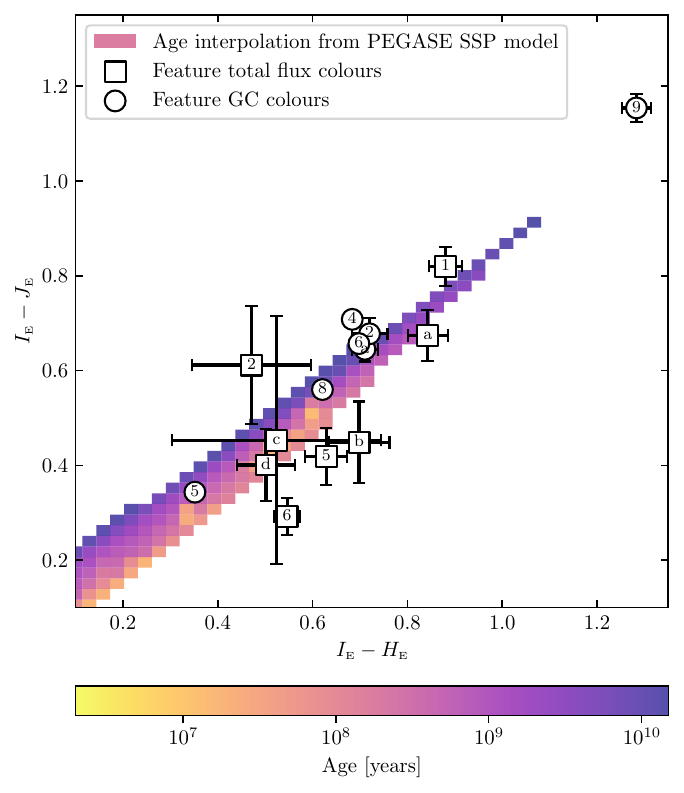}
    \label{subfig:age}
  \end{subfigure}
  \hfill
  \begin{subfigure}[t]{0.495\textwidth}
    \centering
    \includegraphics[width=\textwidth]{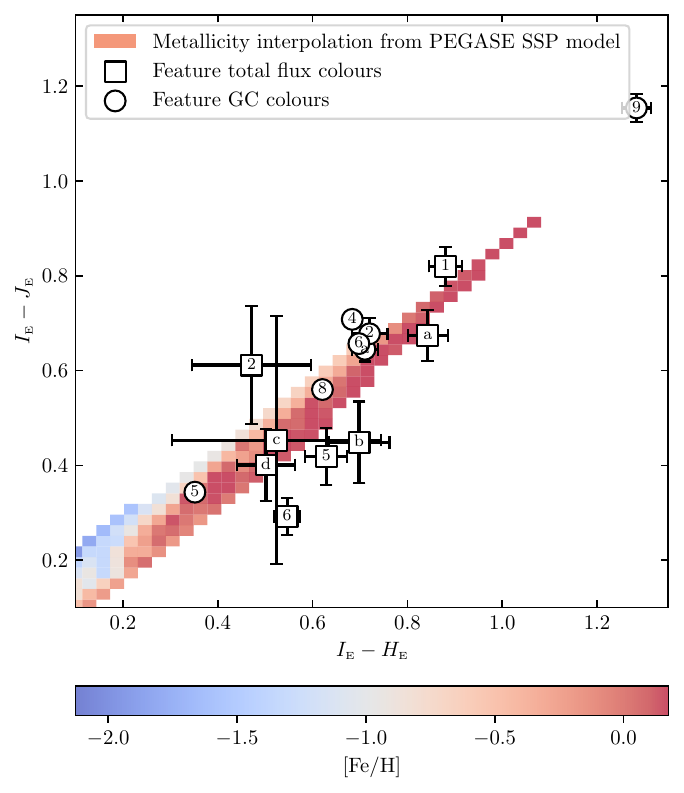}
    \label{subfig:metal}
  \end{subfigure}
  \vspace{-0.8cm}
  \caption{Colour-colour plot of the ERO Dorado features, for both total flux photometry (square labels) and GCs photometry (round labels). The colours are given in magnitude. The interpolated metallicities and ages are obtained using the PEGASE \citep{2004A&A...425..881L} single stellar population (SSP) model. However, the colour-coded area is consistent in age and metallicity across all the other SSP models tested. The feature labels are taken from Fig. \ref{features_colours}. Their magnitude and colours are available in Table \ref{tab:features}.}
  \label{fig:colour-colour}
\end{figure*}

\begin{table*}[h!]
\caption{Photometry of individual features.}
\centering
\renewcommand{\arraystretch}{1.2}
\setlength{\extrarowheight}{2pt}

\begin{tabular}{clccccccccccc}
\hline \hline
\noalign{\vskip 4pt}
ID & Type & $\mu_\sfont{IE}$ & $\IE-\JE$ & $\IE-\HE$ & $(\IE-\JE)_{\rm GC}$ & $(\IE-\HE)_{\rm GC}$ & GC count  \\
(1) & (2) & (3) & (4) & (5) & (6) & (7) & (8) \\
\noalign{\vskip 4pt}
\hline
\noalign{\vskip 4pt}
1 & Stellar stream & $26.13 \pm 0.02$ & $0.82 \pm 0.05$ & $0.88 \pm 0.05$ & \dots & \dots & 1 \\
2 & Stellar stream & $27.18 \pm 0.02$ & $0.61 \pm 0.13$ & $0.47 \pm 0.14$ & $0.68 \pm 0.03$ & $0.72 \pm 0.04$ & 4 \\
3 & Stellar stream & $27.21 \pm 0.02$ & \dots & \dots & \dots & \dots & 0  \\
4 & Plume & $26.65 \pm 0.18$ & \dots & \dots & $0.71 \pm 0.02$ & $0.68 \pm 0.02$ & 8  \\
5 & Tidal tail & $26.43 \pm 0.02$ & $0.42 \pm 0.07$ & $0.63 \pm 0.05$ & $0.34 \pm 0.02$ & $0.35 \pm 0.02$ & 4 \\
6 & Stellar stream & $26.42 \pm 0.02$ & $0.29 \pm 0.05$ & $0.55 \pm 0.04$ & $0.66 \pm 0.02$ & $0.70 \pm 0.02$  & 16 \\
7 & Stellar stream & $27.61 \pm 0.02$ & \dots & \dots  & \dots & \dots  & 1   \\
8 & Plume & $28.00 \pm 0.54$ & \dots & \dots & $0.56 \pm 0.02$ & $0.62 \pm 0.02$  & 5  \\
9 & Isolated plume & $25.87 \pm 0.29$ & \dots & \dots & $1.15 \pm 0.03$ & $1.28 \pm 0.03$  & 3  \\
a & Dwarf galaxy & $25.70 \pm 0.02$ & $0.64 \pm 0.06$ & $0.88 \pm 0.05$ & $0.64 \pm 0.03$ & $0.71 \pm 0.03$ & 5 \\
b & Star-forming region & $25.66 \pm 0.02$ & $0.45 \pm 0.10$ & $0.84 \pm 0.05$ & \dots & \dots & 2 \\
c & Dwarf galaxy & $26.60 \pm 0.02$ & $0.45 \pm 0.27$ & $0.52 \pm 0.23$ & \dots & \dots & 0 \\
d & Dwarf galaxy & $25.96 \pm 0.05$ & $0.40 \pm 0.09$ & $0.40 \pm 0.07$ & \dots & \dots & 1  \\
\noalign{\vskip 4pt}
\hline \hline
\end{tabular}
\label{tab:features}
\tablefoot{Surface brightness, $\mu_\sfont{IE}$, and colours are given respectively in mag\,arcsec$^{-2}$ and in mag. Column (3) is the surface brightness in the $\IE$ band derived from the total flux. Columns (4) and (5) are the colours extracted from the total flux photometry. The surface brightness uncertainties are calculated taking into account both the map used to correct the $\IE$ image and residual background variations. No measure is given for the features which are not well detected in the NIR bands. Columns (6) and (7) are the weighted mean colours of the bright GC candidates contained in each feature. The uncertainties take into account both the \texttt{SExtractor} error estimates and the number of GCs in each feature. No measure is given for the features which have two or less GC candidates. Column (8) is the (non-corrected) number of bright GC candidates for each feature.}
\end{table*}

\begin{figure*}[ht!]
  \centering
  \begin{subfigure}[b]{0.495\textwidth}
    \centering
    \includegraphics[width=\textwidth,trim=0cm 0cm 0cm 0cm, clip]{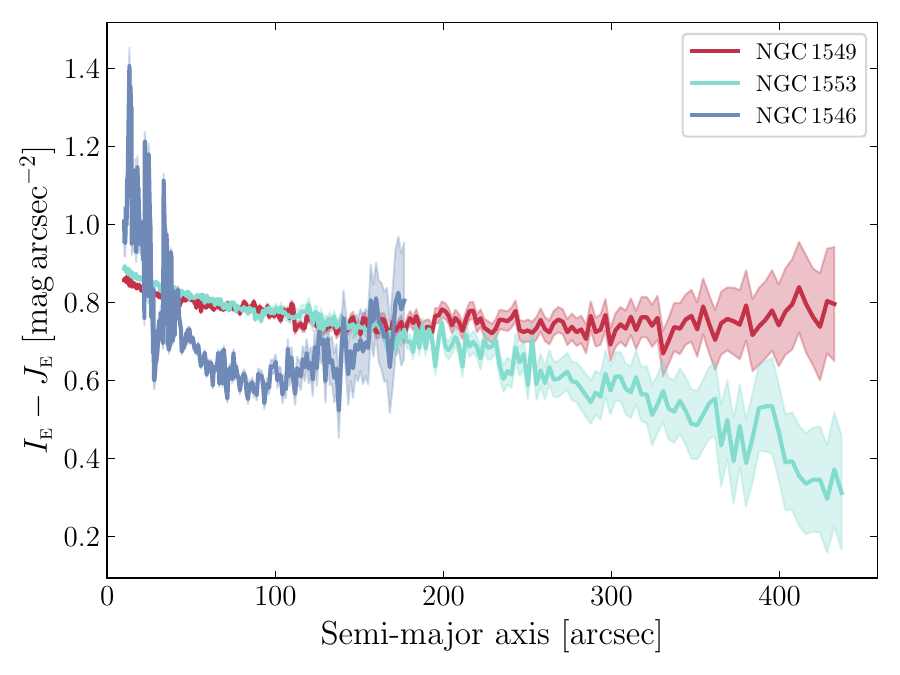}
  \end{subfigure}
  \hfill
  \begin{subfigure}[b]{0.495\textwidth}
    \centering
    \includegraphics[width=\textwidth,trim=0cm 0cm 0cm 0cm, clip]{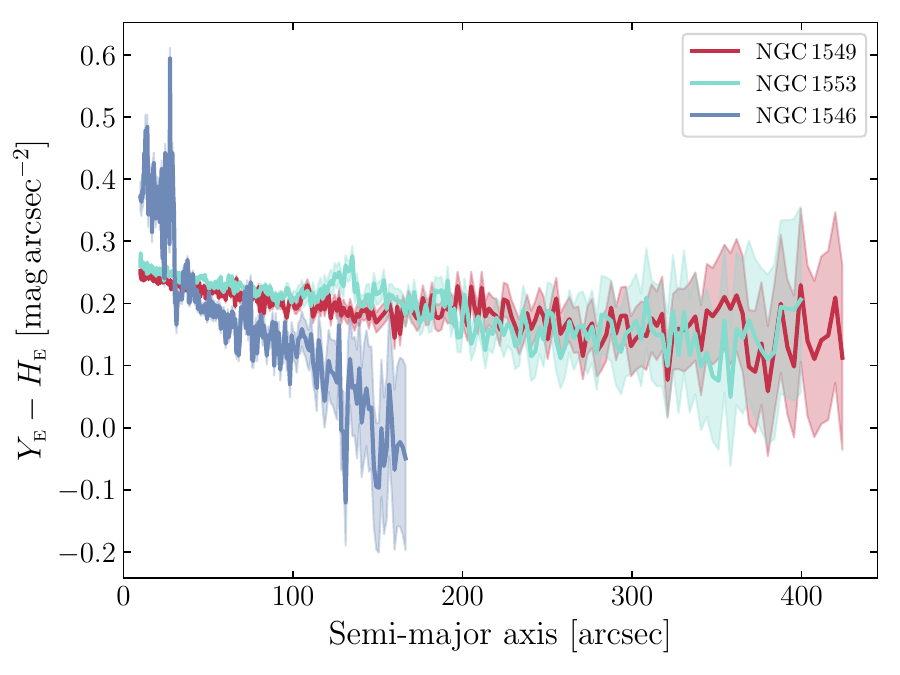}
  \end{subfigure}
  \vskip -5pt
  \caption{Colour profiles for NGC\,1549, NGC\,1553, and
NGC\,1546.  \emph{Left}: $\IE-\JE$. \emph{Right}: $\YE-\HE$. Such plots allow for the analysis of the differences between the galaxy profiles. In particular, the profile of NGC\,1549 is the flattest, while that of NGC\,1546 is the steepest across all colours.}
  \label{fig:colour_profiles}
\end{figure*}

\begin{figure*}[ht!]
  \centering
  \begin{subfigure}[b]{0.495\textwidth}
    \centering
    \includegraphics[width=\textwidth,trim=1cm 0.2cm 0cm 0cm, clip]{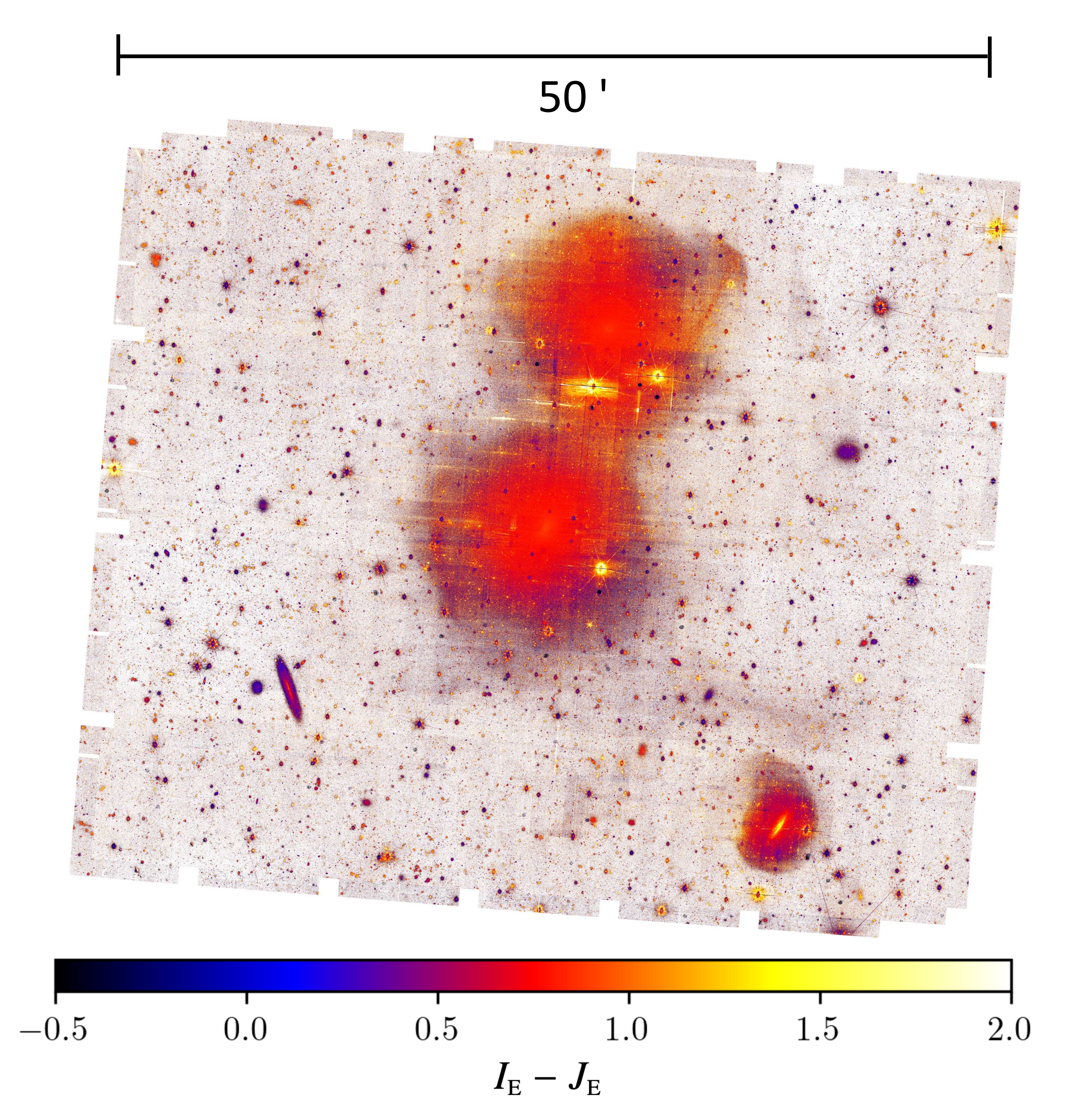}
  \end{subfigure}
  \hfill
  \begin{subfigure}[b]{0.495\textwidth}
    \centering
    \includegraphics[width=\textwidth,trim=1cm 0.2cm 0cm 0cm, clip]{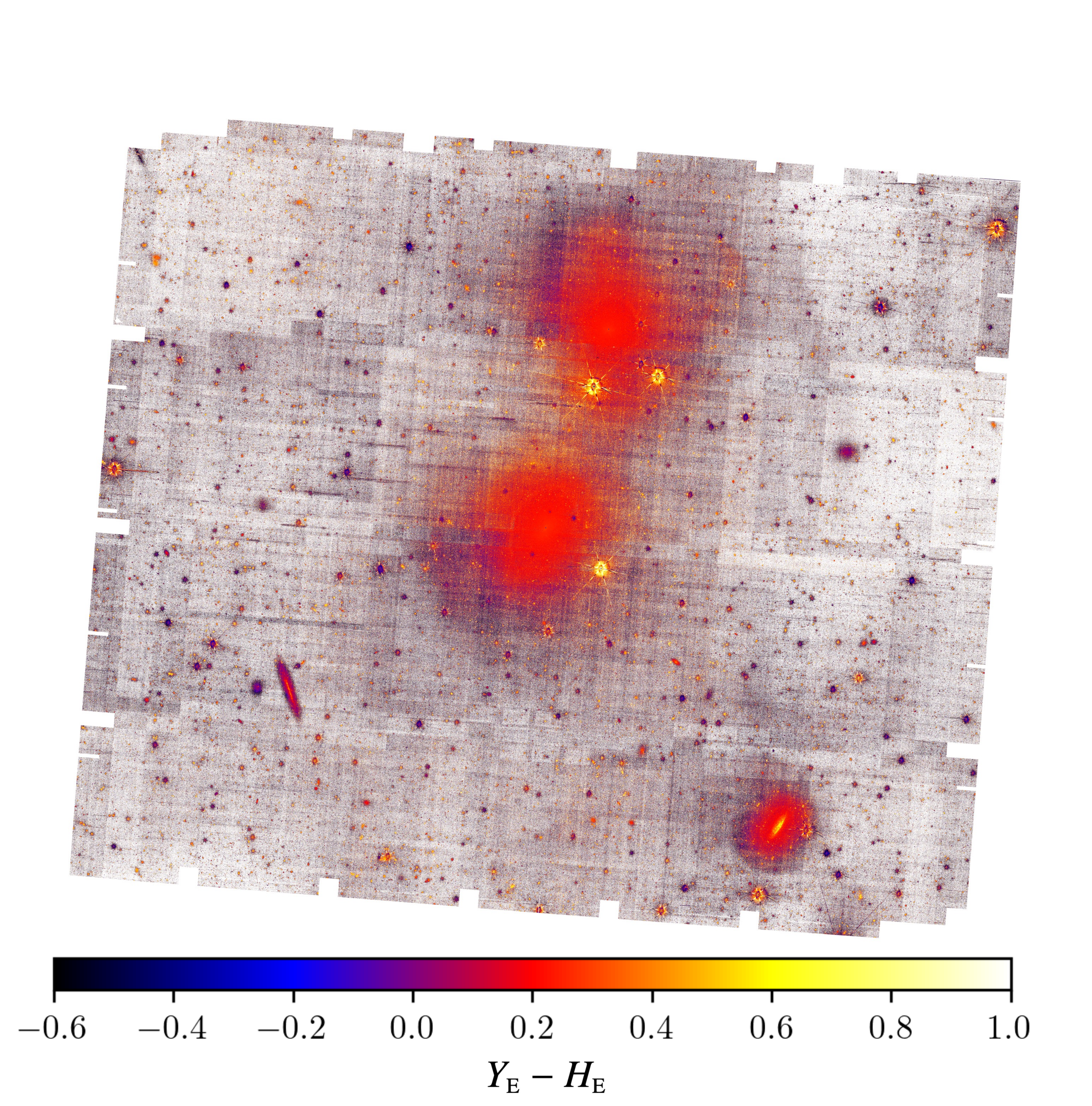}
  \end{subfigure}
  \caption{Colour maps derived from ERO Dorado images.  \emph{Left}: $\IE-\JE$. \emph{Right}: $\YE-\HE$. Such images allow for the qualitative analysis of colour differences between diffuse structures, as well as between central galactic regions and the stellar halo. In those figures, north is up, east is to the left.}
  \label{fig:colour_maps}
\end{figure*}

NGC\,1546 inner features are lobes and a small tidal feature. Further out, there is a shell system and a blue tidal feature (5). As previously mentioned, the blue GCs dominating this structure provoke an overdensity in the distribution (see Fig. \ref{GC_distribution}). This can be the reason why the GC density profile of this galaxy rises significantly at large radii, as seen in Fig. \ref{fig:GC_profiles}. However, this behaviour could also result from the low-number statistics and therefore high uncertainties on this profile. GCs might be missing in the inner galaxy regions because of poor detection in this dusty environment. They could be, for instance, hidden or reddened by the dust and fall below our detection threshold or colour criteria. In the Sect. \ref{merger_history}, we will discuss how the NGC\,1546 morphology and GCs still strongly constrain its history of galactic mergers.

\begin{figure}[htbp!]
  \centering
  \hfill
  \begin{subfigure}[b]{\linewidth}
    \centering
    \includegraphics[width=\textwidth,trim={0cm 0cm 0cm 0cm},clip]{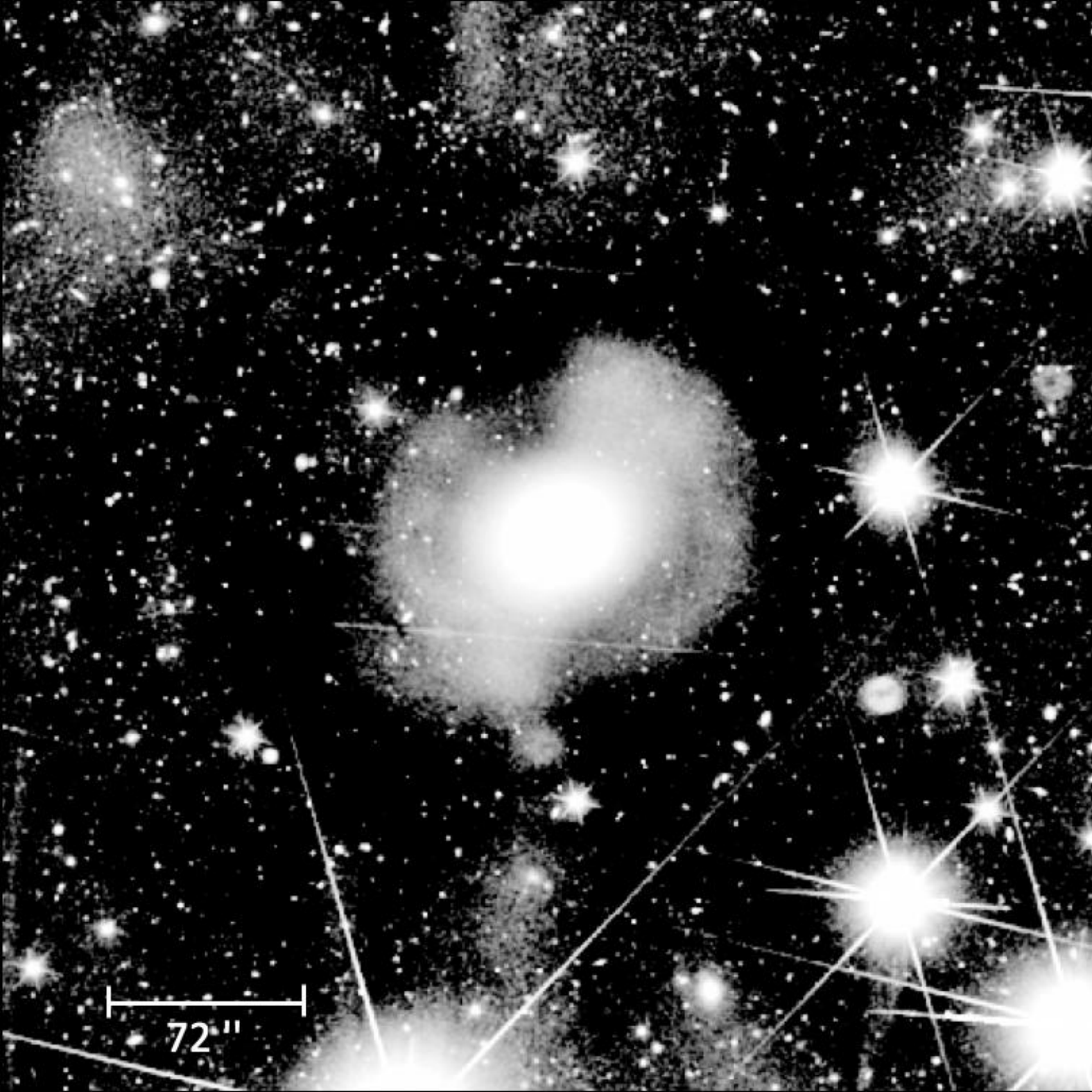}
  \end{subfigure}
  \caption{$\IE$ unsharp masked image of the NGC\,1549 centre, unveiling a bar and/or ``pseudo-spiral arms''. In this figure, north is up, east is to the left.}
  \label{fig:center_NGC1549}
\end{figure}

In the NGC\,1546 ellipse profile, the innermost 40\,arcsec are difficult to interpret due to the presence of arms and strong dust lanes in the disk, making the ellipse profile barely estimable. The tidal tail (5) and the shells are responsible for the bumps located from \ang{;;150} to \ang{;;500} from the centre.

\begin{figure*}[htbp!]
\centering
    \includegraphics[width=\linewidth,trim={0.1cm 0.8cm 0.1cm 0.1cm},clip]{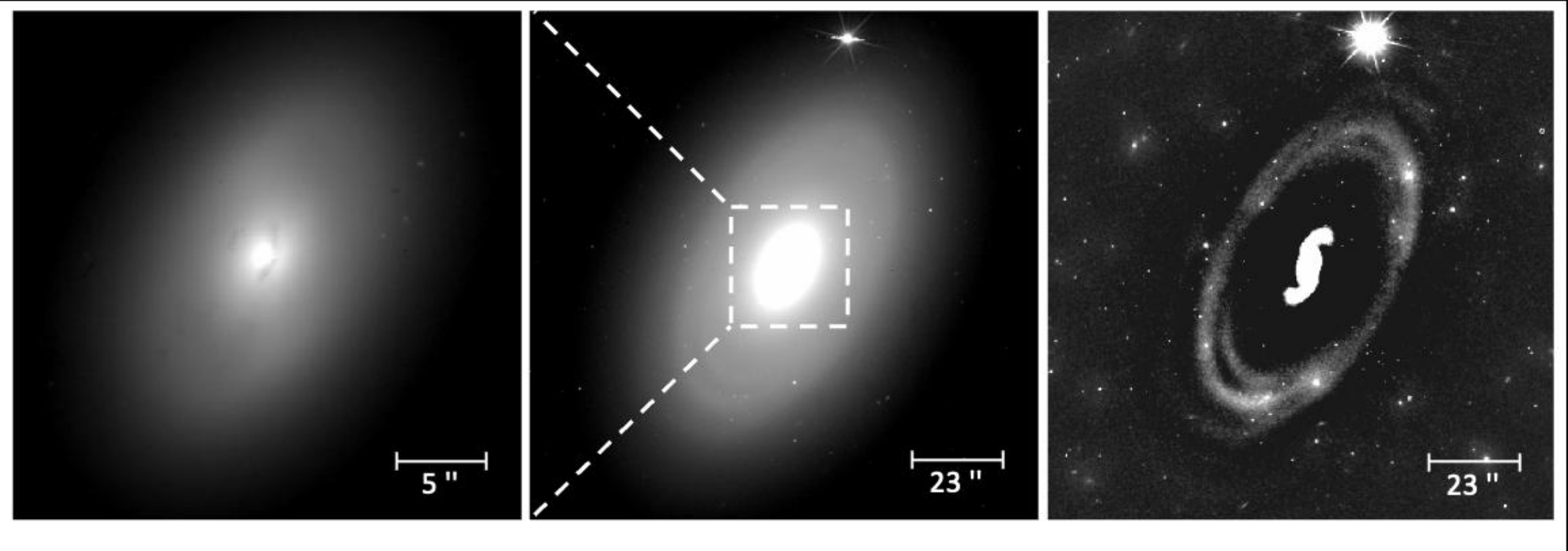}
    \caption{$\IE$ image zooms on the central part of NGC\,1553. In the left panel zoom, a dust lane is visible. \emph{Centre}: Another zoom without any additional processing. \emph{Right}: Unsharp masked corresponding image (using a 10 pixel standard deviation width Gaussian kernel) showing a ring feature composed of different arms and the spiral arms of the very central part. In this figure, North is up, east is to the left.}
  \label{fig:center_NGC1553}
\end{figure*}

\begin{figure}[htbp!]
  \centering
  \hfill
  \begin{subfigure}[b]{\linewidth}
    \centering
    \includegraphics[width=\textwidth,trim={0cm 0cm 0cm 0cm},clip]{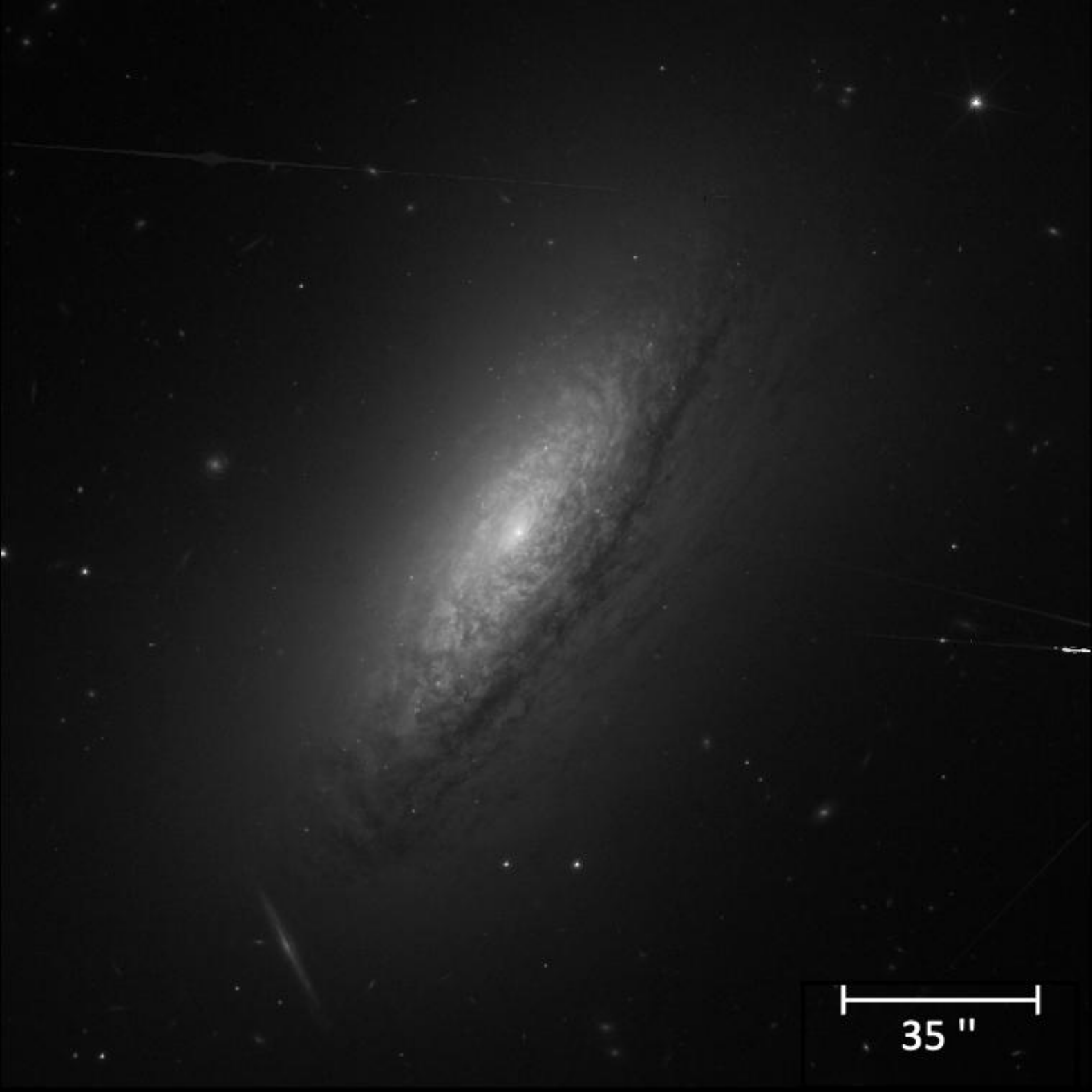}
  \end{subfigure}
  \caption{$\IE$ image of the disk and dust lanes of NGC\,1546, revealing its floculent nature. In this figure, north is up, east is to the left.}
  \label{fig:center_NGC155}
\end{figure}

\begin{figure}[htbp!]
  \centering
  \begin{subfigure}[b]{0.495\linewidth}
    \centering
    \includegraphics[width=\textwidth]{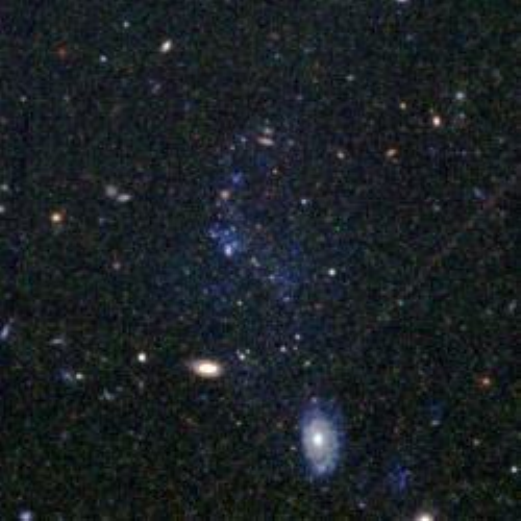} 
  \end{subfigure}
  \hfill
  \begin{subfigure}[b]{0.495\linewidth}
    \centering
    \includegraphics[width=\textwidth]{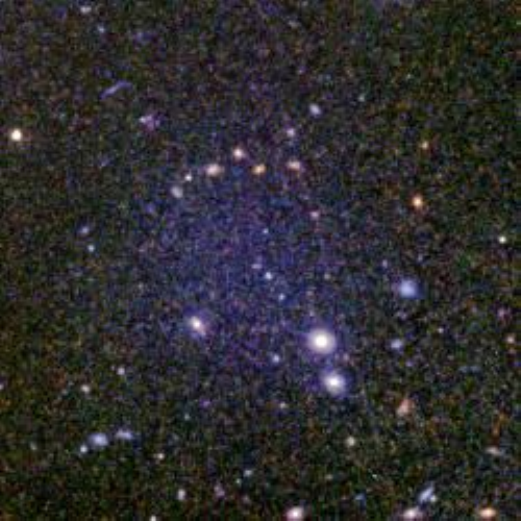} 
  \end{subfigure}
  \caption{Colour ($\IE$, $\YE$, $\HE$) images of two objects of interest. As they overlap with the galaxies with which they are associated, obtaining these images required new cutouts ($13.38\,\textnormal{kpc} \times 13.38\,\textnormal{kpc}$ or $150\arcsec \times 150\arcsec$), the use of \texttt{MTObjects} to mask sources, and the use of the \texttt{GALFIT} software \citep{2002AJ....124..266P} to determine and correct a tilted plane background. Finally, the colour image has been obtained with the \texttt{Aladin} software. \emph{Left}: External star-forming region (possibly belonging to a satellite dwarf irregular galaxy) at the north of NGC\,1553, labelled as (b) in Fig. \ref{features_colours}. \emph{Right}: Faint dwarf satellite galaxy at the north-west of NGC\,1549, labelled as (a) in Fig. \ref{features_colours}. In this figure, north is up, east is to the left.}
  \label{fig:dwarfs}
\end{figure}

\subsubsection{Dwarf galaxies of interest and isolated features}

In addition to the structures clearly associated with the studied galaxies, we note the presence of several features which are isolated or seem linked to some dwarf galaxies. This work presents the first mention of the following isolated features, demonstrating once again \Euclid's contribution to the detection of diffuse structures.

The tidal feature (2) of NGC\,1549, which is not mentioned in the literature, is located to the east of the galaxy's centre. It appears to be connected to the dwarf elliptical LSB galaxy (a) or [CMI2001]\,5012-01 in \cite{2001AJ....121..148C}. Located at ${\rm RA} = \ang{64,0125}$, ${\rm Dec} = \ang{-55,5570}$, this object is presented in the right panel of Fig. \ref{fig:dwarfs}. The stream (2) could be composed of material from the dwarf galaxy (a), which would therefore be undergoing tidal disruption by its host galaxy, NGC\,1549. However, the stream (2) exhibits colours that are not compatible within its uncertainties with those of the dwarf galaxy (labelled `a' in Fig. \ref{fig:colour-colour} and Table \ref{tab:features}). Therefore, this stream has a different metallicity than the dwarf (a), which seems to rule out the possibility of a link between them. Another hypothesis arises when considering the rather red colour of the dwarf galaxy (a) and the similarity of its GC colours to those of its host galaxy, NGC 1549 (seethe bottom-right panel of Fig. \ref{features_colours}). This could suggest that the dwarf galaxy (a) is not composed of low-metallicity material, but rather of material from a massive progenitor. This would make the dwarf galaxy (a) a tidal dwarf located at 16.43\,kpc from the centre of NGC\,1549.

Around NGC\,1553, we detect a  clumpy object (b) at ${\rm RA} = \ang{64,0645}$, ${\rm Dec} = \ang{-55,6977}$ (see left panel of Fig. \ref{fig:dwarfs}). Already mentioned in the catalogue of \cite{2001AJ....121..148C} as a blue LSB dwarf, this object was previously discussed in \cite{2020A&A...643A.176R}, where its detection in UV with the Galaxy Evolution Explorer survey (GALEX) and the presence of H\textsc{ii} regions were highlighted. 

In the \Euclid high-resolution image, its aspect and colour are consistent with a star-formation region. Moreover, Fig. \ref{fig:colour-colour} and Table \ref{tab:features} shows that the object (b) is compatible with low stellar ages and high metallicity. It could then be an outer star-forming region associated with NGC\,1553, or even a tidal dwarf galaxy given its distance to the host centre. Another hypothesis could be that it is an external irregular star-forming dwarf galaxy though finding such an object as a galactic satellite is rare. Spectroscopic data could help to better identify and classify this object.

At a projected distance of approximately 60\,kpc to the west of NGC\,1549, we detect a faint tidal feature (7). Its progenitor might be the dwarf galaxy LEDA\,75104 \citep{1987cspg.book.....A}, which is labelled as (d) Fig. \ref{features_colours}. 

Another faint tidal feature (3) is detected to the north of NGC\,1546. A dwarf galaxy (labelled `c', named SMDG\,J0415044$-$555618 and mentioned for the first time in \citealt{1987cspg.book.....A}) lies along its extension. However, the orientation of its semi-major axis, nearly perpendicular to the stream, challenges the hypothesis that it could be the progenitor of the tidal feature.

To the east of NGC\,1546 and to the south of NGC\,1553, one finds a feature (9) whose origin and nature are uncertain. Given its appearance as a detached tidal feature, we have classified this object as a `plume'. It could also be a tidally disrupting ultra-diffuse galaxy, similar to the case presented in \cite{2023MNRAS.518.2497Z}. In the bottom-right panel of Fig. \ref{features_colours}, we see that the GCs associated with this structure are red. This would be in favour of a tidal origin for this object, as it may have inherited its population of GCs from its progenitor. The flux photometry of this feature is limited by its non-detection in the NIR bands, likely due to contamination sources specific to this wavelength range in the ERO data (a topic that will be discussed in Sect. \ref{contaminants}). The GC photometry places the feature at the furthest position from the stellar population model in the colour-colour diagram (Fig. \ref{fig:colour-colour}), suggesting a complex star-formation history (or a high metallicity not taken into account by our SSP model) for this structure or its progenitor. A more refined stellar population synthesis model, enhanced data processing, and follow-up observations would enable a more in-depth investigation into its nature.

\section{\label{sc:Discussion}Discussion}

\subsection{\label{merger_history}Unveiling the mass assembly history of NGC\,1549, NGC\,1553, and NGC\,1546}

We  explore the origin of the tidal features around the three brightest galaxies analysed in previous sections. One prime driver is to determine whether they originate from ancient mergers or were more recently formed due to the on-going interactions of the massive galaxies in the ERO-D field.

The structures (1), (2), (3), and (7) are too thin for their progenitors to be dynamically hot massive objects, such as the ETGs discussed in this study. Even the larger structure (6) remains too thin to be a deformation of the NGC\,1549 halo. Furthermore, its direction does not point clearly toward any galaxy of the Dorado group. This rules out the possibility that this tidal feature is caused by a current interaction with a massive galaxy. We therefore argue that the tidal features (1), (2), (3), (6), and (7) are likely stellar streams, whose progenitors are lower mass galaxies that have been accreted. On the other hand, structures (4), (5), and (8) have shapes and sizes compatible with a halo disturbance caused by an ongoing interaction. We therefore tentatively classify them as plumes and tidal tails.

Besides the shapes of the structures, further clues on their origin may be derived from their colour. Stellar population models show that \Euclid photometric bands colours of stellar structures not affected by on-going star formation vary much more with metallicity than with age \citep{2025A&A...697A..13K,2025A&A...697A..10S,2025A&A...697A...9H}. Therefore, we make the assumption that the differences in colours between individual tidal features or in the profile of ETGs are primarily due to differences in metallicity.

The flat colour profile of NGC\,1549 suggests a relatively homogeneous mix of metal-rich and metal-poor stellar populations, possibly caused by a major merger \citep[e.g.][]{2009A&A...499..427D,2013ApJ...766..109K,2020ApJ...899L..26S} linked to the largest structure: the north tidal stream (6). Indeed, colour maps (Fig. \ref{fig:colour_maps}) show that it is redder than the halo of NGC\,1549. This argues for a massive progenitor, likely more metal-rich than the NGC\,1549 external parts. Since this galaxy exhibits the highest numbers of GCs, the progenitor of the large stellar stream (6) could have brought its GCs into the body of NGC\,1549.  

The stream (1) belonging to an umbrella-shaped feature is red. We can infer that this metal-rich stellar material comes either from the galactic centre of NGC\,1549 or from an external progenitor more massive than a dwarf galaxy.

\begin{figure}[h!]
  \centering
  \begin{subfigure}[b]{0.495\linewidth}
    \centering
    \includegraphics[width=\textwidth]{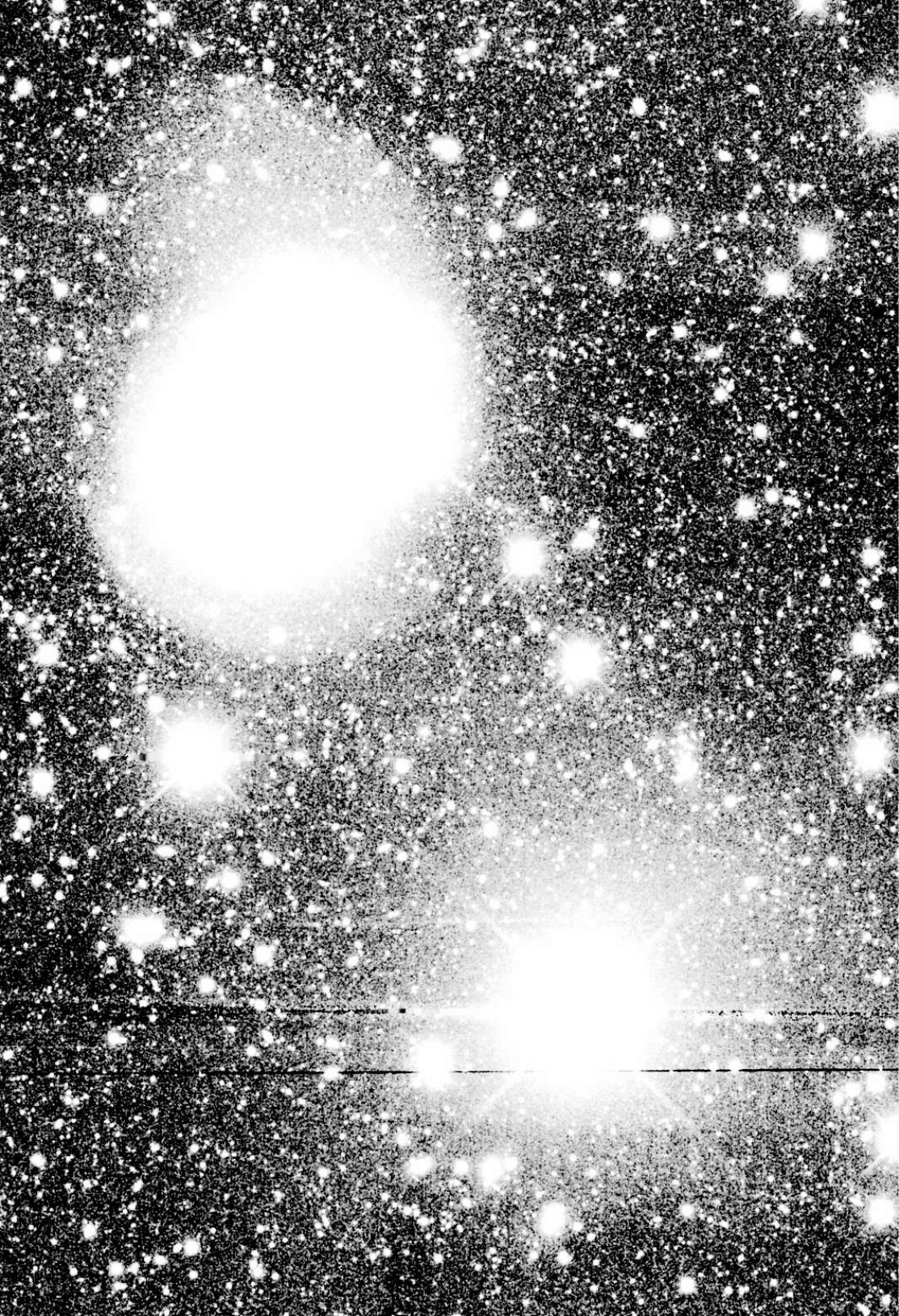}
  \end{subfigure}
  \hfill
  \begin{subfigure}[b]{0.495\linewidth}
    \centering
    \includegraphics[width=\textwidth]{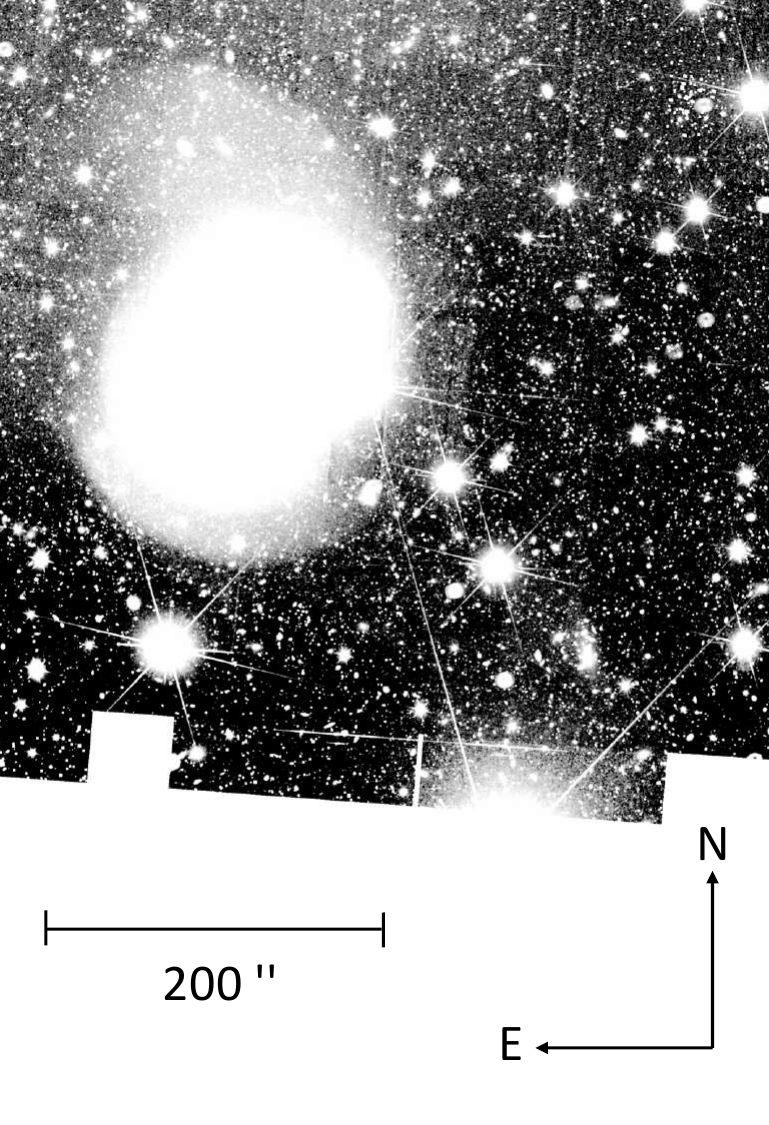}
  \end{subfigure}
  \caption{\emph{Left}: Example of possible false positive in diffuse structure detection with the DECaLS image. The light from a Milky Way star halo seems to form a tidal feature at the south of NGC\,1546. \emph{Right}: Same cutout for the \Euclid $\IE$ image.}
  \label{fig:DESI_Euclid_false_positive2}
\end{figure}

Given the previous considerations on the nature of the tidal features, the only remaining possible tidal features linked to a possible ongoing interaction between NGC\,1549 and NGC\,1553 are their shells and the NGC\,1553 plumes (4) and (8). However, (4) seems oriented towards NGC\,1546, and would therefore be caused by the gravitational attraction of the latter galaxy rather than that of NGC\,1553. As for the shells, simulations \citep[e.g.][]{2013arXiv1312.1643E,2018MNRAS.480.1715P} show that they could not have been produced by the mutual current interaction between NGC\,1553 and NGC\,1549, but rather by  past radial  mergers that occurred in both galaxies. 
A  tracer of a potential on-going interaction between NGC\,1549 and NGC\,1553 is possibly the plume (8) and the symmetric offsets in GC distribution peaks for both galaxies. This  clear lack of evidence of any interaction is expected for massive ETGs, whose structure tends to be stronger than the rotation-supported one of a less massive LTG. An ETG would consequently preferentially form  plumes, or even produce no tidal feature, rather than elongated bridges or tidal tails \citep[e.g.][]{2013LNP...861..327D}.

NGC\,1553 exhibits inner spiral arms as well as traces of possible current and ancient star-forming activity revealed by the multi-wavelength data (see Sect. \ref{sc:Intro} for references). This galaxy could be a former LTG whose star formation was quenched and that has recently transitioned to an ETG following a series of minor mergers, those that are responsible for the formation of the shells observed in its main body \citep[e.g.][]{2023MNRAS.518.3261P,2024MNRAS.529..810R}. 
This is consistent with the colour gradient of NGC\,1553 showing evidence of  material with lower metallicity and thus bluer colour in its outer regions \citep[e.g.][]{2011ApJ...740L..41L,2016A&A...593A..84K,2018A&A...617A..34M}. This last observation is in favour of more recent mergers, since phase-mixing has not yet blended stellar populations of different metallicities. An alternative scenario proposes that NGC\,1553 underwent a transformation into an ETG through mergers, resulting in the loss of its initial angular momentum, followed by the inward migration of material that formed an inner disk at its centre \citep[e.g.][]{2010MNRAS.402.2140S}. This interpretation is supported by the presence of the bar in the centre of NGC\,1553, which could have funnelled material inwards \citep[e.g.][]{2015A&A...584A..90G}. 
In the first scenario, the galaxy would remain a fast rotator, but not in the second. Therefore, a kinematic analysis of the inner and outer stellar populations with integral field units could help distinguishing the two scenarios \citep[e.g.][]{2016ARA&A..54..597C}.

Making a claim about a recent transition from LTG to ETG for NGC\,1549 would be more speculative: aside from the structure at its centre resembling diffuse spiral arms or a bar and visible only with unsharp masking (Fig. \ref{fig:center_NGC1549}), its morphology is that of a pure elliptical galaxy, and the literature \citep[e.g.][]{2020A&A...643A.176R,2021JApA...42...31R,2022A&A...664A.192R} reports evidence of past but not current star formation. However, the mergers responsible for the shells may still have contributed to the dissipative events that led to the cessation of this star formation. An ancient major merger associated to a violent relaxation and phase mixing of the stars would explain its flat colour profile.

NGC\,1546 appears as a hybrid object, consisting of an inner flocculent star-forming disk and a prominent outer diffuse stellar halo. The latter appears  perturbed, making a tidal tail (5) oriented towards  NGC\,1553. An interaction with this massive galaxy is thus the likely origin of this tidal feature. The absence of any form of disturbance in the inner young disk excludes the possibility of a recent major merger. Its diffuse stellar halo could have been assembled by multiple minor mergers, a scenario consistent with the presence of a population of blue GCs at its location. Nevertheless, objects like NGC\,1546 remain intriguing in groups.

The edge-on spiral IC\,2058, whose disc is of a similar size to that of NGC\,1546, shows no detectable stellar halo in \Euclid images. Surprisingly, no warp is observed either, despite its apparent proximity to the massive galaxy NGC\,1553. This challenges the hypothesis that the galaxy is already physically bound to the group \citep[e.g.][]{2002AJ....124.2471I}.

With this interpretation regarding the origin and nature of each tidal feature, we can return to the question of GC clustering for detecting such structures. We notice that GC distribution disturbances are only seen at the locations of suspected major interactions and mergers. Indeed, in the right-hand column of Fig. \ref{GC_distribution}, we observe a blue GC overdensity at the location of tidal feature (5), which is interpreted to indicate an ongoing major interaction, and offsets are detected around NGC\,1549 and NGC\,1553, which are suspected to be in interaction. Thus, although our results indicate that in these \Euclid images, studying GC clustering alone is not sufficient for detecting tidal features, we see that analysing the GC distribution is still valuable for identifying major interactions and mergers. To validate or refute this preliminary conclusion, GC clustering should be analysed on a statistical sample of galaxies with observable debris. Such a study will be made possible by the EWS data.

\subsection{Prospects for LSB detection in the EWS}

Due to the contaminants described earlier, the gain of \Euclid in terms of surface brightness limit is not obvious when comparing the ERO-D fields and DECaLS fields. However, the DECaLS image also suffers from contamination sources, such as CCD gaps, halos from bright stars, larger PSF, and higher sky background. This can lead to false positives in the search for tidal features, as seen in the example of Fig. \ref{fig:DESI_Euclid_false_positive2}. Since the two instruments are not affected by the same systematic effects, a suspected tidal feature in the DECaLS image can be confirmed or rejected in the \Euclid image and vice versa, enabling extremely robust diffuse structure detection. \Euclid's optics are key in this study, as its sharp PSF does not spread Milky Way star light over large radial distances. Moreover, the superior resolution of \Euclid images makes them better suited for determining the nature of the detected tidal features, and in particular the smallest ones (the PSF FWHM is $\ang{;;0.16}$ for \Euclid $\IE$ as explained in \citealt{2024arXiv240513496C}, and $\ang{;;0.82}$ for DECaLS $i$). The DESI Legacy survey, which covers a large portion of the EWS, could serve as a complement to the first \Euclid data releases for a statistical study of tidal features. For the southern hemisphere, they will eventually be replaced by the Legacy Survey of Space and Time (LSST) data from the \textit{Vera Rubin} Observatory.

The average surface brightness of the faintest feature detected in ERO-D (that is to say $28.00\pm0.01\,\text{mag\,arcsec}^{-2}$ in $\IE$ for the plume 8) does not reflect the intrinsic detection capabilities of \Euclid, but rather specific limitations of the ERO-D data. Indeed, we do not detect the features that are fainter than the stray light. This result is still highly promising, as even with the inherent limitations of the ERO-D data set, \Euclid reaches performance comparable to the one of the deepest surveys currently available. Additionally, \Euclid also observes in the NIR, a wavelength range where the LSB realm remains unexplored.

Finally, it is worth noticing that the ERO pipeline used for ERO-D data processing is different from the EWS pipeline. Indeed, the ERO pipeline is optimised for LSB signal preservation on extended objects (even when filling most of the \Euclid FoV, \citealt{2024arXiv240513496C}), whereas the EWS pipeline aims at producing excellent photometry of distant, and therefore small, sources. The methods used, particularly for background subtraction, could therefore vary from the ERO to the EWS pipelines, and the same applies to performance in terms of LSB signal preservation. Comparisons of EWS images to data from ground-based facilities could thus end up showing differences. A systematic study of the effect of the EWS pipeline on LSB objects is necessary for more accurate predictions of tidal features detection limits.

\section{\label{Summary}Summary and conclusions}

\begin{figure*}[htbp!]
        
\includegraphics[width=\textwidth,trim=0cm 0cm 0cm 0cm, clip]{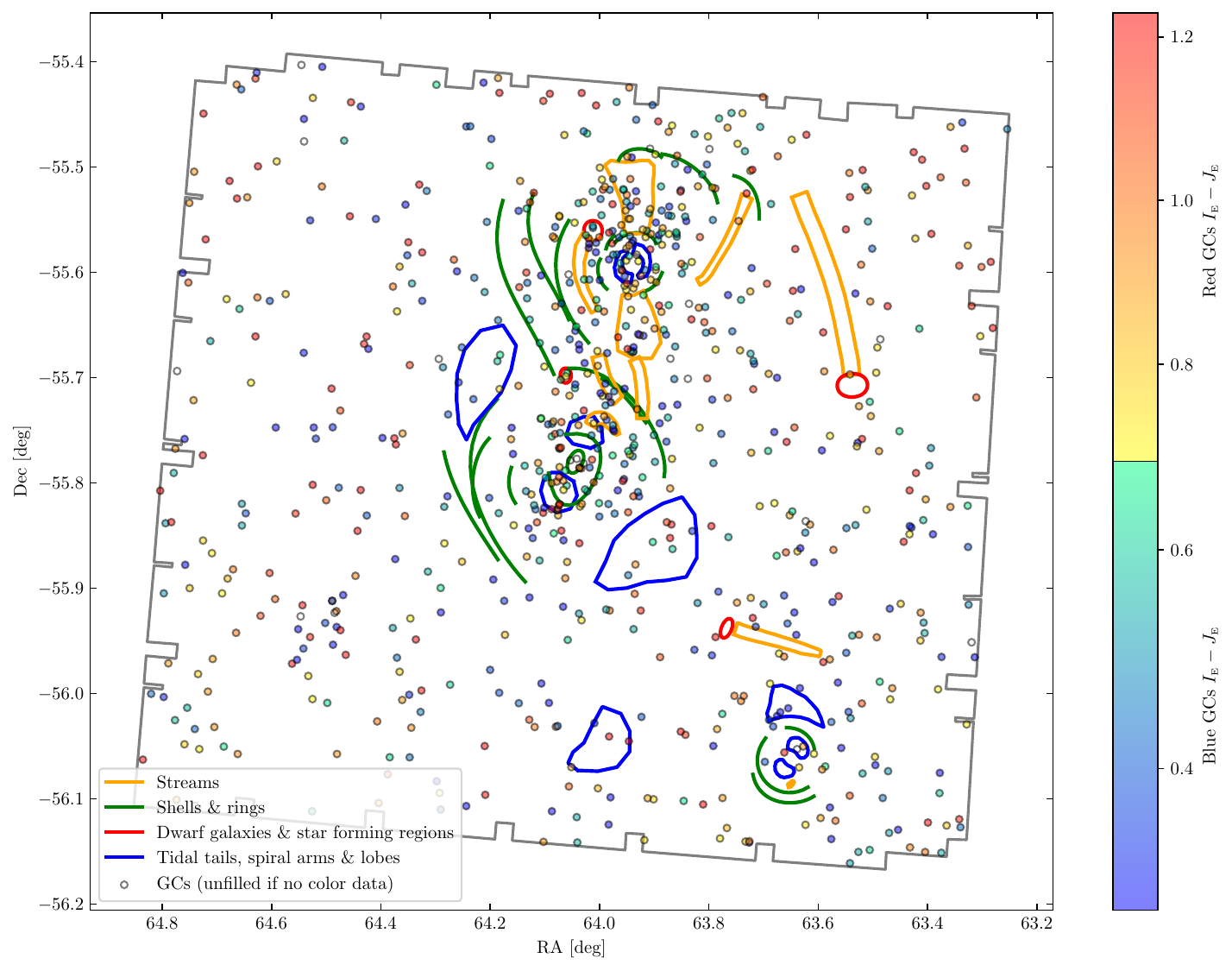}

\caption{
Distribution of the bright GC candidates in the ERO-D field. Each dot is a GC candidate and its colour is scaled  according to its $\IE-\HE$ colour. Outlines of tidal and internal stellar features have been overlaid on the plot.}
\label{distrib_and_features}
\end{figure*}

The \Euclid space telescope provides us with a groundbreaking view of four Dorado group galaxies, NGC\,1549, NGC\,1553, NGC\,1546, and IC\,2058. This image reveals an exceptional environment, rich in past collisions, and potential ongoing interactions. ERO-D allows us to map out an unprecedentedly exhaustive description of their tidal debris systems. The detection of the morphological structures and GCs of these galaxies summarised in Fig. \ref{distrib_and_features} alongside their photometry led to a series of conclusions regarding the mass assembly history of the compact group SCG\,0414$-$5559 formed by these galaxies.

\begin{itemize}
\item NGC\,1549 is an elliptical galaxy surrounded by a wealth of tidal features best explained by a past major merger event. This hypothesis is consistent with the flat colour profile of its stellar halo. An LSB dwarf galaxy is observed in its direction. However, its association with a nearby detected stream appears unlikely due to their difference in colours.
\item NGC\,1553 has a star-forming structure and spiral arms in the centre and innermost ring. It has probably recently transitioned to the ETG regime due to radial  minor mergers, as evidenced by the surrounding shells.
\item NGC\,1546 is a rare specimen of a non-perturbed star-forming disk galaxy with a large disturbed halo. The colour of its tidal tail and its GCs suggest that this halo has been fed by dwarf accretions, then distorted probably due to tidal forces caused by NGC\,1553.
\item IC\,2058 is a pure disk galaxy which does not exhibit any tidal features or other morphological signs of membership to this compact group.
\item The ERO-D data also revealed more isolated features: two stellar streams and a structure that could be the bright end of a larger tidal feature from NGC\,1553 or a tidally disrupted ultra-diffuse galaxy.

\end{itemize}

The ERO-D data were acquired from a single ROS, as will be the case for each stack in the EWS. Hence, this work can be considered as a pathfinder for the EWS researches on diffuse features for galactic merger history studies. We have thus drawn several conclusions regarding the detection in the ERO-D data set and prospects for the EWS, presented below.

\begin{itemize}

\item Several sources of contamination have been identified. In particular, stray light has an impact on the detection of extended and faint stellar structures in the $\IE$ band, while persistence in the NISP detectors introduces additional uncertainties in the photometry of their counterparts in the NIR regime.

\item In the ERO-D image, we have been able to detect tidal features, reaching $\IE\approx28\,\text{mag\,arcsec}^{-2}$. This is comparable to the deepest ground-based wide surveys (e.g. DESI Legacy and the future LSST), which will complement \Euclid data but are affected by different systematic effects. Provided that the aforementioned data limitations are managed and the LSB signal is preserved along the EWS image processing pipeline, EWS data are poised to detect fainter features with better photometry.

\item The GCs detected by \Euclid, their distribution, and their colour within each feature provide additional information that is useful for interpreting the nature of the structures and their progenitors. However, GC clustering alone does not enable the detection of tidal features in a systematic way. Exceptions can be found in cases involving major mergers, with a GC overdensity detected in the same location as a tidal tail, along with GC distribution peak offsets for two galaxies involved in a possible interaction.

\end{itemize}

With images matching the sensitivity of the deepest wide surveys currently available and providing superior resolution, \Euclid enables detailed characterisations of tidal features in Local Universe galaxies (including the smallest structures) and  studies of their GC populations. The EWS will allow for those studies to be extended to statistical samples on the visible and NIR extragalactic sky.

\begin{acknowledgements}
The authors are grateful to the anonymous referee for their constructive feedback, which significantly improved the manuscript.
\AckEC  
\AckERO
\cite{EROcite}

This research has made use of the SIMBAD database (\citealp{simbad}), operated at CDS, Strasbourg, France, the VizieR catalogue access tool (\citealp{2000A&AS..143...23O}), CDS, Strasbourg, France (DOI : 10.26093/cds/vizier), and the Aladin sky atlas (\citealp{aladin1,aladin2}) developed at CDS, Strasbourg Observatory, France and SAOImageDS9 (\citealp{ds9}). This work has been done using the following software, packages and \textsc{python} libraries: \textsc{Numpy} (\citealp{numpy}), \textsc{Scipy} (\citealp{scipy}), \textsc{Astropy} (\citealp{astropy:2018}), \textsc{Scikit-learn} (\citealp{scikit-learn}). The authors are grateful for financial support under the grant WIDERA ExGal-Twin, GA 101158446. E. Sola is grateful to the Leverhulme Trust for funding under the grant number RPG-2021-205. O. Müller is grateful to the Swiss National Science Foundation for financial support under the grant number PZ00P2\_202104. This research was supported by the International Space Science Institute (ISSI) in Bern, through ISSI International Team project \#534. This research makes use of ESA Datalabs (\citealt{Navarro2024}, \url{datalabs.esa.int}), an initiative by ESA’s Data Science and Archives Division in the Science and Operations Department, Directorate of Science.
\end{acknowledgements}

\bibliographystyle{aa}
\bibliography{aa53534-24}

\end{document}